\documentclass[a4paper,usenatbib]{mnras}
\usepackage{newtxtext,newtxmath}
\usepackage[T1]{fontenc}
\usepackage{ae,aecompl}
\usepackage{graphicx}
\usepackage{amsmath}
\usepackage{amssymb}
\usepackage{xcolor}
\usepackage{xspace}

\newcommand{\Te}{{T_{\rm e}}}
\newcommand{\keV}{{\rm keV}}
\newcommand{\pot}[2]{{{#1}\times10^{#2}}}
\newcommand{\Planck}{{\it Planck}\xspace}

\newcommand{\topic}[1]{\vspace{0.75mm}{\noindent{\it #1}}}

\title[SZ temperature scalings]{Relativistic SZ temperature scaling relations of groups and clusters derived from the BAHAMAS and MACSIS simulations}

\author[E. Lee et al.]{
Elizabeth Lee,$^{1}$\thanks{E-mail: elizabeth.lee-2@postgrad.manchester.ac.uk}
Jens Chluba,$^{1}$
Scott T. Kay$^{1}$
and David J. Barnes$^{1,2}$
\\
$^{1}$Jodrell Bank Centre for Astrophysics, Department of Physics and Astronomy, The University of Manchester, Manchester M13 9PL, UK\\
$^{2}$Department of Physics, Kavli Institute for Astrophysics and Space Research, Massachusetts Institute of Technology, Cambridge, MA 02139, USA
}

\date{\vspace{-3mm}Accepted 2020 February 11. Received 2020 February 11; in original form 2019 December 17}
\pubyear{2019}

\begin{document}
\label{firstpage}
\pagerange{\pageref{firstpage}--\pageref{lastpage}}
\maketitle

%---------------------------------------------------------------------------------------
\begin{abstract}
The Sunyaev-Zeldovich (SZ) effect has long been recognized as a powerful cosmological probe. Using the BAHAMAS and MACSIS simulations to obtain $>10,000$ simulated galaxy groups and clusters, we compute three temperature measures and quantify the differences between them. 
The first measure is related to the X-ray emission of the cluster, while the second describes the non-relativistic thermal SZ (tSZ) effect. The third measure determines the lowest order relativistic correction to the tSZ signal, which is seeing increased observational relevance.
Our procedure allows us to accurately model the relativistic SZ (rSZ) contribution and we show that a $\gtrsim 10\%-40\%$ underestimation of this rSZ cluster temperature is expected when applying standard X-ray relations. The correction also exhibits significant mass and redshift evolution, as we demonstrate here.
We present the mass dependence of each temperature measure alongside their profiles and a short analysis of the temperature dispersion as derived from the aforementioned simulations. We also discuss a new relation connecting the temperature and Compton-$y$ parameter, which can be directly used for rSZ modelling. Simple fits to the obtained scaling relations and profiles are provided.
These should be useful for future studies of the rSZ effect and its relevance to cluster cosmology.
\end{abstract}
%---------------------------------------------------------------------------------------

%---------------------------------------------------------------------------------------
\section{Introduction}
%---------------------------------------------------------------------------------------
Galaxy clusters constitute some of the largest structures in our Universe, forming from the highest overdensities of the cosmic web. This makes them excellent probes for cosmology, sensitive to fundamental cosmological parameters, such as the matter density and power spectrum \citep[e.g.,][]{Voit2005,Allen2011,Kravtsov2012,Weinberg2013}, as well as interesting in their own right. These clusters, for our purposes, can be considered as giant pockets of hot ionized plasma, which induce X-ray emission through both bremsstrahlung and line-emission processes \citep[see e.g.][for a review]{Sarazin1986}. They are also observable through the Sunyaev-Zeldovich (SZ) effect \citep{Zeldovich1969,Sunyaev1970}, a unique spectral signature caused by the upscattering of photons from the Cosmic Microwave Background (CMB) by free electrons with temperatures of $\Te\gtrsim 10^7$ K (i.e., $\gtrsim 1$ keV).\footnote{In fact, X-rays can be induced already by plasmas at $\Te\gtrsim 10^5$ K.} 
For reviews of the SZ effect see e.g. \citet{Carlstrom2002} and \citet{Mroczkowski2018}. 

Galaxy clusters comprise of giant dark matter haloes in which baryonic plasma is located -- while some of this gas cools to form galaxies, the majority remains as ionized plasma \citep{Briel1992}, also known as the intracluster medium (ICM). The ICM is often modelled as an isothermal sphere of electrons, allowing for simple mass-temperature relations to be derived. However, both direct measurements and hydrodynamical simulations indicate that clusters are neither isothermal nor spherical \citep[e.g.,][]{Nagai2003, Vikhlinin2009}. As such, instead of directly obtaining the thermodynamic temperature, we obtain volume-averaged temperatures, weighted according to the physical process they derive from. It thus becomes necessary to understand the appropriate weighting of each observable. In particular, it has long been established \citep{Pointecouteau1998,Hansen2004,Kay2008} that X-ray and SZ measurements do give rise to different temperatures once realistic cluster atmospheres are being considered.

The SZ distortion is dominated by the thermal SZ (tSZ) signal \citep{Zeldovich1969}, which gives rise to a redshift and temperature independent spectrum. However, relativistic corrections at typical cluster temperatures lead to both a broadening and drop in magnitude of this signal at fixed $y$-parameter. 
The rSZ effect is caused by the fact that for typical cluster temperature $k\Te \simeq 5\,\keV$ (for mass $M\simeq \pot{3}{14}\,M_{\odot}\,h^{-1}$) the electrons move at a fair fraction of the speed of light ($\varv/c\simeq 0.1-0.2$). In this case, the classical non-relativistic tSZ formula \citep{Zeldovich1969} is no longer sufficient, and higher order temperature corrections become relevant \citep{Challinor1998, Itoh1998, Sazonov1998}. The rSZ effect can be efficiently modelled using {\tt SZpack} \citep{Chluba2012, Chluba2013}; however, accurate estimates for the $y$-weighted temperature are required.
The $y$-weighted temperature is also relevant to precise computations of the SZ power spectra and the interpretation of SZ data from \Planck, as rSZ can cause biases in cosmological parameters such as $\sigma_8$ \citep{Remazeilles2018}. 

In this paper, we examine the differences between three temperature measures; the first is a proxy for the observed X-ray temperature $T_\mathrm{sl}$ \citep[the so-called spectroscopic-like temperature;][]{Mazzotta2004}; the second is a proxy for the Compton-$y$ parameter, $T^m$ (i.e., the mass-weighted temperature), which is closely related to the line-of-sight pressure and Compton-$y$ parameter through a cluster; the third is a measure that allows us to account for the relativistic temperature correction to the tSZ distortion (rSZ), $T^y$ \citep[the $y$-weighted temperature;][]{Hansen2004, Remazeilles2018}. The latter in particular so far has not been studied systematically; the current analyses, both from individual cluster measurements \citep{Hansen2002, Hansen2004, Prokhorov2012, Zemcov2012, Chluba2013} and stacking procedures \citep{Hurier2016,Erler2018}, provide only one measurement of rSZ with significance greater than 3$\sigma$ \citep{Zemcov2012} by relying on the assumption that kSZ is negligible. However, due to the growing sensitivity of planned an ongoing CMB experiments, rSZ is now coming into reach, and future observations with the Simons Observatory and CCAT-prime ought be able to extract this signal more accurately.

Precise SZ power spectrum calculations furthermore depend directly on the clusters' average pressure and $y$-weighted temperature profiles. Cluster pressure profiles have been extensively studied using simulations \citep[e.g.,][]{Nagai2003, Battaglia2010, Battaglia2012} and also have been calibrated against X-ray observations \citep{Arnaud2010, Planck2013}.
The $y$-weighted temperature profiles again have not been studied directly but will affect the precise shape of the relativistic temperature power spectrum \citep{Remazeilles2018}, which could become a novel cluster observable \citep{Basu2019, Remazeilles2019} for future CMB missions similar to {\it CORE} \citep{Melin2018} and {\it PICO} \citep{Hanany2019}.
Here, we carry out a comparative study of various temperature profiles with a particular focus on obtaining a new prescription of the $y$-weighted temperature profiles.

We base our study on BAHAMAS \citep{McCarthy2017,McCarthy2018} and MACSIS \citep{Barnes2017}, two giant hydrodynamical simulations generating over 14,000 haloes of mass $\simeq 10^{13}$ $M_\odot$ to $4 \times 10^{15}$~$M_\odot$ with outputs at redshifts of $z=0$, 0.5, and 1. These allow us to generate temperature--mass relations for each of our temperature measures as well as a detailed understanding of their temperature profiles. To concur with the work of \citet{Barnes2017}, we also consider the effects of restricting our analysis to only the hot and relaxed subsets of clusters within our samples. We will, however, find little change in the conclusions for these cases.

Moreover, since clusters are not isothermal, further corrections to the observed SZ signal must arise. This comes from the understanding that the distortions are caused by electrons of varying temperature along the line of sight, and thus will not be completely modelled by a single temperature. The first corrections to the signal can be found through a temperature moment expansion \citep{Chluba2012,Chluba2013}
and is related to the dispersion of $y$-weighted temperatures within clusters, which we study systematically here. Our results suggest that this dispersion scales at around $\simeq 40$ per cent of the cluster temperature, but overall leads to negligible corrections to the rSZ signal (see Sect.~\ref{sec:SZ_dispersion}).
Finally, we will briefly discuss the relevance of rSZ to determinations of $H_0$ through SZ measurements \citep{Cavaliere1979, Birkinshaw1991, Hughes1998, Reese2002}, showing that it could lead to a systematic shift in the derived $H_0$ values if rSZ is neglected.

The paper is structured as follows: we clarify the mathematical meaning behind each of the considered temperature measures and their purposes in Sect.~\ref{stn:Temperature_Weightings} and describe the simulations used in Sect.~\ref{stn:Simulations}. In Section~\ref{stn:Cluster_Bias}, we discuss the cluster-averaged temperature measures and in Sect.~\ref{stn:Profiles} the profiles across the clusters follow. Finally, we discuss the effects of these temperature measures on common observables in Sect.~\ref{stn:Discussion} and conclude in Sect.~\ref{stn:Conclusion}.

\vspace{-4mm}
%---------------------------------------------------------------------------------------
\section{Definition of the Temperature Measures}
\label{stn:Temperature_Weightings}
%---------------------------------------------------------------------------------------
In this section, we discuss cluster masses and self-similar redshift scaling relations and how they are related to simulation quantities. We will then describe the formulations of our three temperature measures; the spectroscopic-like temperature, a proxy for X-ray temperatures, and the mass and $y$-weighted temperatures, both related to the SZ effect. The mass-weighted temperature will be seen to be a proxy for the integrated electron pressure, or the Compton-$y$ parameter, within clusters, while the $y$-weighted temperature characterizes the precise shape of the SZ distortion. Finally, we will discuss the higher order $y$-weighted temperature moments, and their relationship to the observed SZ signals.

\vspace{-3mm}
\subsection{Formalism}
%---------------------------------------------------------------------------------------
In general, we can define our dark matter haloes to be spheres colocated with the cluster such that the total mass $M_\Delta$ contained within a radius $R_\Delta$ is given by
\begin{equation}
    M_\Delta = \frac{4}{3}R_\Delta^3\;\Delta\;\rho_\mathrm{crit}(z).
\end{equation}
Here, $\rho_\mathrm{crit}$ is the critical density for a flat universe. We set $\Delta=500$ for our main analysis (although a discussion for $\Delta=200$ is presented in Appendix \ref{App:FitValues}). In such a halo, for an isothermal sphere of gas, we can find the temperature to be 
\begin{equation}
	k_\mathrm{B} T_\Delta = \frac{G M_\Delta \mu m_\mathrm{p}}{2 R_\Delta}.
\end{equation}
This is equivalent to the virial temperature and can be used as a reference. As usual, $G$ is the gravitational constant, $m_\mathrm{p}$ the proton mass and $\mu$ the mean molecular weight of the plasma\footnote{The mean molecular mass is set to $\mu=0.59$.}. We note that in this work we always use the true (simulation) mass of clusters rather than any proxy for the observed mass,  \citep[e.g., the hydrostatic mass used in][]{Barnes2017}, which may introduce biases.

\topic{Scaling relations}: Assuming self-similarity, cluster temperatures are a simple function of their mass 
and redshift \citep{Kaiser1986}. We can recall that the critical density of the Universe is
%---------------------------------------------------------------------------------------
\begin{equation}\begin{split}
    \rho_\mathrm{crit} &\equiv \frac{3 H_0^2}{8\pi G} E^2(z)\\
    E(z) &\equiv \frac{H(z)}{H_0} = \sqrt{\Omega_\mathrm{m}(1+z)^3 + \Omega_\Lambda}.
\end{split} \end{equation}
%---------------------------------------------------------------------------------------
Here, $H_0$ is the Hubble constant and the exact form of $E(z)$ is cosmology dependent. From this, a simple geometrical consideration and an assumption of isothermality in the viral sphere gives us that
%---------------------------------------------------------------------------------------
\begin{equation} \label{eqn:TheoryScaling} \begin{split}
    M_\Delta &\propto E^2(z) R_\Delta^3\\
    T_\Delta &\propto E^{2/3}(z) M_\Delta^{2/3}.
\end{split} \end{equation}
%---------------------------------------------------------------------------------------
\topic{Temperature measures}: Since clusters are not isothermal, we must instead define weighted averaged temperatures appropriate to each observable (i.e., X-ray, SZ and rSZ effect). That is,
\begin{equation}
    \langle T \rangle \equiv \frac{\int w T\mathrm{d}V}{\int w \mathrm{d}V},
\end{equation}
where, as we will discuss in the rest of this section, it has been found that for spectroscopic-like, mass-weighted and $y$-weighted temperatures we have $w=n^2\,T^{-\alpha}$ \citep{Mazzotta2004}, $n$ and $n\,T$, respectively, and $\alpha\simeq 0.75$.

\topic{Connection to simulations}: To find all aforementioned quantities from our simulations we must discretize this process. We first ignore all particles with a temperature lower than $10^{5.2}$~K as they make a negligible contribution to the total X-ray or SZ emission \citep[cf.,][]{Barnes2017}. We can then convert our weighted volume integrals, to weighted sums, recalling that $\mu m_\mathrm{p}\,n\,\mathrm{d}V = \mathrm{d}m$.
With this procedure we can compute the various temperature measures discussed below.

\vspace{-3mm}
\subsection{X-ray Temperatures}
%---------------------------------------------------------------------------------------
X-ray emission, from hot clusters ($k_\mathrm{B}T\gtrsim3$~keV),\footnote{This cut-off is in large part due to the the dominance of emission lines rather than bremsstrahlung in the observed X-ray spectra below these temperatures.} is primarily caused by bremsstrahlung radiation within the ICM, and as such has classically been modelled by the emission-weighted temperature. This can be motivated from a simple consideration of the X-ray surface brightness,
%---------------------------------------------------------------------------------------
\begin{equation}
	S_x = \frac{1}{4\pi (1+z)^3}\int n^2(l)\Lambda_{\rm ee}(T(l),Z)\,\mathrm{d}l.
	\label{eq:SX}
\end{equation}
%---------------------------------------------------------------------------------------
Here, $n(l)$, $T(l)$ are the electron density and temperature along line of sight $l$ and $\Lambda_\mathrm{ee}(T,Z)$ is the X-ray emissivity measured by the instrument within the energy band used for the observation; $z$ is the clusters redshift and $Z$ is the metallicity of the ICM.

\topic{Spectroscopic-like temperature}: It has been shown that, due to the non-isothermality of the gas, it is more appropriate to use a modified weighting determined by fitting the X-ray spectrum with a thermal emission model \citep{Mazzotta2004,Vikhlinin2006}. This leads to the spectroscopic-like temperature,
%---------------------------------------------------------------------------------------
\begin{equation}
	T_\mathrm{sl} \equiv \frac{\int n^2\;T^{1-\alpha}\mathrm{d}V}{\int n^2\;T^{-\alpha} \mathrm{d}V}
	\label{eq:TX_sl}
\end{equation}
%---------------------------------------------------------------------------------------
where $\alpha \simeq 0.75$. When compared to observational results, this matches well with data from both {\it Chandra} and {\it XMM-Newton}, provided the temperatures are all sufficiently high, e.g., $k T\gtrsim3.5$ keV. Hydrodynamical simulations have been used to calibrate $T_\mathrm{sl}$ to the observed `X-ray' temperatures and confirm the differences between various X-ray derived temperatures weightings \citep[e.g.,][]{Mathiesen2001,Rasia2014,Biffi2016}.

We also can see that both measures, Eq.~\eqref{eq:SX} and \eqref{eq:TX_sl}, lead to an $n^2$ dependence in the X-ray temperature measurements -- and in general a higher weighting of cooler, denser gas. This indicates (see Sect.~\ref{stn:exploration} and \ref{stn:Profiles}) that the X-ray measurements are far poorer probes of the outskirts of clusters, where the electron density drops significantly, compared to the SZ measurements, which we will see has a linear dependence on $n$. This has also been seen observationally since it requires very long exposures to observe the outskirts of clusters through X-ray emission \citep[e.g.,][]{Simionescu2011}.

%---------------------------------------------------------------------------------------
\vspace{-3mm}
\subsection{SZ Temperatures}
\label{stn:SZFormalism}
%---------------------------------------------------------------------------------------
\topic{Mass-weighted temperature}: The classical tSZ effect gives rise to an intensity distortion that can be written in terms of the Compton-$y$ parameter as \citep{Zeldovich1969}:
%---------------------------------------------------------------------------------------
\begin{equation}
	\Delta I_{\nu} \approx I_0 y \frac{x^4 {\rm e}^x}{({\rm e}^x-1)^2} \left(x\frac{{\rm e}^x+1}{{\rm e}^x-1}-4 \right)\equiv I_0 y g(x).
\end{equation}
%---------------------------------------------------------------------------------------
The spectral function $g(x)$ for the tSZ effect is defined here implicitly. To characterize the photon energy we use $x = h\nu /k_\mathrm{B} T_\mathrm{CMB}$, where $T_\mathrm{CMB}$ is the temperature of the CMB, $k_\mathrm{B}$ the Boltzmann constant. The intensity normalization constant furthermore is
%---------------------------------------------------------------------------------------
\begin{equation} \label{eqn:I_0}
	I_0 = \frac{2(k_\mathrm{B} T_\mathrm{CMB})^3}{(h c)^2} = 270.33 \left[\frac{T_\mathrm{CMB}}{2.7255\mathrm{K}}\right]^3 \mathrm{MJy/sr}.
\end{equation}
%---------------------------------------------------------------------------------------
The Compton-$y$ parameter, as previously mentioned, is directly related to the integrated electron pressure, $P_\mathrm{e}$, along the line of sight and is typically written as,
%---------------------------------------------------------------------------------------
\begin{equation} 
\label{eqn:Compton_y}
	y \equiv \int \frac{k_\mathrm{B} T}{m_\mathrm{e} c^2}\mathrm{d}\tau
	= \sigma_\mathrm{T} \frac{k_\mathrm{B}}{m_\mathrm{e} c^2}\int n T \mathrm{d}l
	= \frac{\sigma_\mathrm{T}}{m_\mathrm{e} c^2}\int P_\mathrm{e} \mathrm{d}l.
\end{equation}
%---------------------------------------------------------------------------------------
Here, $\tau$ is the Thomson scattering optical depth and all the other constants have their usual meaning.

The second equality in Eq.~\eqref{eqn:Compton_y} leads to an expression for the mass-weighted temperature, when we extend this formalism to a volume-averaged, rather than a line-of-sight, integral:
%---------------------------------------------------------------------------------------
\begin{equation} \label{eqn:Tm}
	T^m \equiv \frac{\int n T\mathrm{d}V}{\int n \mathrm{d}V} = \frac{\int T\mathrm{d}m}{\int\mathrm{d}m}.
\end{equation}
%---------------------------------------------------------------------------------------
Here $m$ now refers to the mass of the electron gas. We can see that the volume averaged Compton $y$ parameter is
\begin{equation} \label{eqn:Ydefn}
    Y = \sigma_\mathrm{T} \frac{k_\mathrm{B}}{m_\mathrm{e} c^2}\int n T \mathrm{d}V \propto M\;T^m
\end{equation}
where $M$ denotes the total gas mass, i.e., Y is the total thermal energy of the gas.

\topic{$y$-weighted temperature}: Including the rSZ corrections, the SZ distortion is no longer temperature/mass independent and the dimensionless signal has to be expressed as $S(\nu) = \Delta I/I_0 = y  f(\nu, T_\mathrm{e})$. Since the temperature varies within each cluster, we can then consider a temperature moment expansion about some pivot temperature $\bar{T}_\mathrm{e}$ for each cluster, as detailed in \citet{Chluba2013} and \citet{Remazeilles2018}, to obtain
%---------------------------------------------------------------------------------------
\begin{equation} 
\label{eqn:mom_exp}
	S_{\ell m}(\nu) \simeq f(\nu, \bar{T}_\mathrm{e}) y_{\ell m} + f^{(1)} (\nu, \bar{T}_\mathrm{e}) y^{(1)}_{\ell m}+ \frac{1}{2} f^{(2)} (\nu, \bar{T}_\mathrm{e}) y^{(2)}_{\ell m}
\end{equation}
%---------------------------------------------------------------------------------------
to second order in $\Delta T_\mathrm{e} = T_\mathrm{e} - \bar{T_\mathrm{e}}$. For more generality, we have expressed $S(\nu)$ using spherical harmonic coefficients, introducing $y^{(k)}_{\ell m} = [\Delta T_\mathrm{e}^k y]_{\ell m}$, where $ [X]_{\ell m}$ denoted the spherical harmonic expansion of $X$. These will become relevant for the $y^2$-weighted temperature measure introduced below. Here, the derivatives of the SZ signal are $f^{(k)} = \partial^k_T f(\nu, T)$ and, if applied to isothermal clusters, one has $y^{\mathrm{iso},(k)}_{\ell m} = \Delta T_\mathrm{e}^k y_{\ell m}$.

Equation~\eqref{eqn:mom_exp} motivates the use of a pivot temperature that eliminates the first-order term in $S(\nu)$. We therefore introduce the $y$-weighted temperature, by requiring $\int y^{(1)}_{00}\mathrm{d}V = 0$, i.e.,
%---------------------------------------------------------------------------------------
\begin{equation} \label{eqn:Ty}
	T^y \equiv \frac{\int [T y]_{00}\mathrm{d}V}{\int y_{00}\mathrm{d}V}=      \frac{\int y T \mathrm{d}V}{\int y \mathrm{d}V}
	= \frac{\int n T^2 \mathrm{d}V}{\int n T \mathrm{d}V}.
\end{equation}
%---------------------------------------------------------------------------------------
Thus, setting $\bar{T}_\mathrm{e} = T^y$, removes the first-order correction to the volume average of $S_{00}$, yielding
%---------------------------------------------------------------------------------------
\begin{equation} 
\label{eqn:mom_exp-II}
\langle S_{00}(\nu)\rangle \simeq f(\nu, \bar{T}_\mathrm{e}) \langle y_{00}\rangle + \frac{1}{2} f^{(2)} (\nu, \bar{T}_\mathrm{e}) \langle y^{(2)}_{00} \rangle.
\end{equation}
%---------------------------------------------------------------------------------------
As shown below, even the second-order term can become relevant for our simulation clusters. This is consistent with previous studies \citep{Chluba2013}, but here we derive explicit scaling relations.

In \citet{Planck2016Map}, the assumption that $f(\nu, \bar{T}_\mathrm{e}) \simeq f(\nu, 0)$, or equivalently that the observed signals are well-modelled by the classical tSZ distortion, was used. However, in \citet{Remazeilles2018}, it was shown that due to rSZ this is insufficient. Relativistic corrections will lead to a lower amplitude of the SZ signal at fixed $y$-parameter as well as broadening of the SZ signal, which causes a miscalibration and underestimation of the true Compton-$y$ values for each cluster. In Sect.~\ref{stn:Ysz_M_rel}, we find that this results in a $\simeq 10-20$ per cent correction to the derived $y$-parameters for typical clusters, and thus is worth quantifying further.

\topic{Higher order temperature moments}: While using the $y$-weighted temperature removes the first-order correction to the SZ signal, higher order terms proportional to $y^{(k)}_{\ell m}$ remain. We thus define the volumetric $y$-weighted temperature moments as
%---------------------------------------------------------------------------------------
\begin{equation}
	T_k^y = \frac{\int \Delta T_\mathrm{e}^k y \mathrm{d}V}{\int y \mathrm{d}V} = \frac{\int y (T-T^y)^k \mathrm{d}V}{\int y \mathrm{d}V}.
\end{equation}
%---------------------------------------------------------------------------------------
From this we see that\footnote{In the work \citet{Chluba2013}, a different definition for the SZ temperature moments is used. First, they take the mass-weighted temperature moments $T_k^m$, so that their moments are weighted by $n \mathrm{d}V$ rather than $y \mathrm{d}V$. Furthermore, they use dimensionless moments $\omega^{(k)} = T_{k+1}^m/(T^m)^{k+1}$. In the limit of many moments, the definitions in terms of $T^m$ and $T^y$ are equivalent and yield the same result.} $T_0^y = 1$ and $T_1^y = 0$. While we could theoretically expand to arbitrarily many orders of $\Delta T$, in this paper we will consider only the lowest order correction, i.e., $T_2^y$. We can see that this is closely related to the intrinsic variance of the electron temperature within the cluster gas. To match the dimensionality of the $y$-weighted temperature, we will later discuss $\sigma(T^y) = (T_2^y)^{1/2}$ instead, which provides a proxy for the standard deviation of temperature variation within clusters. 

The higher order temperature moments further change the detailed shape of the SZ signal, and thus may cause additional biases to SZ measurements if omitted \citep{Chluba2013}.  We will see that from simulations this standard deviation is around $\simeq 40$ per cent of the cluster temperature (Sect.~\ref{stn:vol_yw_mom}); however, overall this is likely to only lead to a $\lesssim 0.5$ per cent correction in $y$ (see Sect.~\ref{sec:SZ_dispersion}).

\topic{Compton-$y$ power spectra}: As discussed in \citet{Remazeilles2019}, to correctly calculate the tSZ power spectrum, we need temperature profiles and in particular the $y$-weighted temperature profiles. They show that for the tSZ power spectrum one requires a $y^2$-weighted  or $C_\ell^{yy}$-weighted temperature as a pivot. This demands that for each multipole $\ell$, $\langle y_\ell^* y_\ell^{(1)}\rangle = 0$, for an isotropic homogeneous, spherical cluster. For an isothermal temperature profile for each cluster, this yields
%---------------------------------------------------------------------------------------
\begin{equation}
	k\bar{T}^{yy}_{\rm e,\ell} = \frac{\langle k T_\mathrm{e}(M,z)|y_\ell|^2\rangle}{\langle |y_\ell|^2\rangle}
	= \frac{C_\ell^{T_\mathrm{e},yy}}{C_\ell^{yy}}.
\end{equation}
%---------------------------------------------------------------------------------------
This assures only second-order terms in $\Delta T_\mathrm{e}$ remain in the theoretical tSZ power spectrum, $C_\ell^{tSZ}(\nu) \propto |y_{\ell m}|^2$. With the outputs from this work we can improve the calculation by using explicit temperature profiles and their Fourier transforms for the computation of the relativistic temperature power spectra.

\vspace{-4mm}
\section{Simulations}
\label{stn:Simulations}
\begin{table}
\caption{Cosmological parameters used in the BAHAMAS and MACSIS simulations.}
\begin{tabular}{l |@{}c |@{}c |@{}c |@{}c |@{}c |@{}c |@{}c }
    \hline
    Simulation & $\Omega_\Lambda$ & $\Omega_\mathrm{m}$ & $\Omega_\mathrm{b}$ & $\sigma_8$ & $n_\mathrm{s}$ &$h^\dagger$\\
    \hline
    BAHAMAS & 0.6825 & 0.3175 & 0.0490 & 0.8340 & 0.9624 & 0.6711 \\
    MACSIS  & 0.6930 & 0.3070 & 0.0482 & 0.8288 & 0.9611 & 0.6777 \\
    \hline \noalign{\smallskip}
    \multicolumn{7}{l}{$^\dagger$ where $h \equiv H_0/(100 \,\mathrm{km\,s}^{-1}\,\mathrm{Mpc}^{-1})$ }
\end{tabular}
\label{tab:SimulationCosmology}
\end{table}

We use a combined sample of clusters from the BAHAMAS and MACSIS simulations, both of which we explain in more detail below. From the BAHAMAS project \citep{McCarthy2017}, we obtain $> 14,000$ haloes with masses $M_{500} \geq 10^{13}$ $M_\odot$. However, these simulations provide a limited numbers of high mass clusters. These are supplemented by the compatible MACSIS project \citep{Barnes2017}, which generated 390 clusters with $M > 10^{15}$ $M_\odot$. The MACSIS simulations were designed to match the hydrodynamical properties of the BAHAMAS simulations and use compatible cosmologies (see Table \ref{tab:SimulationCosmology}).

We note that there is a small redshift discrepancy between the BAHAMAS sample at $z=0.5$ and the MACSIS sample at $z=0.46$. However, since the redshift dependence of our quantities are slight (as we discuss below) this requires no correction. Further we acknowledge there is a mismatch in cosmological parameters, however, we believe that this has little effect on our measured values, and again, is left unadjusted.

In this section, we highlight the key properties of these simulations and discuss how we combine the samples. We also discuss the subsamples used within the work of \citet{Barnes2017} for hot and relaxed clusters, and define the versions we will explore later in this paper. We also explain the core excision procedure used for X-ray observations and how it is recreated in simulations.

\vspace{-3mm}
\subsection{BAHAMAS simulation}
The BAHAMAS simulation \citep{McCarthy2017,McCarthy2018} is a calibrated version of the model used in the cosmo-OWLS simulations \citep{LeBrun2014}. Following this work, the BAHAMAS simulation consists of a 400 Mpc/$h$ periodic box. For the simulations used in this paper, the cosmological parameters used are consistent with those from \Planck 2015 \citep{Planck2016SZ}, and can be seen in Table \ref{tab:SimulationCosmology}. 
%The BAHAMAS simulations consist of both a Dark Matter only simulation and a baryonic simulation; the latter is the one used throughout this work and hereafter will be referred to as the `BAHAMAS simulation'.

The full run has $2\times 1024^3$ particles, yielding a dark matter mass of $m_\mathrm{DM} = 4.5\times10^9$ $M_\odot/h$ and initial baryon particle mass of $m_\mathrm{gas}=8.1\times10^8$ $M_\odot/h$. The Plummer equivalent gravitational softening length was fixed to 4~kpc/h in comoving units for $z > 3$ and in physical coordinates thereafter. The simulations were run with a version of the smoothed particle hydrodynamics code \texttt{\textsc{p-gadget3}}, which was last publicly discussed in \citet{Springel2005} but has since been greatly modified to include new subgrid physics as part of the ambitious OWLS project \citep{Schaye2010}. The feedback calibration was set to match the observed gas mass fraction of groups and clusters and galaxy stellar mass function at $z = 0$ \citep[see][for details]{McCarthy2017}.

\vspace{-3mm}
\subsection{MACSIS simulation}
As already mentioned, to extend the BAHAMAS simulations to higher mass haloes the MACSIS project  \citep[described in detail in ][]{Barnes2017} was developed. This entails a sample of 390 massive clusters. To obtain this number of massive clusters, with current computational resources, the MACSIS sample was generated using a zoomed simulation technique from a very large volume Dark Matter only simulation. This 'parent' simulation was a periodic cube with a side length of 3.2 Gpc. The cosmological parameters were taken from the \Planck 2013 results combined with baryonic acoustic oscillations, WMAP polarization and high multipole moments experiments \citep{Planck2014} and are summarised in Table \ref{tab:SimulationCosmology}.

The MACSIS sample was then selected by finding all haloes with a Friends-of-Friends (FoF) mass, $M_\mathrm{FoF} > 10^{15}\;M_\odot$, and grouping them into logarithmically spaced bins of width $\Delta \mathrm{log}_{10}(M_\mathrm{FoF}) = 0.2$. The bins with masses above $10^{15.6}$ $M_\odot$ had less than 100 haloes each and all were selected. The other bins were further subdivided, each into 10 logarithmic bins, from each of which 10 haloes were randomly selected -- this ensured the sample is not biased to low masses by the steep slope of the mass function.

These selected clusters were then re-simulated using the zoomed simulation technique \citep{Katz1993,Tormen1997} to recreate the chosen sample at an increased resolution compared to the parent simulation. Both a DM only and full gas physics resimulation was then carried out. The latter, which we use in this work,  had a dark matter mass of $m_\mathrm{DM} = 4.4\times10^9$ $M_\odot/h$ and gas particle inital mass of $m_\mathrm{gas}=8.0\times10^8$ $M_\odot/h$. The softening length was fixed as in the BAHAMAS simulation. The simulations were again run with the same version of the smoothed particle hydrodynamics code \texttt{\textsc{p-gadget3}}. The resolution and softening of the zoom re-simulations were deliberately chosen to match the values of the periodic box simulations of the BAHAMAS project. \citet{Barnes2017} further shows that the MACSIS clusters reproduce the observed mass dependence of the hot gas mass, X-ray luminosity and SZ signal at redshift $z = 0$ and $z=1$

\vspace{-3mm}
\subsection{Combined sample}
\begin{table}
\caption{Selected halo counts with $M_{500}>10^{13}$ $M_\odot$, and with a mass cut between the BAHAMAS and MACSIS samples at the given values. } \centering
\begin{tabular}{c | c c c}
    \hline
    Redshift & BAHAMAS & MACSIS & $M_\mathrm{500,cut}/M_\odot$\\
    \hline
    0        & 14333 & 295 & $ 6.03\times 10^{14}$ \\
    $0.5/0.46^a$ & 10791 & 263 & $ 3.55\times 10^{14}$ \\
    1        & 6344  & 186 & $ 2.00\times 10^{14}$  \\
    \hline \noalign{\smallskip}
    \multicolumn{4}{l}{$^a$ that is, 0.5 for BAHAMAS and 0.46 for MACSIS.}
\end{tabular}
\label{tab:RedshiftSampleSizes}
\end{table}

We combine these simulations to allow for clear comparison with the work in \citet{Barnes2017}, taking only haloes with $M_{500}>10^{13} M_\odot$. Further we take a mass cut at each redshift, as detailed in Table \ref{tab:RedshiftSampleSizes}, above which we take only MACSIS haloes and below which we take only BAHAMAS haloes. The final halo counts at each redshift are detailed there. Haloes are identified in both simulations through the friends-of-friends method described in \citet{McCarthy2017}. The centre of these haloes is taken to be the minimum of the local gravitational potential, and any sub-haloes lying outside a given characteristic radius, $R_\Delta$, are ignored.

\vspace{-3mm}
\subsection{Core Excised Averages}
It is a common technique in X-ray observations to exclude the central regions of clusters to reduce the scatter in X-ray properties. These core-excised quantities are often considered to be better mass proxies \citep{Pratt2009}. Within simulations, this can have an added effect of reducing the potential impact of the central (more uncertain) physics inside the cores. In the work of \citet{Barnes2017}, the excluded region is that of $r<0.15 R_{500}$.

Theoretically it would be possible to core excise all of our volume averaged quantities, not just the X-ray calculations. However, it can be seen that while $T_\mathrm{sl}$ has a large correction under core-excision -- raising the temperatures increasingly at higher masses, but undergoing a more complex increase across the entire mass range -- both $T^y$ and $T^m$ undergo very minimal modifications [the mean corrections are $(T_\mathrm{CE}-T_\mathrm{full})/T_\mathrm{CE}=-0.011\pm0.065$ and $-0.003\pm0.015$ for each measure respectively\footnote{These are the values for the volume average over $R_{500}$; over $R_{200}$ instead, arguably a more applicable volume for SZ measurements, these corrections reduce to $-0.010\pm0.048$ and $-0.004\pm0.009$ respectively.}].

In general, SZ measurements are flux and resolution limited and the full volume average is taken (since taking core excised values would be difficult in practice). We will thus use the full volume averages for $T^m$ and $T^y$, but the core excised values for $T_\mathrm{sl}$.

\vspace{-3mm}
\subsection{Hot and Relaxed Sub-samples}
\label{stn:HotRelaxed}
\begin{table}
\caption{Selected halo counts with $M_{500}>10^{13}$ $M_\odot$, and with a mass cut between the \textsc{bahamas} and \textsc{macsis} samples at the given values for the Hot and Relaxed samples. } \centering
\begin{tabular}{c | c c c c}
    \hline
    Redshift & \textsc{bahamas} & \textsc{macsis} & $M_\mathrm{500,cut}/M_\odot$ & $M_\mathrm{500,min}/M_\odot$\\
    \hline
    \multicolumn{4}{c}{Hot Sample}\\ 
    \noalign{\smallskip}
    0            & 271 & 295 & $6.03\times 10^{14}$ & $2.29\times 10^{14}$ \\
    $0.5/0.46^a$ & 87  & 263 & $3.55\times 10^{14}$ & $2.09\times 10^{14}$ \\
    1            & 4   & 186 & $2.00\times 10^{14}$ & $1.91\times 10^{14}$ \\
    \hline
    \multicolumn{4}{c}{Relaxed Sample}\\ 
    \noalign{\smallskip}
    0        & 165 & 188 & $6.03\times 10^{14}$ & $2.29\times 10^{14}$ \\
    $0.5/0.46^a$ & 50 & 178 & $3.55\times 10^{14}$ & $2.09\times 10^{14}$ \\
    1        & 3  & 126 & $2.00\times 10^{14}$ & $1.91\times 10^{14}$ \\
    \hline\noalign{\smallskip}
    \multicolumn{4}{l}{$^a$ that is, 0.5 for \textsc{bahamas} and 0.46 for \textsc{macsis}.}
\end{tabular}
\label{tab:HotRelaxedSampleSize}
\end{table}

As previously noted, the models for the X-ray temperatures, all rely on continuum emission, while at low cluster temperatures the effects of spectral lines begin to seriously affect the observed X-ray spectra. Accordingly, following the analysis of \citet{Mazzotta2004}, we note that the spectroscopic-like temperature is validated only for higher temperatures. This motivates the use of a Hot Sample, where $T_\mathrm{sl}$ is a more reliable proxy for the X-ray emission. To avoid biases, we introduce a mass cut by finding the minimal mass that fulfills $T_\mathrm{sl}(M)\geq 3.5\,\keV$ -- this ensures that the maximal temperature at a given mass is $T_\mathrm{sl}(M)\gtrsim 3.5\,\keV$.\footnote{In the work of \citet{Barnes2017}, they take the smaller sample of all clusters with $T_\mathrm{sl} >5$~keV} These cuts arise at $\mathrm{log}_{10}(M_{500})=14.36$, 14.32, 14.28 $M_\odot$ for $z=0$, 0.5 and 1 respectively, with the results summarized in Table~\ref{tab:HotRelaxedSampleSize}.

The final sample is a relaxed sub-sample of these Hot clusters. Although there are many ways to define a relaxed halo \citep[see e.g.][]{Neto2007,Duffy2008,Klypin2011,Dutton2014,Klypin2016,Barnes2017CEagle}, in this paper we follow the criteria used in \citet{Barnes2017}, that is
\begin{equation*}
	X_\mathrm{off} < 0.07;\; f_\mathrm{sub} < 0.1 \; \mathrm{and}\; \lambda < 0.07,
\end{equation*}
where $X_\mathrm{off}$ is the distance offset between the point of minimum gravitational potential in a cluster and its centre of mass, divided by its virial radius; $f_\mathrm{sub}$ is the mass fraction within the virial radius that is bound to substructures and $\lambda$ is the spin parameter for all particles within $R_{200}$. It should be noted that, as in \citet{Barnes2017}, this is not a small sample of the most relaxed objects, but instead a simple metric to remove those that are significantly disturbed.

\begin{figure*}
    \includegraphics[width=0.8\textwidth]{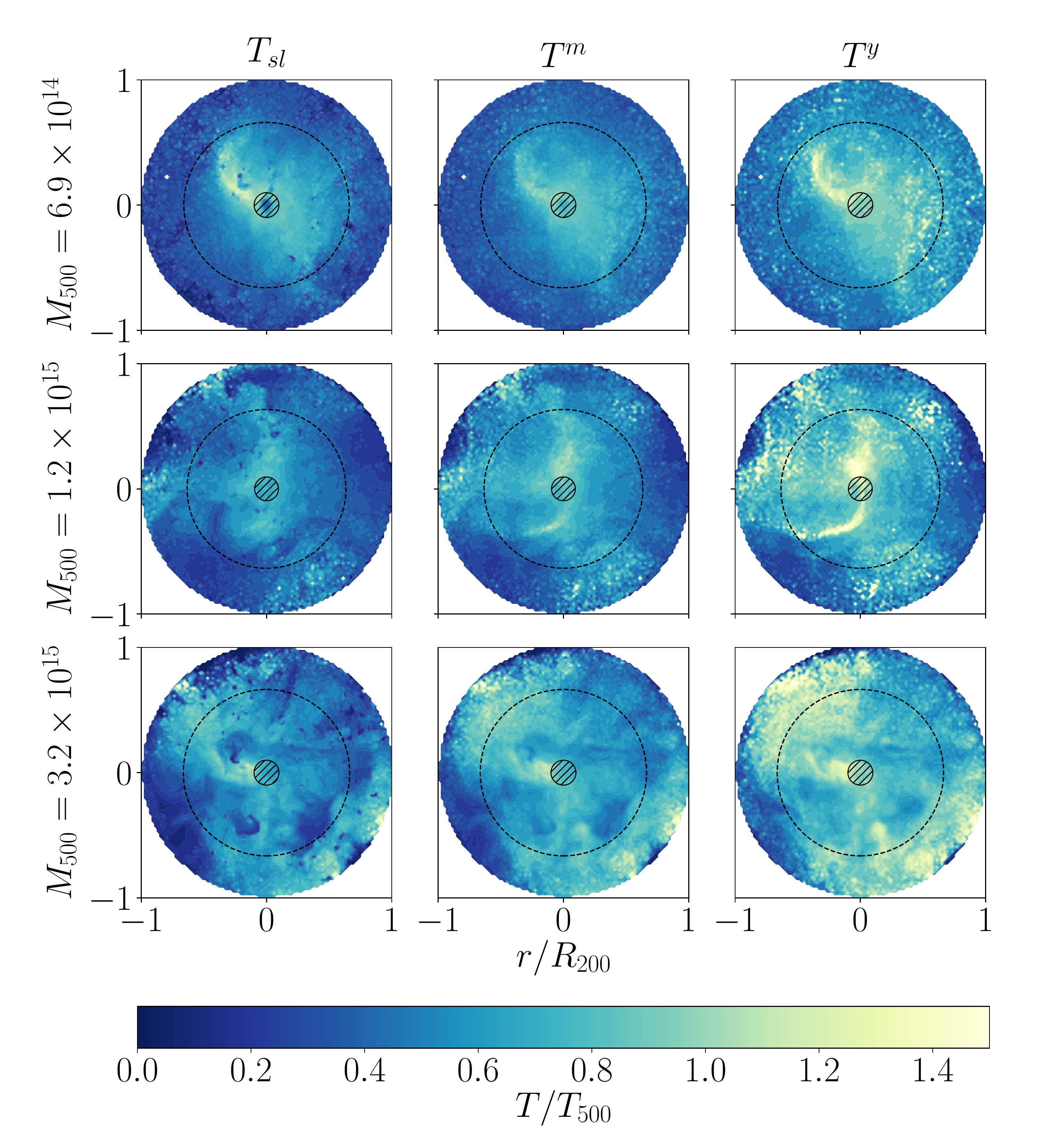}
    \caption{A comparison of the projected temperatures through a range of clusters at $z=0$, relative to $T_{500}$ for that cluster. These projections are taken within spheres of radius $R_{200}$ about the cluster centre of potential. From left to right we see $T_\mathrm{sl}$, $T^m$ and $T^y$, and from top to bottom clusters of various masses. Since these are just the projections for single clusters, they are subject to variations from the median expected behaviours. To guide the eye, on each plot a dotted line at $R_{500}$ has been drawn, alongside a hatched region at $0.15\times R_{500}$, which would be the core-excised region. These clusters have been chosen with $T_\mathrm{sl,500}>3.5$ keV so that it is an appropriate proxy for the X-ray temperature.}
    \label{fig:Slices}
\end{figure*}

\vspace{-3mm}
\section{Cluster Temperature Scalings}
\label{stn:Cluster_Bias}
%\vspace{-2mm}
To understand the cluster-wide, i.e., volume-averaged temperatures, it is instructive to first consider the contributions to each temperature measure, given by each part of the cluster. These lead to variations between the temperature measures calculated over spheres of regions $R_{500}$ (as typical for X-ray measurements) and $R_{200}$ (a proxy for the viral radius and arguably more applicable for SZ measurements). In this work we will present all our figures with respect to the $R_{500}$ sphere, but tabulate all our fits for both regions in the Appendix. In this section we will discuss both of these elements, and present our results for the volume-averaged temperature measures from the simulations. These allow us to generate both temperature--mass scaling relations as well as some temperature--temperature scaling relations. Finally, we will discuss the volume-averaged values for $\sigma(T^y)$, the standard variation of $T^y$ within clusters.

\vspace{-4mm}
\subsection{Causes for differences in temperature measures}
\label{stn:exploration}
From an illustrative point of view, we can examine the different temperature measures over clusters through the projected temperatures in a selection of clusters. These, as can be seen in Figure \ref{fig:Slices}, give us an indicative understanding of various features (e.g., shocks, outflows, sub-haloes and filamentary behaviours) that might exist within haloes under each temperature measure. 

While we generally see that $T^y>T^m>T_\mathrm{sl}$ \footnote{This can be seen especially in the features, but is generally evident in the slightly brighter overall colours of the halos from left to right.}, it is also the case that at larger radii, $T^y$ is more susceptible to the structures within the haloes. This can be seen by the increase in visibility of features in the haloes from the $T_\mathrm{sl}$ to the $T^y$ projections. This can be understood fairly simply: $T_\mathrm{sl}$ depends on the square of the local density, so in regions of high density -- i.e., the core of the cluster or in dramatic substructures, this will be clearly visible. On the other hand, $T^y$ depends on both the local temperature and density (i.e., the local pressure), so it is more affected by areas of diffuse, but warm gas, and thus highlights shocks. This particularly weights the observed temperatures in the outer regions, e.g., $R_{500} \to R_{200}$ which is barely probed by the X-ray temperatures (as reflected in $T_\mathrm{sl}$). We see that $T^m$ typically lies between these other two temperature measures.

\vspace{-4mm}
\subsection{Volume-averaged temperature measures}
First, we will discuss the difference between the temperature measures averaged over spheres of $R_{500}$ and $R_{200}$, and then quantify the temperature relations to the cluster masses, and the covariance between these values for each temperature measure. We will then discuss both the temperature--temperature fits and the fits for the Hot and Relaxed subsamples.

\vspace{-4mm}
\subsubsection{The effect of averaging over volumes of radii $R_{500}$ or $R_{200}$}
\begin{figure}
    \centering
    \includegraphics[width=1.\linewidth]{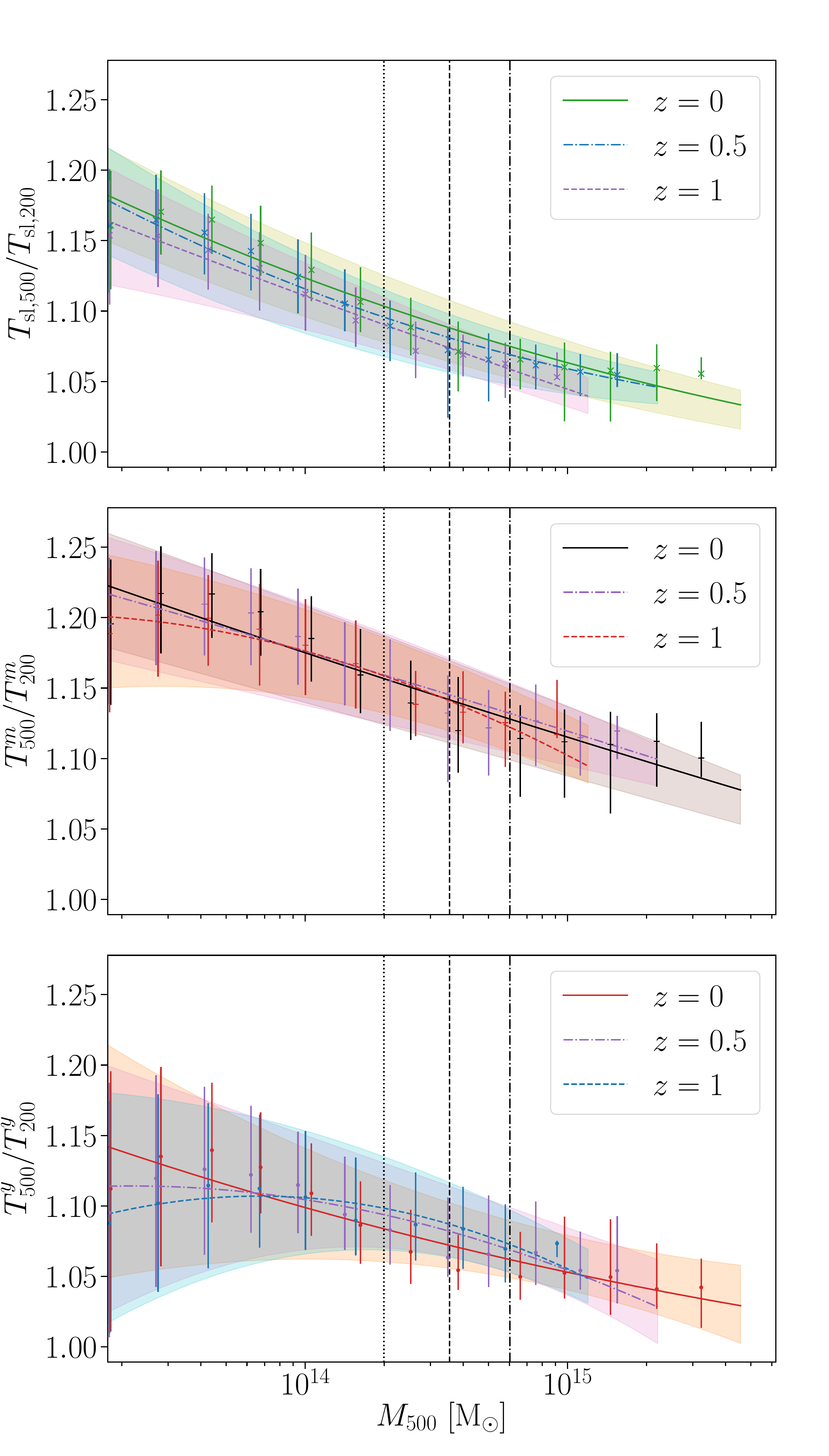}
    \caption{A comparison of the temperature measures depending on whether they are calculated over a sphere of radius $R_{200}$ or $R_{500}$ against the mass of the same clusters. We also display the redshift dependence of the same. The vertical dot-dashed, dashed and dotted lines here depict the mass cut offs between the BAHAMAS and MACSIS samples. The datapoints and errors show the median, 84$^\mathrm{th}$ and 16$^\mathrm{th}$ values for various mass bins, while the solid line and shaded regions demonstrate the best fits (discussed in Section~\ref{TMassRels}) for the same.}
    \label{fig:200v500}
\end{figure}

It is important to determine the difference between averaging over spheres of radii $R_{200}$ and $R_{500}$. X-ray measurements, in particular, are almost always taken over $R_{500}$, and as such $R_{500}$ values are those commonly used in the literature. However, it can be argued that $R_{200}$, as a better proxy for the virial radius, should also be widely considered. Since $R_{500}$ generates a smaller region, it encapsulates only the hotter core with less of the cooler outskirts of the cluster. As such, regardless of temperature measure, it returns a higher temperature than that obtained within $R_{200}$.

This can be seen graphically in Figure \ref{fig:200v500}. Here, we have plotted the fractional variation between $R_{500}$ and $R_{200}$ values. These appear to be predominantly redshift independent; while there are variations between each redshift, they are all within the scatter. Secondly we see that for all measures the differences between the two measures become smaller at higher masses.
This may in fact be an averaging effect due to the the distribution of temperatures in clusters (see Section \ref{stn:Profiles}), and the mass-dependent changes to the profiles and thus the fall-off of temperatures nearer the outskirts of clusters. These will lead to the averaged effects that can be seen here.

We see in general that the changes to $T^m$ are the most acute, followed by $T_\mathrm{sl}$, with $T^y$ undergoing the smallest corrections. However, this is still a sizeable effect: $\simeq 10$ per cent at $M_{200}=10^{14}$ $M_\odot$ ($\simeq 20$ per cent for $T^m$). This indicates that this should potentially be considered in more detail for future SZ measurements.
 
For the rest of the paper we will use $R_\mathrm{500}$ to reproduce the results commonly cited in cluster papers -- the analysis has also been carried out across a radii of $R_\mathrm{200}$ with few qualitative variations. The full tabulated numerical results can be found in Appendix \ref{App:FitValues}.

\vspace{-4mm}
\subsubsection{Temperature--mass scaling relations}
\label{TMassRels}
%---------------------------------------------------------------------------------------
\begin{figure}
    \includegraphics[width=1.0\linewidth]{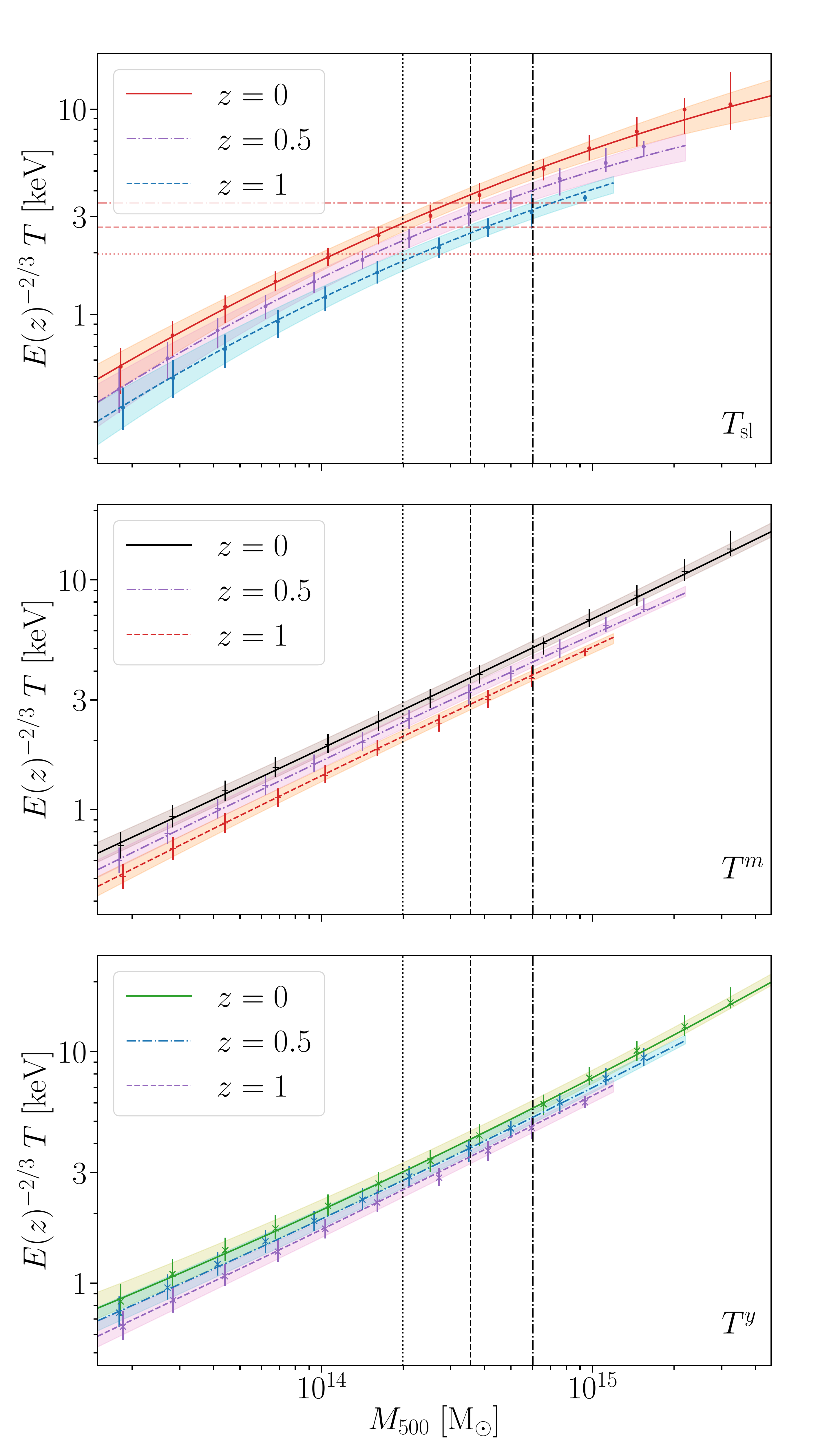}
    \caption{A comparison of the three temperature measures at three different redshifts. The plotted points show the medians of the binned data, with the error bars demonstrating their 16$^\mathrm{th}$ and 84$^\mathrm{th}$ percentiles. The solid lines show the fits to the data, with the shaded regions showing the 68 per cent confidence region. The horizontal lines in the top panel show the 3.5 keV cutoff for the reliability of $T_\mathrm{sl}$ as a proxy for the X-ray temperature.}
    \label{fig:T_M_Tsplit}
\end{figure}
%---------------------------------------------------------------------------------------
\begin{figure}
    \includegraphics[width=1.0\linewidth]{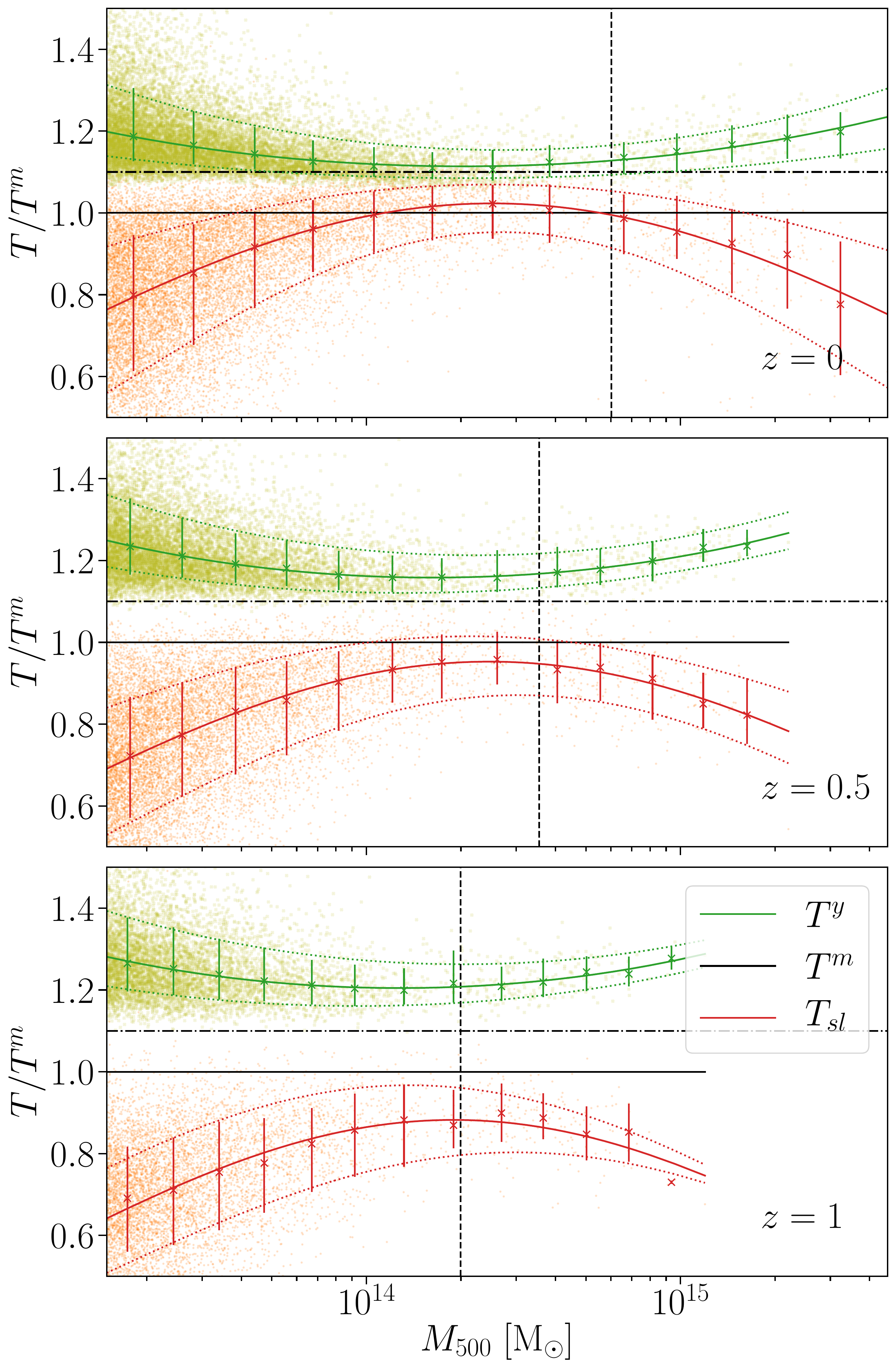}
    \caption{A comparison of the three temperature measures on a cluster by cluster basis. Here we consider the temperature measures with respect to $T^m$. The solid lines indicate the line of best fit of the data sets, while the dotted lines show the 16$^\mathrm{th}$ and 84$^\mathrm{th}$ percentiles. The horizontal dot-dashed line lies at $T/T^m=1.1$ to guide the eye. In the case against $T^y$ we can see that the minimum values of the means lie at 1.11, 1.16 and 1.21 for $z=0$, 0.5 and 1 respectively.}
    \label{fig:TvmT_againstMass}
\end{figure}

%---------------------------------------------------------------------------------------
In Figs.~\ref{fig:T_M_Tsplit} and \ref{fig:TvmT_againstMass}, we display the temperature mass scaling relationships for our three temperature measures at each redshift. Figure \ref{fig:T_M_Tsplit} shows the redshift dependence of each temperature measure individually, relative to self-similar scaling -- i.e., scaling out  $E(z)^{2/3}$; while Figure \ref{fig:TvmT_againstMass} shows the the results divided through by $T^m$, the mass-weighted temperature, so that the variations between the three measures are more visible. We see, first, that the spread in the data is far larger for $T_\mathrm{sl}$ than for $T^y$ or $T^m$. This furthers the common observation that the SZ signal, $Y_\mathrm{SZ}$, provides a tighter mass proxy than the X-ray signal. \footnote{Though, of course, this has many factors, and generally relies on the accurate calibration of the SZ mass relation.}

In Figure \ref{fig:T_M_Tsplit}, we can see that in general, the redshift variation of each temperature measure is similar to the self-similar relation -- i.e., $T \propto E(z)^{2/3}$. In particular, while with increasing redshift $T^y$ falls a little at low masses and has a slightly steeper mass dependence, overall the $y$-weighted temperature is consistent within the intercluster variation with self-similar evolution. The mass-weighted temperature shows more departures from self-similarity, and $T_\mathrm{sl}$ shows the greatest departure from this $E(z)^{2/3}$ scaling. The spectroscopic-like temperature both falls in magnitude and has increasing curvature, indicating that at the highest masses, the differences under redshift evolution are magnified.

From a physical point of view, this can be understood since at higher redshifts, the haloes have had a shorter cooling time, leading to denser cooler gas, and thus a lower $T_\mathrm{sl}$. However, the pressure of the gas is largely fixed to match the potential wells of the haloes themselves (as they are roughly in hydrostatic equilibrium) and reduces the redshift-dependent $T^y$, which is less affected by the evolution of the clusters themselves.

In Figure \ref{fig:TvmT_againstMass}, we can see that $T^y$ has a larger magnitude than $T^m$ and $T_\mathrm{sl}$, while the latter two are at points consistent, with $T^m$ higher at both higher and lower redshifts. Furthermore, we see hints of a strong cluster by cluster correlation in the values of $T^y$ and $T^m$, from the $\gtrsim 10-20$ per cent shift between these two values. This may be a consequence of the calibration scheme used in defining the spectroscopic-like temperature, which is focused on clusters at low redshifts with masses $M_{500}\simeq 10^{14}\;M_\odot$, but more work would have to be done to fully analyse this effect. In fact, with respect to $T_\mathrm{sl}$ we can see that there is a correction for $T^y$ of $\gtrsim 10$ per cent (or $\gtrsim 40$ percent) at $z=0$ ($z=1$), increasing greatly to both higher and lower masses with equality around $2.3\times 10^{14}$~$M_\odot$ ($1.8\times 10^{14}$~$M_\odot$). We also find that the differences between these three temperature measures increase strongly with redshift. We see that at $z=0$, for instance, $T^m$ and $T_\mathrm{sl}$ lie within each other's uncertainties, while by $z=1$ they are clearly separated. This means that accounting for these corrections will become even more important when considering distant clusters, which are typically those more easily probed through the SZ signal.

We can find in general that our data are well modelled by a 3-parameter fit, which corresponds to a quadratic equation in log--log space. We will express our values as 
%---------------------------------------------------------------------------------------
\begin{equation} \label{eqn:MassScalingRelation}
	E(z)^{-2/3}\;T = A\;\left(\frac{M}{M_\mathrm{fid}}\right)^{B + C\; \mathrm{log}(M/M_\mathrm{fid}).}\;\mathrm{keV},
\end{equation}
%---------------------------------------------------------------------------------------
where $M_\mathrm{fid}=3\times 10^{14}h^{-1} \mathrm{M}_\odot$. Hence, a self-similar fit around $M\simeq M_\mathrm{fid}$, would be given by $B=2/3$. By simply examining these fit values\footnote{These fits are for the median of the distributions, in Appendix \ref{App:FitValues} the fits to the 84$^\mathrm{th}$ and 16$^\mathrm{th}$ percentiles of the data set can be found to clarify the cluster-to-cluster spread in temperatures.}, as tabulated in Table \ref{tab:M5_medianfits}, we can immediately see the differences between the three temperature measures. Here, we have also tabulated the scatter about the best fit relation by calculating the root mean squared dispersion across all the haloes according to
\begin{equation}
    \sigma_{\mathrm{log_{10}}T}=\sqrt{\frac{1}{N}\sum_{i=1}^N[\mathrm{log}_{10}(T_i/T_\mathrm{fit})]^2},
\end{equation}
where $i$ indexes all the haloes at a given redshift and $T_\mathrm{fit}$ is the value given by the best fit at the mass, $M_i$, associated with the halo.

\begin{table}
\caption{Best fit values for the medians of each temperature measure at each redshift. The errors are determined through bootstrap methods. The fit parameters correspond to those described in Equation \eqref{eqn:MassScalingRelation}.} \centering
\begin{tabular}{lrrrr}
$M_{500}$& \multicolumn{1}{c}{A} & \multicolumn{1}{c}{B} & \multicolumn{1}{c}{C} & \multicolumn{1}{c}{$\langle \sigma_{\mathrm{log_{10}}T}\rangle$}\\
\hline\hline
\multicolumn{5}{l}{$z = 0.0$} \\
\hline
$T^y$ 		  & $4.763^{+0.015}_{-0.015}$ & $0.581^{+0.003}_{-0.002}$ & $0.013^{+0.001}_{-0.001}$  & $0.2707\pm 0.0014$ \\
\noalign{\smallskip}
$T^m$ 		  & $4.248^{+0.013}_{-0.012}$ & $0.565^{+0.003}_{-0.002}$ & $0.002^{+0.001}_{-0.001}$  & $0.2861\pm 0.0012$ \\
\noalign{\smallskip}
$T_\mathrm{{sl}}$ & $4.295^{+0.023}_{-0.025}$ & $0.514^{+0.012}_{-0.013}$ & $-0.039^{+0.005}_{-0.005}$ & $0.323 \pm 0.005$ \\
\hline\hline
\multicolumn{5}{l}{$z = 0.5$} \\
\hline
$T^y$ 		  & $4.353^{+0.019}_{-0.020}$ & $0.571^{+0.006}_{-0.006}$ & $0.008^{+0.002}_{-0.002}$  & $0.2521\pm 0.0016$\\
\noalign{\smallskip}
$T^m$ 		  & $3.702^{+0.013}_{-0.013}$ & $0.546^{+0.005}_{-0.004}$ & $-0.006^{+0.002}_{-0.001}$ & $0.2523\pm 0.0019$\\
\noalign{\smallskip}
$T_\mathrm{{sl}}$ & $3.474^{+0.027}_{-0.025}$ & $0.483^{+0.023}_{-0.028}$ & $-0.051^{+0.008}_{-0.010}$ & $0.350 \pm 0.007$\\
\hline\hline
\multicolumn{5}{l}{$z = 1.0$} \\
\hline
$T^y$ 		  & $3.997^{+0.021}_{-0.020}$ & $0.593^{+0.004}_{-0.004}$ & $0.009^{+0.001}_{-0.001}$  & $0.2438\pm 0.0016$\\
\noalign{\smallskip}
$T^m$ 		  & $3.237^{+0.015}_{-0.017}$ & $0.558^{+0.004}_{-0.005}$ & $-0.005^{+0.001}_{-0.001}$ & $0.2142\pm 0.0018$\\
\noalign{\smallskip}
$T_\mathrm{{sl}}$ & $2.754^{+0.036}_{-0.035}$ & $0.478^{+0.015}_{-0.014}$ & $-0.053^{+0.004}_{-0.004}$ & $0.401 \pm 0.004$\\
\hline
\end{tabular}
\label{tab:M5_medianfits}
\end{table}

In particular, we see, as previously observed in Figs.~\ref{fig:T_M_Tsplit} and \ref{fig:TvmT_againstMass}, that $T^y$ appears to be systematically higher than $T^m$, which itself lies above $T_{\rm sl}$. The gradients of these three temperature measures seem to match this same pattern. Finally we note that $T^y$ always have a positive curvature, while $T_\mathrm{sl}$ has a strong negative curvature and $T^m$ seems to develop curvature at higher redshifts. Further, we note that none of these are consistent with hydrostatic equilibrium scalings, which would have $B=2/3$ and $C=0$. While $T^y$ has the closest gradients to this value for hydrostatic equlibrium, even at the highest cluster masses the curvature is not sufficient for $T^y$ to match this scaling.

\vspace{-3mm}
\subsubsection{Covariance of fits}
\label{App:FitQuality}

\begin{figure}
    \centering
    \includegraphics[width=0.96\linewidth]{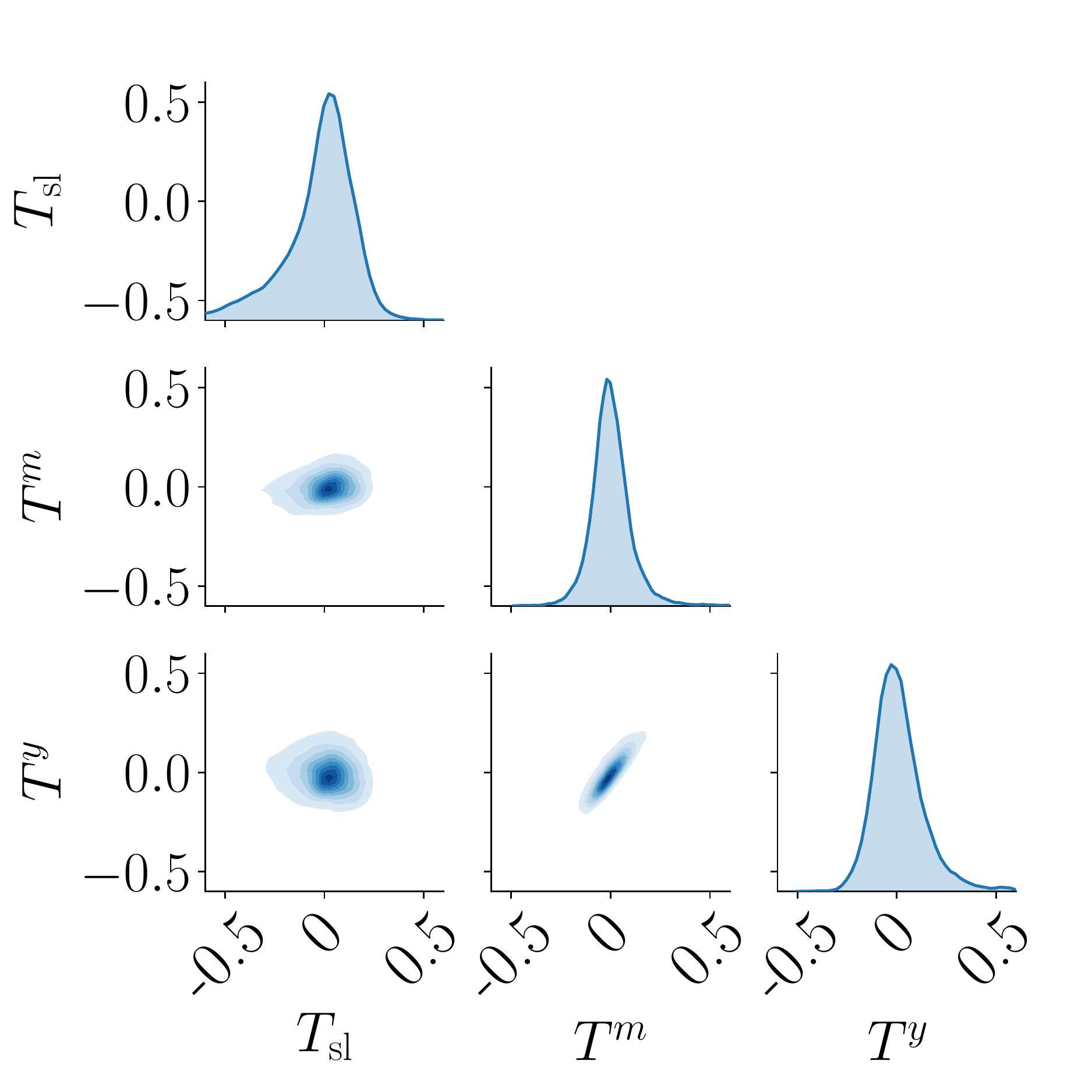}
    \caption{A representation of the covariance of $\mathrm{log}(T_\mathrm{data}(M)/T_\mathrm{fit}(M))$ for the three temperature measures at $z=0$. That is, a comparison of the overall distributions abound the line of best fit for each temperature measure. The diagonal parts show the overall distributions for each measure, while the lower triangle shows the contours of these covariances.}
    \label{fig:trianglefits}
\end{figure}

It is now instructive to understand the spread of cluster temperatures about the best fits of the temperature measures as displayed in the previous section. In Fig.~\ref{fig:trianglefits}, we show the covariances at $z=0$ of the quantity $\mathrm{log}(T_\mathrm{data}(M)/T_\mathrm{fit}(M))$ for each of the three temperature measures. Here $T_\mathrm{fit}(M)$ is the tabulated best-fit value, while $T_\mathrm{data}(M)$ refers to the calculated temperature measure for each cluster. We find that this behaviour is replicated well for $z=0.5$ and 1.0.

We can immediately see from the diagonal part that, while $T^m$ is almost normally distributed in the log--log space (that is, log-normally), the other two temperature measures have visible skews. This is most apparent for $T_{\rm sl}$, which skews to higher temperatures with a long tail to lower temperatures, while the $y$-weighted temperature measure seems only gently skewed to lower temperatures -- thus being almost log-normally distributed in the log--log space.

Furthermore, from the lower triangle we can see the correlations between the temperature measures within each cluster -- in particular the strong interdependence between $T^y$ and $T^m$. This indicates that on a cluster-by-cluster basis the difference between the $y$-weighted and mass-weighted temperatures are maintained. However, the spectroscopic-like temperature seems to be distributed independently of the other two measures.

This strong correlation in the values of $T^m$ and $T^y$ motivates the exploration of temperature--temperature scaling relations -- and moreover, since these two temperatures define the complete SZ signal, they motivate a volume averaged $Y-T^y$ scaling relation. This allows for a self-calibration of the relativistic corrections to the SZ signal, from measurements of the SZ signal itself.

\vspace{-3mm}
\subsubsection{$Temperature$-$temperature$ scaling relations}
\label{stn:T_T_scalings}
%---------------------------------------------------------------------------------------
As an alternative to a temperature--mass relations, we can consider temperature--temperature scaling relations. These lead to a predominantly mass-independent conversion between temperature measures. We see that a similar fitting formula [to that in equation \eqref{eqn:MassScalingRelation}] can be used, replacing $M_\mathrm{fid}$ with $T_\mathrm{fid} = 5\;\mathrm{keV}$,
%---------------------------------------------------------------------------------------
\begin{equation} \label{eqn:TvTscalingRelation}
	T = A\;\left(\frac{T_\mathrm{rel}}{T_\mathrm{fid}}\right)^{B + C\; \mathrm{log}(T_\mathrm{rel}/T_\mathrm{fid}).}\;\mathrm{keV}.
\end{equation}
%---------------------------------------------------------------------------------------
Since we have already discussed that the cluster temperature is often a good mass proxy we will not discuss these fits in much detail here as they take a very similar form to those against the mass, although the full tables fitting the temperature relations with respect to $T_\mathrm{rel}=T^m$ and $T_\Delta$ can be found in Appendix~\ref{App:FitValues}. While it is true that, due to the covariance of $T^y$ and $T^m$, we find that the spread in the fits of $T^y$ against $T^m$ are smaller than those against $M_{500}$, this effect is minimal.

\begin{table}
\caption{Best fit values for the medians of each temperature measure against $T_{500}$ at each redshift. The errors are  determined through bootstrap methods. The fit parameters correspond to those described in Equation \eqref{eqn:TvTscalingRelation}.} \centering
\begin{tabular}{lrrr}
$T_\mathrm{rel}=T_{500}$& \multicolumn{1}{c}{A} & \multicolumn{1}{c}{B} & \multicolumn{1}{c}{C}\\
\hline\hline
\multicolumn{4}{l}{$z = 0.0$} \\
\hline
$T^y$ 	    	  & $4.812^{+0.014}_{-0.013}$ & $0.889^{+0.003}_{-0.003}$ & $0.041^{+0.002}_{-0.002}$\\
\noalign{\smallskip}
$T^m$		      & $4.289^{+0.011}_{-0.011}$ & $0.873^{+0.004}_{-0.004}$ & $0.021^{+0.002}_{-0.002}$\\
\noalign{\smallskip}
$T_\mathrm{{sl}}$ & $4.293^{+0.023}_{-0.022}$ & $0.825^{+0.018}_{-0.018}$ & $-0.049^{+0.010}_{-0.011}$\\\hline\hline
\multicolumn{4}{l}{$z = 0.5$} \\
\hline
$T^y$ 		      & $4.964^{+0.021}_{-0.019}$ & $0.868^{+0.006}_{-0.005}$ & $0.026^{+0.004}_{-0.003}$\\
\noalign{\smallskip}
$T^m$ 		      & $4.247^{+0.013}_{-0.013}$ & $0.835^{+0.005}_{-0.004}$ & $-0.006^{+0.003}_{-0.002}$\\
\noalign{\smallskip}
$T_\mathrm{{sl}}$ & $4.039^{+0.021}_{-0.022}$ & $0.804^{+0.010}_{-0.011}$ & $-0.093^{+0.005}_{-0.006}$\\
\hline\hline
\multicolumn{4}{l}{$z = 1.0$} \\
\hline
$T^y$ 		      & $5.108^{+0.024}_{-0.024}$ & $0.875^{+0.005}_{-0.005}$ & $0.020^{+0.003}_{-0.003}$\\
\noalign{\smallskip}
$T^m$ 		      & $4.196^{+0.017}_{-0.016}$ & $0.846^{+0.004}_{-0.006}$ & $-0.010^{+0.003}_{-0.003}$\\
\noalign{\smallskip}
$T_\mathrm{{sl}}$ & $3.681^{+0.029}_{-0.029}$ & $0.807^{+0.015}_{-0.015}$ & $-0.118^{+0.009}_{-0.009}$\\
\hline
\end{tabular}
\label{tab:TvT5_medianfits}
\end{table}
%---------------------------------------------------------------------------------------

A shortened selection of the fits against $T_{500}$ can be found in Table \ref{tab:TvT5_medianfits}. First, we see that $T^y$ is always the closest temperature measure to $T_{500}$, the temperature assuming the cluster is an isothermal sphere \citep[agreeing with][]{Kay2008}. However, we can see that there is significant curvature in all of these fits alongside the gradient of the temperature measures being significantly lower than that for $T_{500}$, indicating further that the assumption of isothermality often used in SZ cluster calculations is inaccurate. In fact, we find that while $T_{500}$ is an overestimate of $T^y$ for the most massive clusters, it becomes an underestimate for lower mass, cooler, clusters, particularly at higher redshifts. This is likely due to the increased AGN feedback effects driving gas from these lower mass systems. This would lead to a decreased $T_{500}$ (which is mass dependent) compared to the $y$-weighted temperature.

\begin{table}
\caption{Best fit values for the medians, 84$^\mathrm{th}$ and 16$^\mathrm{th}$ percentiles of $T^y$ to $Y_{500}$ at each redshift. The errors are  determined through bootstrap methods. The fit parameters correspond to those described in Equation \eqref{eqn:YscalingRelation}.} \centering
\begin{tabular}{lrrr}
$T^Y-Y_{500}$& \multicolumn{1}{c}{A} & \multicolumn{1}{c}{B} & \multicolumn{1}{c}{C}\\
\hline\hline
\multicolumn{4}{l}{$z = 0.0$} \\
\hline
median & $5.017^{+0.012}_{-0.011}$ & $0.3749^{+0.0014}_{-0.0018}$ & $0.0044^{+0.0003}_{-0.0004}$\\
\noalign{\smallskip}
84 & $5.375^{+0.019}_{-0.019}$ & $0.3654^{+0.0021}_{-0.0021}$ & $0.0043^{+0.0005}_{-0.0005}$\\
\noalign{\smallskip}
16 & $4.732^{+0.012}_{-0.012}$ & $0.3796^{+0.0020}_{-0.0018}$ & $0.0046^{+0.0005}_{-0.0004}$\\
\hline\hline
\multicolumn{4}{l}{$z = 0.5$} \\
\hline
median & $5.745^{+0.020}_{-0.020}$ & $0.3707^{+0.0043}_{-0.0038}$ & $0.0034^{+0.0009}_{-0.0008}$\\
\noalign{\smallskip}
84 & $6.096^{+0.027}_{-0.027}$ & $0.3612^{+0.0035}_{-0.0039}$ & $0.0033^{+0.0008}_{-0.0009}$\\
\noalign{\smallskip}
16 & $5.423^{+0.021}_{-0.020}$ & $0.3772^{+0.0033}_{-0.0029}$ & $0.0038^{+0.0007}_{-0.0006}$\\
\hline\hline
\multicolumn{4}{l}{$z = 1.0$} \\
\hline
median & $6.639^{+0.037}_{-0.041}$ & $0.3693^{+0.0054}_{-0.0065}$ & $0.0016^{+0.0011}_{-0.0013}$\\
\noalign{\smallskip}
84 & $7.029^{+0.048}_{-0.047}$ & $0.3555^{+0.0054}_{-0.0052}$ & $0.0008^{+0.0011}_{-0.0010}$\\
\noalign{\smallskip}
16 & $6.254^{+0.036}_{-0.036}$ & $0.3837^{+0.0056}_{-0.0052}$ & $0.0038^{+0.0011}_{-0.0011}$\\
\hline
\end{tabular}
\label{tab:Y5_Tfits}
\end{table}

\topic{Volume-averaged Y relations}: As already noted, $T^m$ forms a strong proxy for the volume averaged $y$-parameter, $Y$. Since this relates to the amplitude of the SZ signal, while the shape is dependent on $T^y$, it is instructive to consider the scaling of $T^y$ with respect to $Y$. This gives us a self-calibrated scaling relationship to determine the rSZ signal. We use a fit similar to equations \eqref{eqn:MassScalingRelation} and \eqref{eqn:TvTscalingRelation},
\begin{equation} \label{eqn:YscalingRelation}
    T^y = A\;\left(\frac{Y}{Y_\mathrm{fid}}\right)^{B + C\; \mathrm{log}(Y/Y_\mathrm{fid}).}\;\mathrm{keV},
\end{equation}
where we take $Y_\mathrm{fid}= 0.0003\;\mathrm{Mpc}^2$. These results are shown in Table \ref{tab:Y5_Tfits} -- we also tabulate the $Y$--$M$ relationship in Appendix \ref{App:FitValues}. It is interesting to observe that while we have used 3-parameter fits, there is significantly less curvature in all of these fits to that seen in our mass-temperature and temperature--temperature relations. Further, combining equations \eqref{eqn:TheoryScaling} and \eqref{eqn:Ydefn}, we could expect from self-similarity, $T \propto (Y\,E)^{2/5}$. Although, this proportionality depends on the relationship between $T^m$ and $T_\Delta$, we can see that in Table \ref{tab:Y5_Tfits}, $B$ lies close to the expected value $B=2/5$.

We note that there is no explicit redshift dependence in these fits -- since we would expect from self-similarity both $T^y$ and $Y$ to scale with $E(z)^{2/3}$. However, we do see distinct redshift evolution in our fit parameters; in particular in the normalisation factor, $A$, which seems to almost scale $\propto E(z)^{1/2}$ (similar to, but above, the self-similar prediction), increases dramatically towards higher redshifts. We see a similar but smaller decrease in the gradient to higher redshifts. However, overall this dependence shows that at higher redshifts it becomes increasingly important to consider the relativistic corrections to the SZ signal.

\subsubsection{Hot and Relaxed Samples}
%---------------------------------------------------------------------------------------
\begin{table}
\caption{Best fit values for the medians of each temperature measure for the Hot and Relaxed Samples against $M_{500}$ at each redshift. The errors are  determined through bootstrap methods. The fit parameters correspond to those described in Equation \eqref{eqn:MassScalingRelation}, taking $C=0$.}  \centering

\begin{tabular}{lrr|rr}
&\multicolumn{2}{c}{Hot Sample}&\multicolumn{2}{c}{Relaxed Sample}\\
$M_{500}$& \multicolumn{1}{c}{A} & \multicolumn{1}{c}{B} & \multicolumn{1}{c}{A} & \multicolumn{1}{c}{B} \\
\hline\hline
\multicolumn{4}{l}{$z = 0.0$} \\
\hline
$T^y$ & $4.693^{+0.028}_{-0.028}$ & $0.633^{+0.009}_{-0.010}$ & $4.635^{+0.035}_{-0.036}$ & $0.626^{+0.010}_{-0.010}$\\
\noalign{\smallskip}
$T^m$ & $4.174^{+0.023}_{-0.025}$ & $0.598^{+0.019}_{-0.010}$ & $4.147^{+0.033}_{-0.034}$ & $0.593^{+0.019}_{-0.013}$\\
\noalign{\smallskip}
$T_\mathrm{{sl}}$ & $4.117^{+0.064}_{-0.053}$ & $0.531^{+0.043}_{-0.055}$ & $4.206^{+0.049}_{-0.051}$ & $0.531^{+0.051}_{-0.014}$\\
\hline\hline
\multicolumn{4}{l}{$z = 0.5$} \\
\hline
$T^y$ & $4.335^{+0.030}_{-0.027}$ & $0.597^{+0.017}_{-0.016}$ & $4.329^{+0.040}_{-0.034}$ & $0.598^{+0.018}_{-0.016}$\\
\noalign{\smallskip}
$T^m$ & $3.677^{+0.018}_{-0.021}$ & $0.561^{+0.011}_{-0.011}$ & $3.681^{+0.021}_{-0.024}$ & $0.561^{+0.011}_{-0.011}$\\
\noalign{\smallskip}
$T_\mathrm{{sl}}$ & $3.433^{+0.034}_{-0.033}$ & $0.457^{+0.023}_{-0.099}$ & $3.445^{+0.036}_{-0.037}$ & $0.455^{+0.025}_{-0.098}$\\
\hline\hline
\multicolumn{4}{l}{$z = 1.0$} \\
\hline
$T^y$ & $3.984^{+0.029}_{-0.030}$ & $0.611^{+0.016}_{-0.016}$ & $3.974^{+0.035}_{-0.035}$ & $0.610^{+0.020}_{-0.018}$\\
\noalign{\smallskip}
$T^m$ & $3.235^{+0.019}_{-0.023}$ & $0.586^{+0.008}_{-0.011}$ & $3.228^{+0.023}_{-0.024}$ & $0.581^{+0.012}_{-0.013}$\\
\noalign{\smallskip}
$T_\mathrm{{sl}}$ & $2.745^{+0.036}_{-0.049}$ & $0.469^{+0.017}_{-0.037}$ & $2.767^{+0.036}_{-0.043}$ & $0.473^{+0.017}_{-0.027}$\\
\hline
\end{tabular}

\label{tab:Hot_Relaxed_medians}
\end{table}
%---------------------------------------------------------------------------------------
Finally, it is useful to consider the behaviours of the Hot and Relaxed samples, as defined in Section \ref{stn:HotRelaxed}, for which the median fits are found in Table \ref{tab:Hot_Relaxed_medians}. Here, we have fitted both the hot and relaxed samples with a simple 2 parameter model\footnote{Since we ultimately find little difference between these values and those for the whole combined sample, these 2 parameter fits allow for comparison with other fits found in previous studies.}, or equivalently, we have taken Equation \eqref{eqn:MassScalingRelation}, setting $C=0$.

While we can see variations in the medians between the Hot and relaxed samples, we also find that the 16$^\mathrm{th}$ and 84$^\mathrm{th}$ percentiles are wider for the relaxed sample, so that these two samples give fits that lie within each other's cluster-to-cluster variance. Further, they agree well with the 3-parameter combined sample fits for both $T^m$ and $T^y$, though the fit can be found to be less appropriate for the spectroscopic-like temperature due to the strong curvature in the $T_\mathrm{sl}$ combined sample fits.

We can also find that, while the relaxed fits' larger 68 per cent error region for $T^y$ and $T^m$ seems to be well centred over the errors predicted by the complete combined sample fits, for $T_\mathrm{sl}$  these extend to higher temperatures, indicating that Relaxed clusters are more likely to have higher spectroscopic-like temperatures. We can understand this as $T_\mathrm{sl}$ is largely driven by the denser central region, and since more spherical (i.e., more relaxed) clusters are more likely to have a larger region for the same given mass, they are likely to lead to higher observed values for $T_\mathrm{sl}$.

%---------------------------------------------------------------------------------------
\begin{figure}
    \includegraphics[width=1.0\linewidth]{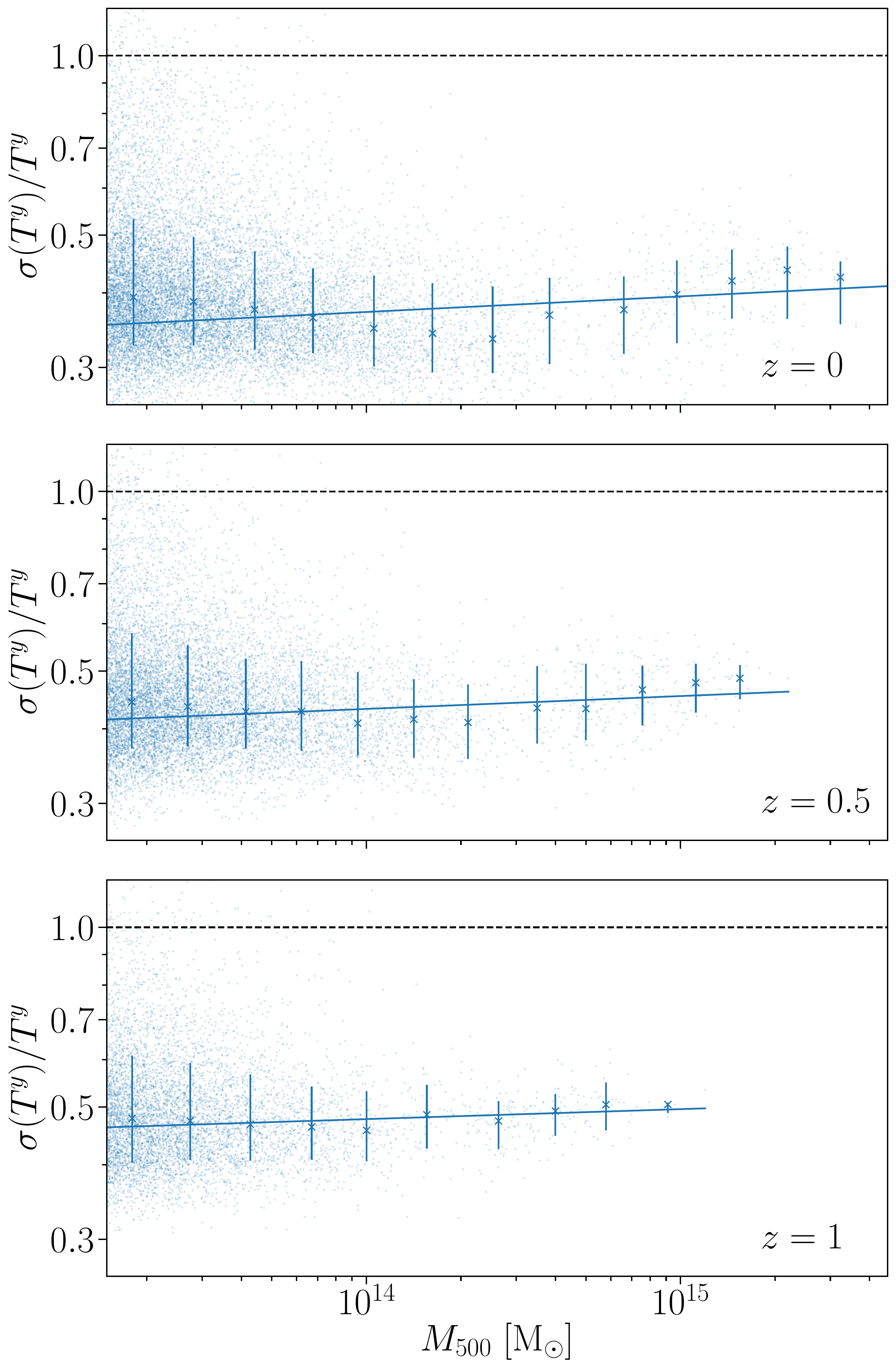}
    \caption{The redshift evolution of $\sigma(T^y) = \sqrt{T^y_2}$ with respect to $T^y$, as a function of $M_{500}$. The scattered points show the whole dataset, the error bars show the same data binned and the straight line shows the 2 parameter best fit of the data.}
    \label{fig:mom_zsplit}
\end{figure}
%---------------------------------------------------------------------------------------
\vspace{-4mm}
\subsection{$y$-weighted Temperature Dispersion}
\label{stn:vol_yw_mom}
%---------------------------------------------------------------------------------------
As noted in Section \ref{stn:SZFormalism}, the second moment of the $y$-weighted temperature ($T^y_2$) is a measure of the variance of the temperature distribution within the cluster\footnote{Recall that this is different from the distribution between clusters at each temperature, and as such is a measure of the intrinsic temperature variation within clusters rather than the variation between different clusters.}. Here we discuss $\sigma(T^y)$, the standard deviation and its comparison to $T^y$. We recall that under a temperature moment expansion about $T^y$, the leading order correction is proportional to $[\sigma(T^y)]^2$ [see Eq.~\eqref{eqn:mom_exp-II}].

In Figure \ref{fig:mom_zsplit}, we explore $\sigma(T^y)/T^y$ and can see that, while there is a small variation of the values across the mass range, they are well approximated by a power law (i.e., straight lines in the log--log space) -- which are tabulated in the Appendix (Table C5\footnote{Found in the supplementary online material.}). Generally we can see that, at higher redshifts, $\sigma(T^y)/T^y$ increases and that at all redshifts increases slightly with increased mass, approximately scaling as $\sigma(T^y)/T^y \simeq 0.39\,(1+z)^{0.34}\,[M_{500}/M_{\rm fid}]^{0.022}$. 
Since this redshift evolution closely matches the evolution of $T^y$ with respect to $T_{500}$, it may be that $\sigma(T^y)$ is mainly dependent on $T_{500}$ or equivalently the potential well of the cluster rather than the specifics of the substructure. That is, the variation in $\sigma(T^y)/T^y$ with redshift with respect to mass, is dominated by the near self-similar redshift evolution of $T^y$. It is also possible that there is an effect of clusters thermalizing over time, since this would explain the increase in variance for larger clusters and clusters at higher redshifts. However, since there are no clear differences between the dispersion of relaxed sample and the combined sample there is little evidence either way.
In Section \ref{stn:mom_prof}, we will explore the radial profiles of these values to see that these clusters see almost constant values across the whole width of the clusters, so that the overall dispersion is indicative of the dispersion at each point in the cluster.

Generally we find that the data spread is small, with around $\simeq 68$ per cent of the values for $\sigma(T^y)$ lying at around 40 percent of the overall temperature. However, we do see a characteristic small dip in the values of $\sigma(T^y)$ in the middle of our mass range ($\simeq 2-3\times 10^{14}$ $M_\odot$ at $z=0$). One possibility is that as the masses increase from $\simeq 10^{13}--10^{14}\;M_\odot$, the systems become more resilient to AGN feedback due to the increased potential well. As the masses increase further, the temperature variance is likely to increase again, due to the clusters still thermalizing (i.e., they are still forming). The exact details however are not explored in this paper.

%---------------------------------------------------------------------------------------
\vspace{-3mm}
\section{Cluster temperature profiles}
\label{stn:Profiles}
%---------------------------------------------------------------------------------------
In this section we discuss various cluster temperature profiles. 
To find analytic averages of our temperature profiles (to discern between each in a quantitative manner) we refer to the fits suggested by \citet{Vikhlinin2006_chandra}
%---------------------------------------------------------------------------------------
\begin{equation} \label{eqn:VikFits}\begin{split}
    T_\mathrm{tot}(r) &= T_0\,t_\mathrm{cool}(r)\,t(r)\\
    t_\mathrm{cool}(r) &=\frac{x_\mathrm{cool}+T_\mathrm{min}/T_0}{x_\mathrm{cool}+1}\\
    t(r) &= \frac{x_t^{-a}}{[1+x_t^b]^{c/b}}.
\end{split}\end{equation}
%---------------------------------------------------------------------------------------
Here we defined $x_t=r/r_t$ and $x_\mathrm{cool} = (r/r_\mathrm{cool})^{a_\mathrm{cool}}$. $T_\mathrm{cool}(r)$ accounts for the temperature decline of the central region of most clusters, while $t(r)$ acts as a broken power law with a transition region, to model the area outside this central region.\footnote{This model has 8 fit parameters \{$T_0$,$r_\mathrm{cool}$,$a_\mathrm{cool}$,$T_\mathrm{min}$,$r_t$,$a$,$b$,$c$\}, and requires fitting data within the 'core excised region' to allow the fit to access the central cooler region.}

There are two methods of generating profiles for our simulation measurement, each intuitive in different manners. From a simulation perspective, it is natural to consider a full radial profile, where we bin the particles in spherical shells from the centre of the cluster, and volume average the particles within each bin. However, from an observational standpoint, it is perhaps more relevant to consider the line-of-sight profiles, which we will here refer to as cylindrical profiles. In the next section we will discuss these radial profiles, as they are those normally discussed of the literature, while an exploration of the cylindrical profiles can be found in Appendix \ref{stn:cy6prof}.

In most observational work, the observed line-of-sight profiles are deprojected to generate radial profiles, and radial profiles are compared in the literature. However, we can see from our cylindrical profiles that care must be taken in this deprojection process, as the different weighting in each temperature measure, lead to complicated variations in the behaviour of the radial and cylindrical profiles. We will then finally discuss the profiles derived for the $y$-weighted temperature moments in Section \ref{stn:mom_prof}.

\vspace{-4mm}
\subsection{Radial Profiles}
%---------------------------------------------------------------------------------------
\begin{figure}
    \includegraphics[width=1.0\linewidth]{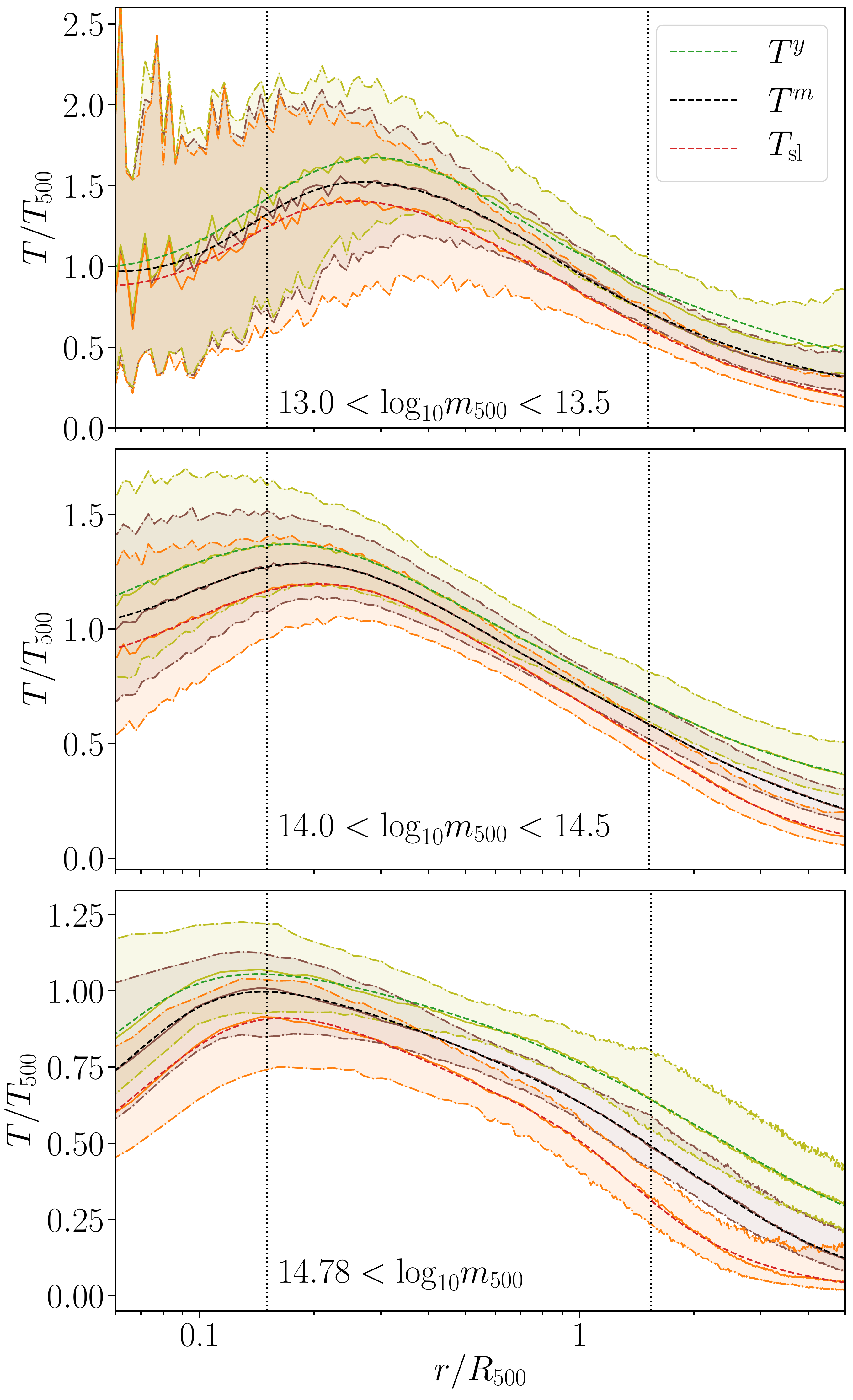}
    \caption{The radial profiles of the three different temperatures across 3 different mass bins -- note here $m_{500}=M_{500}/\mathrm{M}_\odot$. As is standard the temperatures have all been scaled by $T_{500}$ for the same cluster, and the radii have been scaled by $R_{500}$. The vertical dotted lines indicate the core region (0.15 $R_{500}$) and virial radius ($R_{200}$) respectively. The solid lines show the median values at each radial bin across the clusters and the shaded region the 68 per cent confidence region. The dotted lines show the fits using the Vikhlinin model.}
    \label{fig:radprof_massbin}
\end{figure}
%---------------------------------------------------------------------------------------

%---------------------------------------------------------------------------------------
\begin{figure}
    \includegraphics[width=1.014\linewidth]{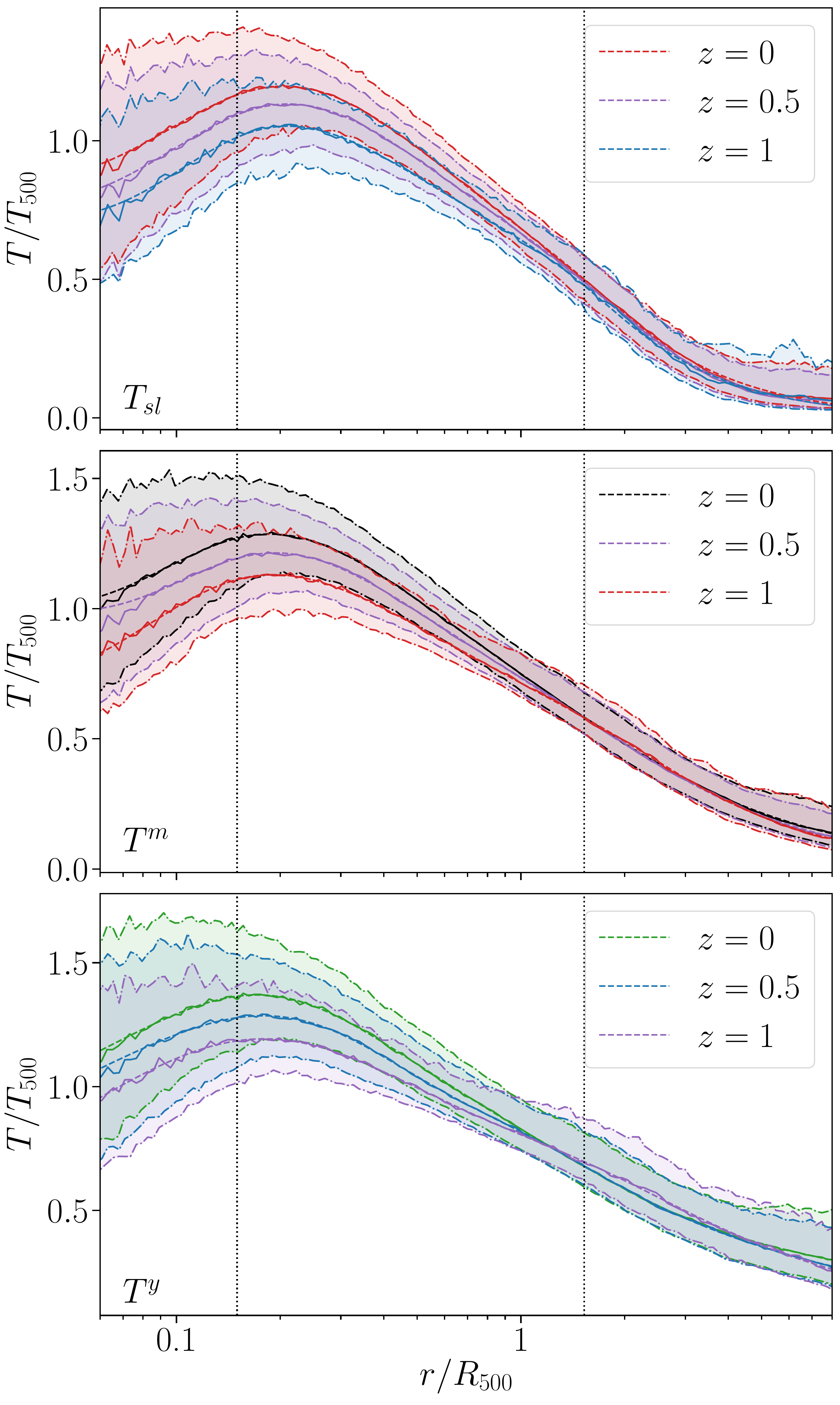}
    \caption{The radial profile evolution across z for the clusters of masses $M_{500} = 13.5-14\times 10^{14}$ $M_\odot$ clusters. This is indicative of the evolution of all of the clusters, through the profile fits for the others can be found in Appendix \ref{App:FitValues}. This figure is otherwise arranged as in Figure \ref{fig:radprof_massbin}. Recall also that $T_{500}$ is defined to be redshift dependent, so is removing the $E^{2/3}(z)$ dependence. Note, that these are the same clusters traced over the redshift evolution, so they would appear to have lower masses at higher redshifts.}
    \label{fig:radprof_redshift}
\end{figure}
%---------------------------------------------------------------------------------------

In Figure \ref{fig:radprof_massbin} we display the radial profiles at $z=0$ where we have sorted the clusters into 5 mass bins (three of which are graphically displayed); the $5^\mathrm{th}$ bin (lowest panel of figure) corresponds to the selection of clusters from the MACSIS sample, hence the uneven bin width. In Figure \ref{fig:radprof_redshift}, we show the redshift evolution of the clusters with $13.5 \leq \mathrm{log}_{10}(M_{500}/\mathrm{M}_\odot)\leq 14$, which are indicative of the variation of all mass bins. The median fits of all of these quantities can be found in Appendix \ref{App:FitValues}.

First, we can see in Figure \ref{fig:radprof_massbin} that  $T^y$ is once again systematically larger than $T^m$ which is in turn larger than $T_\mathrm{sl}$. Further we can see that this increase appears systematically larger at larger radii. This is in agreement with our previous observations that the $y$-weighted temperature is more attuned to the affects of larger radii.

We can further see that these differences are enhanced at higher masses \citep[see also][]{Henson2017,Pearce2019}. For instance, we can see that at higher masses $T_\mathrm{sl}$ developes a defined downwards turn between $R_{500}$ and $R_{200}$ where the density falls and thus the contribution to the temperature drops markedly. We also note that as masses increase, the initial peak in the temperature shifts to smaller radii; that is that the cooled central region of clusters (which generates the cooling flow) becomes proportionally smaller for higher mass clusters. This indicates that the highly variable inner regions of the clusters will have a smaller effect on the overall temperatures in higher mass clusters than smaller. 

Considering the redshift evolution as seen in Figure \ref{fig:radprof_redshift}, we see that all of the temperature measures evolve self-similarly in the outskirts of clusters ($r\gtrsim R_{500}$) while the interior appears to heat up comparatively from high to low redshift. This indicates that there is some true increase in temperature in the centre of clusters not explained by self-similar evolution, as redshift decreases. The differences between the three temperature measures are very small, largely dominated by the overall scaling of the three volume averaged temperature measures.

%---------------------------------------------------------------------------------------
\vspace{-2mm}
\subsection{Profiles of $y$-weighted Temperature Moments}
%---------------------------------------------------------------------------------------
\label{stn:mom_prof}
\begin{figure}
    \includegraphics[width=1.0\linewidth]{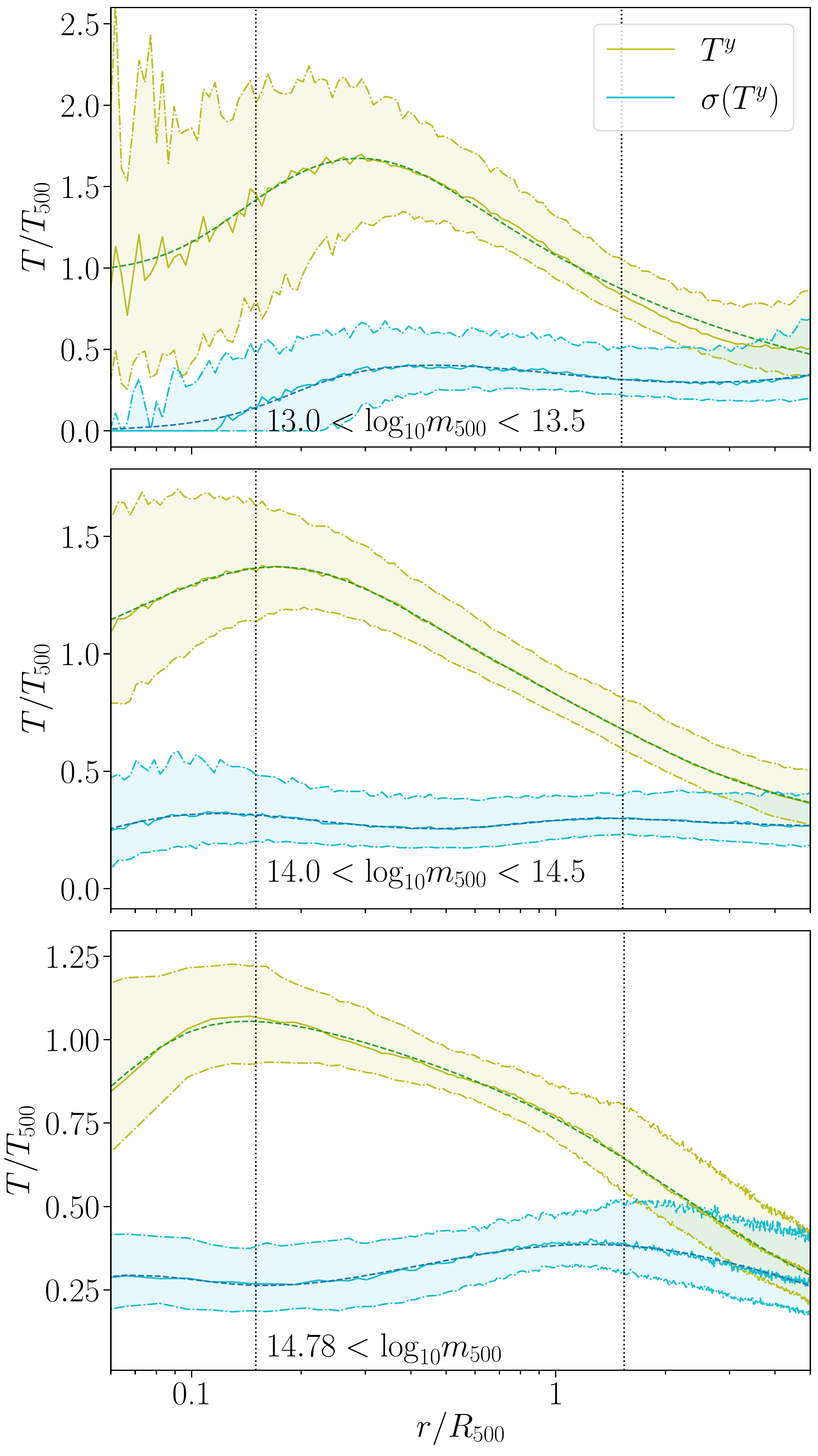}
    \caption{The radial profiles of $T^y$ and the first moment, $\sigma(T^y) = \sqrt{T^y_2}$. This figure is arranged as in Figure \ref{fig:radprof_massbin}.}
    \label{fig:radmom_massbin}
\end{figure}

We find that the radial and cylindrical profiles for $\sigma(T^y)$ behave very similarly across all masses and redshifts, in that they are approximately constant with respect to $T_{500}$. This can be seen in Figure \ref{fig:radmom_massbin}.\footnote{These are defined, identically to the temperature weightings, as the averaged values over each spherical shell.} This approximate mass independence matches what we observe in Figure \ref{fig:mom_zsplit} where we see that $\sigma$ is a roughly constant fraction of $T^y$. Furthermore, we see that under redshift evolution $\sigma(T^y)(r)/T_{500}$ remains roughly constant, suggesting that the variation in $\sigma(T^y)$ seen in Section \ref{stn:vol_yw_mom} is due to the variation of $T^y$ against $T_{500}$ rather than reflective of an increase in temperature dispersions within clusters at higher redshifts.

However, the values are not entirely constant, we can see that at higher masses $\sigma(T^y)$ rises at higher radii, implying that as the temperatures fall the variation in the temperature increase. This makes sense if we suppose the outskirts of clusters to contain clumpy substructure, leading to cool and hot regions at the same radii -- this could also be related to the cluster asphericities causing similar hot and cool effects in the spherically averaged shells. Similarly we can see that the variation falls off in the central regions of the clusters, implying that the central region (as modelled in the simulations) are approximately isothermal and we see little variation.

%---------------------------------------------------------------------------------------
\section{Implications for Cosmology}
\label{stn:Discussion}
%---------------------------------------------------------------------------------------
In this section, we will discuss the effects these different temperature measures have on determining $Y_\mathrm{SZ}$, and the further effects of the higher order moments on the determination of the $y$-weighted temperature from examining the spectral shape. Finally, we will discuss the effect of these corrections and related `corrections' to the radial profiles and their impacts on the common method to determine $H_0$ through the SZ effect -- this will give us an indicative view of the magnitude of the necessary corrections.

%---------------------------------------------------------------------------------------
\subsection{Effect on $Y_\mathrm{SZ}-M$ relation}
\label{stn:Ysz_M_rel}
\begin{figure}
    \includegraphics[width=0.99\linewidth]{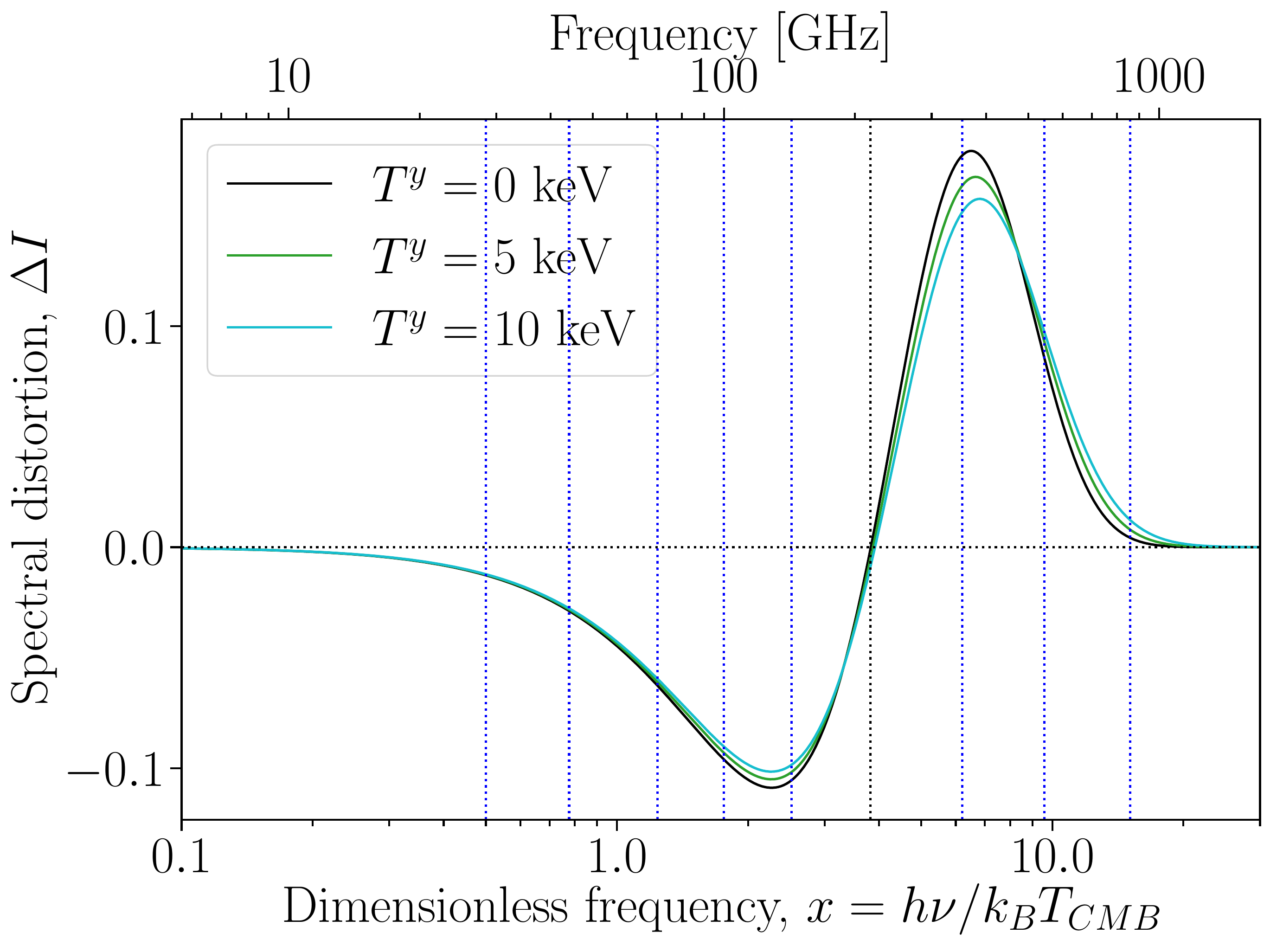}
    \caption{A comparison of the different spectral shapes determined by $T_\mathrm{e}=T^y$ given $T^y = 0$, 5 or 10 keV. The blue vertical lines mark the \Planck bands, with the black line at 217 GHz to show the expected mean of the distribution. These plots were made with \texttt{SZpack}, taking $y=10^{-4}$ and $\tau = 0.01$.}
    \label{fig:SZDistortions}
\end{figure}

%---------------------------------------------------------------------------------------
First we recall that, as mentioned in Section \ref{stn:SZFormalism}, to second order in $\Delta T$ we can express the SZ signal as
%---------------------------------------------------------------------------------------
\begin{equation}
    S(\nu) = y f(\nu, T_\mathrm{e}) + y^{(1)} f^{(1)}(\nu,T_\mathrm{e})+\frac{1}{2} y^{(2)} f^{(2)}(\nu,T_\mathrm{e}).
\end{equation}
%---------------------------------------------------------------------------------------
By setting the pivot temperature $T_\mathrm{e}=T^y$, when we take the volume averages we can find that
%---------------------------------------------------------------------------------------
\begin{equation}
	\Delta I \propto Y f(\nu, T^y) + \frac{1}{2} Y^{(2)}_{T^y} f^{(2)}(\nu,T^y).
\end{equation}
%---------------------------------------------------------------------------------------
Here, $Y$ is the volume integrated $y$-parameter and $Y^{(2)}_{T^y} = Y\,T_2^y = Y\,[\sigma(T^y)]^2$ relates to the temperature dispersion. In \citet{Remazeilles2018}, it is explained that in the analysis of \citet{Planck2016Map} $f(\nu,\Te) \simeq f(\nu, 0)$ is implicitly assumed. As mentioned above this leads to an underestimation of the deduced $y$-parameter and also biases the tSZ power spectrum amplitude.
\citet{Remazeilles2018} characterize the correction to $C^{yy}$ (which is $\propto S^2$) showing that for \Planck it scales as $C^{yy}_\ell(T_\mathrm{e})/C^{yy}_\ell(0) \simeq 1+0.15[k_\mathrm{B}T_\mathrm{e}/5\,\mathrm{keV}]$, where the electron temperature should be the $y$-weighted temperature.
Hence, we could approximate the correction to the SZ signal around the maximum at $\nu \simeq 353\,{\rm GHz}$ as
%---------------------------------------------------------------------------------------
\begin{equation}
    \frac{f(353\,{\rm GHz}, T_\mathrm{e})}{f(353\,{\rm GHz}, 0)} \simeq 1-0.08\left[\frac{k_\mathrm{B}T^y}{5\,\mathrm{keV}}\right],
\end{equation}
%---------------------------------------------------------------------------------------
which can also be seen in Fig.~\ref{fig:SZDistortions}, where we have plotted the observed distortions we would expect from our scaling relations given $T_\mathrm{e} = T^y = 0$, 5 and $10$~keV. In the presence of foregrounds, this was found to give a reasonable estimate for the effect of rSZ on the \Planck $y$-analysis \citep{Remazeilles2018}.

%---------------------------------------------------------------------------------------
%---------------------------------------------------------------------------------------

When folded into the analysis of $Y_\mathrm{SZ}$, for \Planck this leads to a systematic mismatch between the observed relativistic temperature distortions and the magnitude of the integrated pressure from $Y_\mathrm{SZ}$. This leads to the calculation that:
\begin{equation} \label{eqn:yM_relation_corr}
    \frac{Y(T^y)}{Y(T^y=0)} \simeq 1 + 0.08 \left[\frac{k_\mathrm{B}T^y}{5\,\mathrm{keV}}\right].
\end{equation}
The temperatures here refer to those assumed in the analysis of the spectral shape. We have established above that, for a given mass, the spectroscopic-like temperature underestimates the $y$-weighted temperature in a mass-dependent way by $\gtrsim 9--38$ per cent. As such, these relativistic corrections lead to even larger errors in the calculations of $Y_\mathrm{SZ}$ than X-ray measurements alone would suggest, especially for hotter clusters or clusters at higher redshifts.

In \citet{Remazeilles2018}, a standard X-ray temperature mass relation was used, indicating $\Te \simeq 5--7~\keV$ to be a typical cluster temperature relevant to tSZ power spectrum measurements. Using our $T^y--M$ relations, we expect this typical temperature to increase to $\simeq 6-9\,\keV$, which could further help reduce apparent differences in the deduced hydrostatic mass bias seen in various SZ observables \citep{Remazeilles2018}. For refined estimates our new $T^y-M_{500}$ (i.e., the true mass) relations should thus be very useful.

As previously discussed, we can also consider the $T^y--Y$ scaling relations to fully calibrate the SZ signal within SZ measurements. That is, we could consider the SZ signal explicitly as a function of $Y$, by defining $f(\nu,Y)=f(\nu,T^y(Y))$, such that $\Delta I \propto Y\,f(\nu,Y)$. This form of self-calibrated scaling allows for an X-ray independent calculation of the relativistically corrected SZ signal, which could theoretically be confirmed by direct checks of the shape of the signal.

\subsubsection{Comparison to other temperature--mass scaling relations}

In \citet{Remazeilles2018}, they use a temperature--mass scaling relationship derived from \citet{Arnaud2005} of
\begin{equation}
    k_\mathrm{B}T^X_\mathrm{e} \simeq 5\; \mathrm{keV}\;\left(\frac{E(z) M_{500}}{3\times 10^{14}h^{-1}M_\odot}\right)^{2/3}
    \label{eq:totally_wrong_TeX_M}
\end{equation}
for estimates. 
\citet{Arnaud2005} used 10 nearby relaxed galaxy clusters with masses ranging between $(0.8--8)\times10^{14}\;M_\odot$. This is a form consistent with the results seen in \citet{Barnes2017}, although the latter extends this work to higher masses, which fit the simulated hydrostatic mass to the simulated observed spectroscopic X-ray temperature using the BAHAMAS and MACSIS simulations. 
Equation~\eqref{eq:totally_wrong_TeX_M} can now be replaced with our $T^y-M_{500}$ relation from simulations to avoid conversion issues.

It is commonly known that there is a hydrostatic mass bias between X-ray derived masses and the true total mass of clusters \citep[e.g.,][]{Rasia2006,Rasia2012,Nagai2007,Meneghetti2010,Nelson2014,Shi2015,Biffi2016,Barnes2017CEagle,Ansarifard2019} -- which can in particular be seen in comparisons of the X-ray and weak lensing derived masses of clusters. Weak Lensing, as a probe of the depth of the gravitational well, gives a closer result to the true mass of clusters than X-ray observations. This underestimate of the hydrostatic model is due to the limitations of the assumption of hydrostatic equilibrium within clusters. In particular, the mass biases calculated to occur from the MACSIS and BAHAMAS simulations have been discussed in e.g., \citet{Henson2017}. Generally, this mass bias is considered to be $M_\mathrm{spec} \simeq (1-b)M_\mathrm{total}$ with $b\simeq 0.2$, although in fact, this bias is both mass and redshift dependent (\citealp[e.g.,][]{Henson2017,Pearce2019}; \citealp[but see also][]{Ansarifard2019}).

However, the temperature--temperature scalings discussed in Section \ref{stn:T_T_scalings} will hold entirely independently of the mass measured of a given  cluster. As such, any of these scaling relationships measured to obtain the X-ray temperatures (at high temperatures where $T_\mathrm{sl}$ is an appropriate proxy for the spectroscopic X-ray temperature) can be adjusted by the $\gtrsim10-40$ per cent conversion discussed before between $T_\mathrm{sl}$ and $T^y$.

We furthermore note that for $T^y(M)$ we currently can only rely on numerical simulations, as no accurate direct measurements of this variable exist. In computation of the rSZ effect, the scaling relations given in Table~\ref{tab:M5_medianfits} and \ref{tab:Y5_Tfits} should thus be most useful and directly applicable in computations of the SZ power spectra, e.g., using {\tt Class-SZ} \citep{Bolliet2018}.

\subsubsection{Corrections from temperature dispersion}
\label{sec:SZ_dispersion}
While we have focussed on the leading order rSZ correction, the 2$^\mathrm{nd}$ order correction due to the temperature dispersion is also worth discussing. As previously previously noted, the volume averaged dispersion is significant, scaling with the cluster temperatures, i.e., $\sigma(T^y) \simeq 0.4\,T^y$. However, as we argue now, at the current level of precision this rSZ correction remains negligible.

Using the asymptotic expansions \citep[e.g.,][]{Sazonov1998, Chluba2012}, we can express the fully relativistic SZ signal at low temperatures as:
\begin{equation}
\nonumber
    f(\nu,\theta) \simeq  \left(Y_0(\nu)+\theta\,Y_1(\nu)+\theta^2\,Y_2(\nu)+\theta^3\,Y_3(\nu)+\ldots\right),
\end{equation}
where we note that these $\theta = k_\mathrm{B}T_\mathrm{e}/(m_\mathrm{e}c^2)$, that is, the dimensionless temperature.\footnote{In our range of interest, i.e., temperatures 1--10~keV, $\theta$ assumes values $\simeq 2\times 10^{-3}-2\times 10^{-2}$.} This allows us to directly calculate an approximation for the signal associated with the second-order corrections, $f^{(2)}(\nu,\theta)\simeq (2\,Y_2(\nu)+6\,\theta\,Y_3(\nu)+\ldots)$. As such we can express the full signal, with second order corrections as,
\begin{equation}
\nonumber
    S(\nu) \simeq y\left(Y_0(\nu)+\theta\,Y_1(\nu)+\theta^2\left(1+\left[\frac{\sigma(T^y)}{T^y}\right]^2\right)\,Y_2(\nu)+\ldots\right).
\end{equation}
Now, $Y_2(\nu)$ has an effect on broadening the SZ signal and pushing it to slightly higher frequencies -- a full explanation of the functions can be found in \citet{Chluba2012}. In particular, at 343~GHz (the frequency most applicable for determining the SZ signal magnitude in \Planck), $Y_2(343\;\mathrm{GHZ})/Y_0(343\;\mathrm{GHZ}) \simeq 70)$. Assuming a cluster temperature of 5 keV, one has $\theta\simeq 0.01$ and with $\sigma(T^y)/T^y \equiv 0.4$ we find a $\simeq 70\times (0.01)^2 \times (0.4)^2 \simeq 0.1$ per cent correction to the overall SZ signal stemming from the average intracluster temperature-dispersion.

It is worth noting that since the radial $\sigma(T^y)$ is constant even as the temperature changes (see Figure \ref{fig:radmom_massbin}), this correction accordingly will be larger proportionally near the outskirts of clusters. However, these outskirts also correspond to lower temperatures -- which would both make the signal itself harder to detect, but also damp further the corrections from the temperature dispersion.
More work must be done to see how different feedback models effect these values of $\sigma(T^y)$ -- and thus to see if there is any possibility of them giving detectable results. We also mention that the intercluster temperature variations, relating to the shape of the mass-function, should also be carefully considered.

%---------------------------------------------------------------------------------------
\vspace{-2mm}
\subsection{Applications to the determination of $H_0$}
\label{H0application}

It has long been established that $H_0$ can be determined through a combination of X-ray and SZ measurements \citep[e.g.,][]{Birkinshaw1979,Reese2004,Jones2005,Bonamente2006,Kozmanyan2019}. While these are generally less precise than those calculations from the CMB \citep[e.g.,][]{Planck2018CC} or direct measurements \citep[e.g.,][]{Riess2019}, as the systematics in the approach are being accounted for, they are becoming both increasingly competitive and complementary.

The general approach for this is as follows \citep[see also][]{Bourdin2017,Kozmanyan2019}. From the X-ray data, the density and temperature profiles can be constrained [i.e., $n_\mathrm{e}(r)$ and $T_\mathrm{sl}(r)$], and from the SZ data the pressure profile, $P_{\rm e}(r)$ can be constrained through the measurements of $y$ assuming the distortion is wholly non-relativistic. This allows for a second temperature profile to be calculated, $T^m(r) = \eta_\mathrm{T}P_\mathrm{e}(r)/n_\mathrm{e}(r)$. By assuming these two temperature profiles are equal, i.e., $T^m(r)\equiv T_{\rm sl}(r)$, this allows for a measurement of $\eta_\mathrm{T}$, which can be found to depend on (among other variables) the angular diameter distance, $d_A$. As such, $\eta_\mathrm{T} \propto d_A^{-1/2}\propto H_0^{1/2}$, which provides a way to obtain $H_0$ estimates. 

Now, in this consideration, we already have two issues, the first is the estimation of the $P_\mathrm{e}$ which, as discussed above, will be underestimated due to the omission of relativistic effects [exactly as in Eq. \eqref{eqn:yM_relation_corr}]. The second is the concordance of $T_\mathrm{sl}(r)$ and $T^m(r)$, which, as can be seen in Figure \ref{fig:radprof_massbin}, is not an accurate assumption. We see that, if $T^m(r)>T_\mathrm{sl}(r)$, this method leads to an underestimation of the temperature. As such, these two corrections counteract one another, and we must determine which one is dominant.
The two temperature profiles furthermore have slightly different shapes, which will additionally bias the derived value for the $H_0$ parameter. However, we do not go into more detail here.
Overall, we can express the correction due to rSZ as,
\begin{equation}
    \frac{H_0,\rm corr}{H_0} \simeq \left[\frac{P_0}{P_\mathrm{corr}}\right]\left[\frac{T^m}{T_\mathrm{sl}}\right],
\end{equation}
where $P_0$ is the pressure calculated assuming there are no relativistic corrections. For instance, to estimate the effect, at $T^y = 5$ keV, we have already determined that $P_\mathrm{corr} \simeq 1.08 P_\mathrm{0}$. We can also use our previous profile fits to estimate the mismatch in the $T^m(r)$ and $T_\mathrm{sl}(r)$ profiles. Since $T^y = 5$~keV corresponds to a $M_{500} \simeq 5.0\times 10^{14}$ $M_\odot$, we can see this correction as $T_\mathrm{sl}(r) \simeq 0.92 \, T^m(r)$. In this specific case, the two corrections match well and cancel each other, but we can expect that generally not to hold.

%---------------------------------------------------------------------------------------
\begin{figure}
    \includegraphics[width=1.0\linewidth]{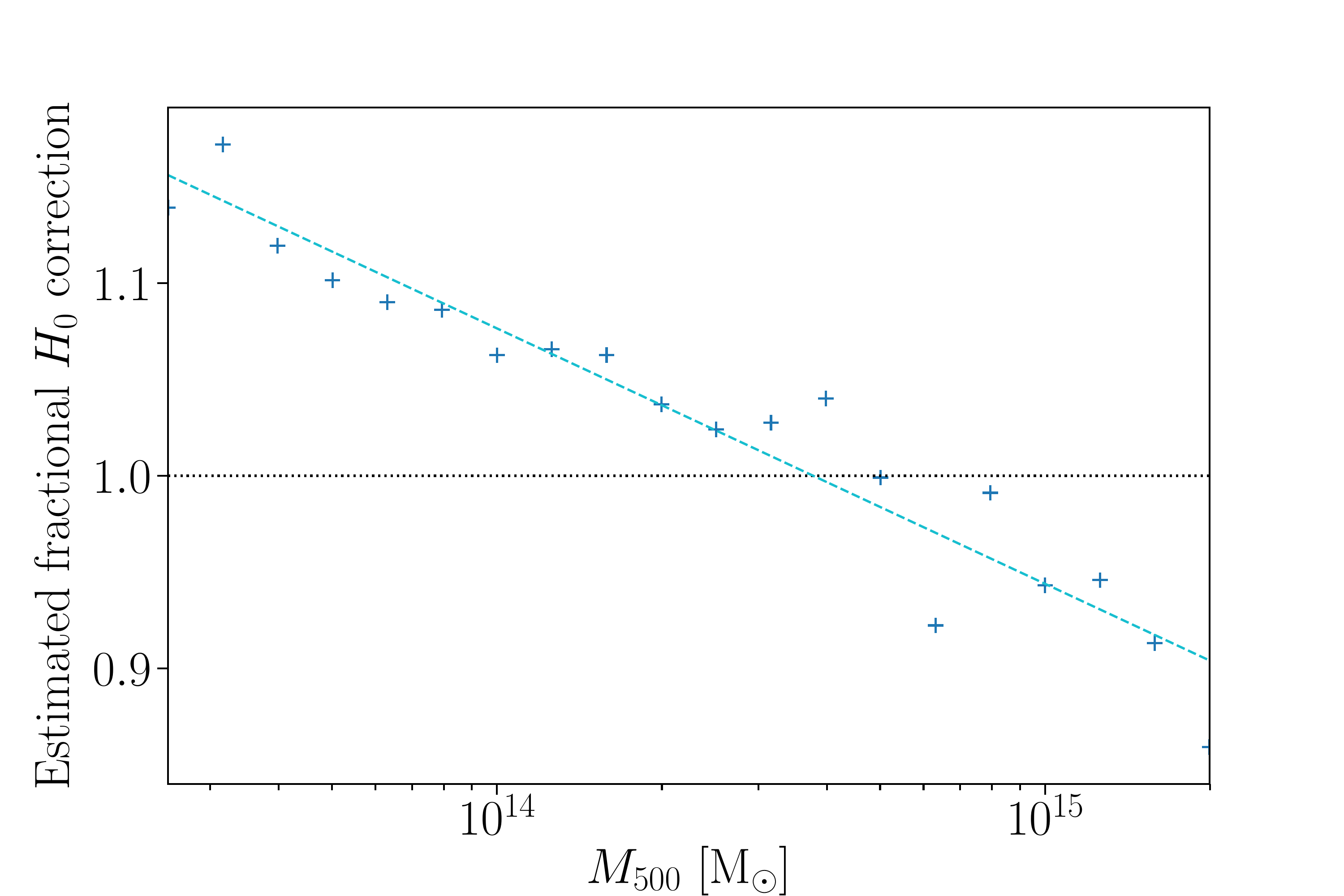}
    \caption{An indicative plot of the potential magnitudes of the corrections to $H_0$. The dashed line is merely to guide the eye. It should be noted that this is not a full or complete accounting of the corrections, merely a indication of the necessity of carrying out these two opposing corrections. These corrections are measured as $H_\mathrm{0,corr}/H_{0,0}$. Note: No errors are quoted as this is a fast calculation and a true representation of the errors would require an in depth study of the various interlocking factors.}
    \label{fig:H0_corr}
\end{figure}
%---------------------------------------------------------------------------------------

In Figure \ref{fig:H0_corr}, we ran a calculation of the indicative correction over. While this is not a full or complete accounting of the rSZ corrections, this exercise indicates that these corrections have the potential to swing by $\simeq 10$ per cent in either direction, tending to higher values of $H_0$ for lower masses and smaller values for higher masses. 
In, for instance, \citet{Kozmanyan2019} the median of the observed sample of clusters lies at $M_{500} = 7.3\times 10^{14}$ $M_\odot$, which would indicate a potential overestimation of $\simeq 4$ per cent (i.e., naively shifting the derived value of $H_0$ to $\simeq 64\pm3$).
This indicates a potentially sizeable correction in the deduced values of $H_0$; however, it is not clear which way this correction will ultimately fall, and a more careful analysis of the effect should be undertaken, in particular focusing on the assessment of the error budget.

At lower masses, we note that this effect will be dominated by the profiles of the spectroscopic-like temperature -- which, below masses of $\simeq 2.8 \times 10^{14}\,M_\odot$ is no longer a good probe of the observed X-ray signal. Furthermore, these calculation are all at $z=0$, while at higher redshifts $E(z)^{-2/3}\, T^y$ will remain almost constant with mass and the higher order corrections may increase; however, the behaviours of the profiles are harder to predict.
The exact details of this correction should be studied more carefully, including an in depth comparison of the different radial profiles from $T^m$ and $T_\mathrm{sl}$.

%---------------------------------------------------------------------------------------
\vspace{-4mm}
\section{Conclusions}
\label{stn:Conclusion}
%---------------------------------------------------------------------------------------
The importance of rSZ corrections is increasing with growing sensitivity of future CMB experiments. To incorporate the expected effects on SZ observables reliable temperature--mass scaling relations and temperature profiles are required.
Here, we have greatly extended the works of \citet{Pointecouteau1998,Hansen2004,Kay2008} to classify, in detail, the three temperature measure $T_\mathrm{sl}$, $T^m$, and $T^y$ across the mass ranges allowed through the combined BAHAMAS and MACSIS simulations. We find differences $\simeq 10-40$ per cent between the three temperature measures, with a general trend that $T_\mathrm{sl}< T^m < T^y$. The differences increase to both higher redshifts, and when the temperature measures are determined over the virial radius (i.e., $R_{200}$), as opposed to the more commonly (and less applicable for SZ measurements) used radius, $R_{500}$ (i.e., Figures \ref{fig:TvmT_againstMass} and \ref{fig:T_M_Tsplit}). 
We find that $T^y$ scales almost self-similarly, i.e., $\propto E(z)^{2/3}$, out to $z=1$, while $T_\mathrm{sl}$ and $T^m$ both undergo significant evolution relative to this `expected' scaling. Hence, for higher mass clusters, and clusters at higher redshifts (e.g., those detected in \Planck), $T_\mathrm{sl}$ is an increasingly poor proxy for $T^y$, or equivalently, the rSZ signal will be larger than X-ray measurements would imply.
Our analysis also suggests that the $y$-weighted temperature is a better proxy for cluster mass, a possibility that could be used for self-calibration of cluster masses using rSZ measurements.

We find a strong correlation between $T^y$ and $T^m$, with $T^y \gtrsim 1.1 \,T^m$ at $z=0$. While this correction is more complex for $T_\mathrm{sl}$, we none-the-less find that $T^y \gtrsim 1.09\,T_\mathrm{sl}$ at $z=0$, with similarity around $M_{500} \sim 2.3\times 10^{14}$ $M_\odot$ ($T_\mathrm{sl} \simeq 3.0$ keV) and these values diverging increasingly to both higher and lower masses, or equivalently temperatures (see Figure \ref{fig:TvmT_againstMass}). We find, moreover, that these corrections depend very little of the nature of the cluster, i.e., whether they are relaxed or not. 
This strong correlation leads to tight scaling relations between $Y$, the volume averaged compton-$y$ parameter, and $T^y$ [see Eq.~\eqref{eqn:YscalingRelation}]. This relationship can be used to calibrate the relativistic corrections to the SZ signal, from the signal itself. This allows for an estimate of the rSZ signal in, for instance, the \Planck SZ whole sky maps and in computations of the SZ power spectra, e.g., using {\tt Class-SZ} \citep{Bolliet2018}.

On average our findings suggest that X-ray derived temperatures underestimate the level of the rSZ by $\simeq 10-40$ per cent.
For instance, we can estimate a correction for the averaged temperature of clusters in the \Planck maps calculated in \citet{Hurier2016,Remazeilles2018}. These papers determined them to be $T^X = 6.8$ keV or $T^X \gtrsim 5$ keV respectively, which would naively lead to $T^y = 8.4$~keV or $T^y \gtrsim 5.7$~keV, a correction $\gtrsim15$ per cent in both cases. 
These differences will also affect the expected value for the sky-averaged SZ contribution, as calculated in, e.g., \citet{Hill2015}. There a X-ray temperature--mass scaling relation was used to determine the size of the relativistic corrections, finding a value of $k\Te \simeq 1.3~\keV$. This value could increase if our $T^y--M$ relation is used. Given that in particular low-mass haloes ($M\lesssim 10^{13}\,M_\odot$) contribute to the average SZ signal, the differences in this prediction are further amplified by redshift-evolution, likely leading to another increase of the expected value, although they may be mediated by the true spectroscopic temperature in such regimes being poorly modelled by the spectroscopic-like temperature. Measurements of the sky-averaged rSZ effect with future CMB spectrometers \citep{Chluba2019, Kogut2019} could lead to interesting constraints to feedback models and thus deserves more attention.

The profiles of these three radial temperature measures show similar trends (see Figure \ref{fig:radprof_massbin}). 
These differences will be very important when interpreting and combining future X-ray and high-resolution SZ profile measurements \citep[e.g.,][]{Ameglio2009MNRAS, Morandi2013}.
From these projected profiles, it will also be possible \citep[see][]{Remazeilles2018} to calculate a corrected power spectrum for the tSZ effect, which could play a role in reducing the tension between $\sigma_8$ found with \Planck and the SZ measurements. An understanding of the differences between the three profiles could also be useful for quantifying conversions between the observed X-ray and SZ signals -- in particular an understanding of the different behaviour of $T_\mathrm{sl}(r)$ and $T^m(r)$, which are commonly taken to be identical. These differences can lead to various miscalculations where these are used interchangeably, for instance in the SZ-derived $H_0$ as discussed in Section \ref{H0application}.

The intracluster temperature dispersion is found to be almost mass independent (at around $\sigma(T^y) \simeq 0.4\, T^y$, see Figure \ref{fig:mom_zsplit}), but increases slightly towards higher redshifts as a result of cluster evolution. However, we find that this adds little modification $\lesssim 0.5$ per cent to the SZ signal. Larger effects due to temperature dispersion could arise from intercluster temperature variation, which directly relate to the shape of the halo mass function; however, an estimation of this correction is beyond the scope of this paper.

While we have presented a classification of all three temperature measures and the $y$-weighted temperature dispersion, further work must be done to establish the independence of these results from the simulations (i.e., BAHAMAS and MACSIS) used. Through comparisons to other simulations it will be possible to assess the robustness of these results with respect to feedback models and other aspects of the gas physics used to generate these clusters. In particular, it would be interesting to understand how variations of the microphysics between simulations may lead to differences in the calculated intracluster temperature dispersion, $\sigma(T^y)$ and $T^y-Y$ or $T^y-M$ relations. All these could potentially be used to learn about the dynamical state of the cluster.

Extracting the rSZ signals with future CMB experiments still presents a challenge \citep{Basu2019, Chluba2019}. However, there is work to be done to establish the utility of rSZ quantities across a variety of cluster models and simulations. Further, the significant temperature differences from using the more appropriate temperature measures ($T^y$ rather than $T^X$), compounded with corrections from the temperature dispersion effects (and higher order terms to be considered in future works), will lead to improvements in the ability to interpret the rSZ signal.

{\small
\section*{Acknowledgements}
The authors would like to thank the referee for their helpful comments, Ian McCarthy for use of the BAHAMAS simulations and Francesca Pearce for useful discussions regarding the MACSIS and BAHAMAS samples.
EL was supported by the Royal Society on grant No RGF/EA/180053.
JC was supported by the Royal Society as a Royal Society University Research Fellow at the University of Manchester, UK.
This work was also supported by the ERC Consolidator Grant {\it CMBSPEC} (No.~725456) as part of the European Union's Horizon 2020 research and innovation program.
}

{\small
\bibliographystyle{mnras}
\bibliography{TemperatureBias}
}

\appendix
\section{Analysis of the mass dependence of the quality of the fits}
\begin{figure}
    \centering
    \includegraphics[width=1.\linewidth]{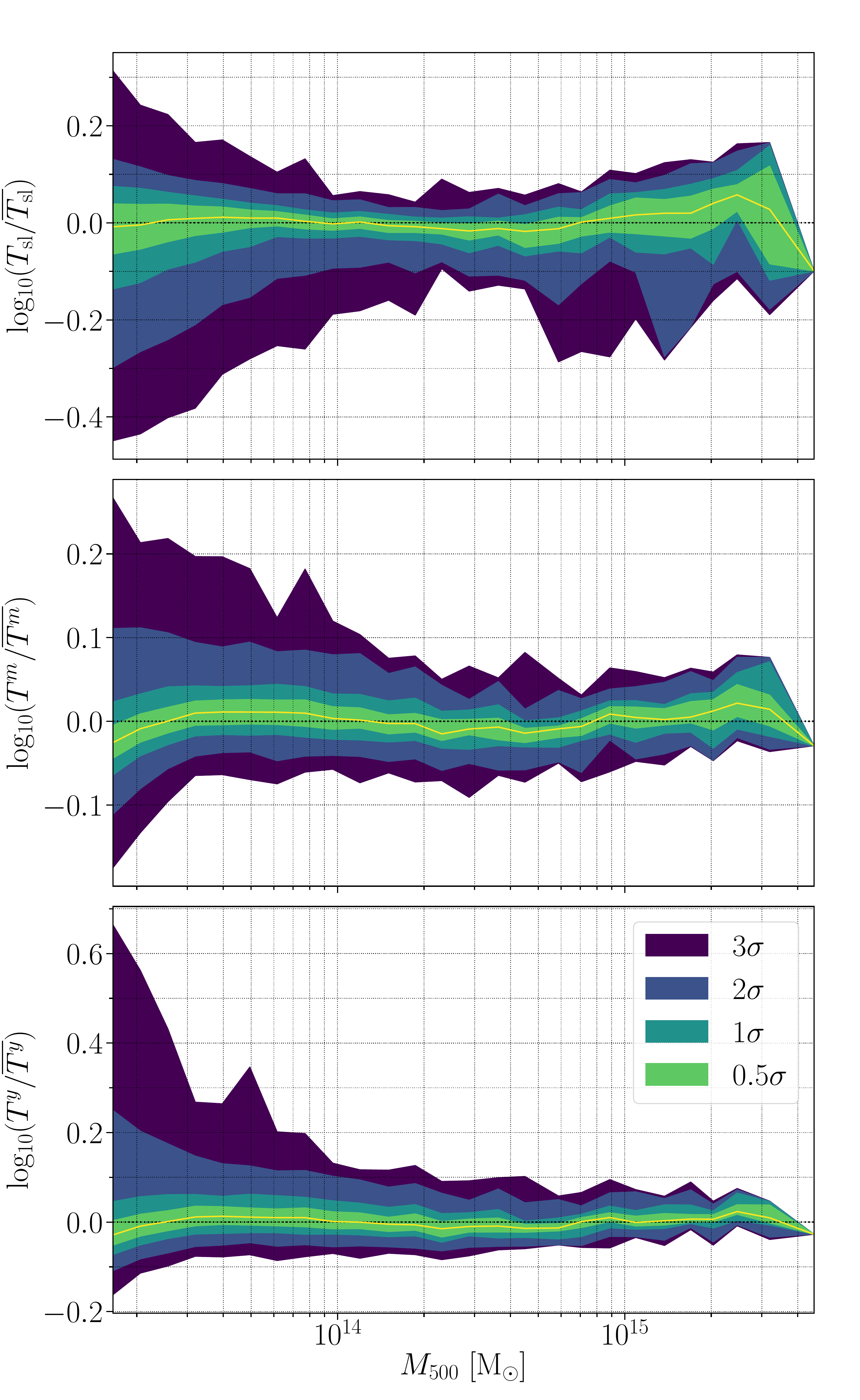}
    \caption{A graphical depiction of the spread of data around the $T$--$M_{500}$ fits at $z=0$. The shaded regions here are the percentile regions associated with the $\sigma$ values, were the data spread normally about the mean, that is at the 0.2--99.8, 2.5--97.5, 16--84 and 31--69 percentiles for the 3, 2, 1, and 0.5$\sigma$ regions labelled. The medians of the data are plotted in yellow.}
    \label{fig:ErrorResiduals}
\end{figure}

Although we can see from Section \ref{App:FitQuality} the skewness of the quality of fits as a whole, across all the data, it is instructive to consider how the quality of the fit varies over the mass range of the samples. This can be seen graphically in Figure \ref{fig:ErrorResiduals}.

Here we have plotted the contours for the percentiles associated with what would be the 0.5, 1, 2, and 3$\sigma$ confidence regions were the data normally distributed against its line-of-best-fit. 

The first thing to note is that there is a change over in data set at $M_{200}\simeq10^{15}$ $M_\odot$, on the left is the BAHAMAS data and on the right the MACSIS. This is of note simply because the data in the MACSIS set is less dense than that in the BAHAMAS set, and this will contribute to the increased errors we see to the right of the graph -- the errors are driven by lack of data as much as by the intrinsic scatter.

Secondly we see, especially in $T^y$, some anomalous results at low masses, skewing the 2 and $3\sigma$ contours dramatically. In $T^m$, we can see that the data is in fact roughly normally distributed across the entire mass range, with roughly constant errors -- this skew at low masses appears to be the only changing factor. In fact, the $3\sigma$ region outside of this skew is, if anything, underrepresented compared to a normal distribution -- that is, indicating smaller tails in the distribution that would be expected. This may, however, be simply a limitation in the number of clusters in each mass bin to be considered.

In $T^y$, however, we see this low-mass skew continued strongly in $2\sigma$ but still present to an extent across the entire range. This corroborates the long tail seen in the distribution of $T_\mathrm{sl}$ in Figure \ref{fig:trianglefits} -- however, it is worth noting that the skew appears to decrease to higher masses. A similar, but opposite, phenomena is seen in the $T_\mathrm{sl}$ contours, were we see a persistent and strong skew in the data to lower temperatures. This indicates that although the fits model well the median and, even the $1\sigma$ variations, it would be inappropriate to consider this data as normally distributed.

\section{Cylindrical Profiles}
\label{stn:cy6prof}
\begin{figure}
    \includegraphics[width=1.0\linewidth]{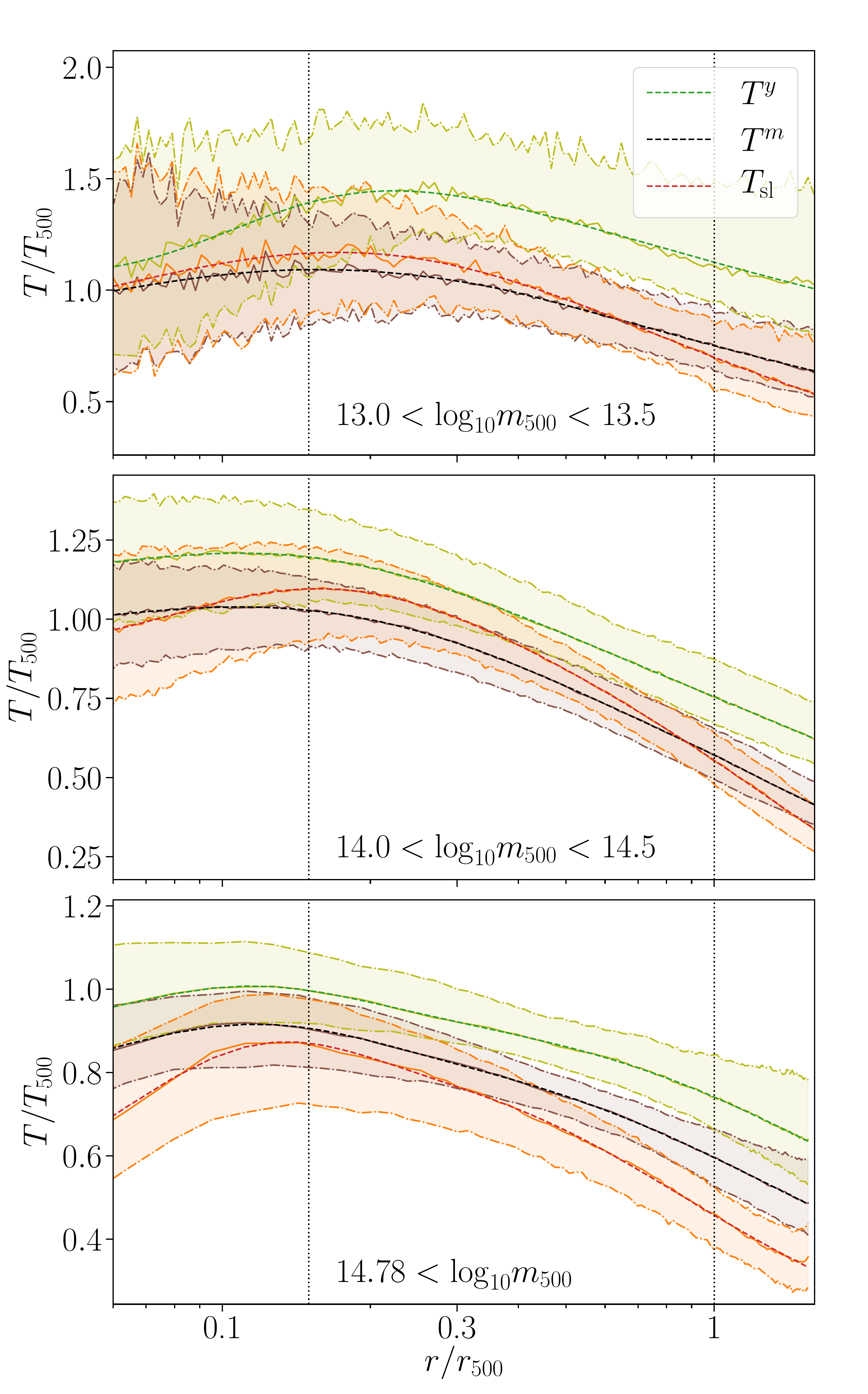}
    \caption{The cylindrical profiles of the three different temperatures across the 5 different mass bins. This figure is arranged as in Figure \ref{fig:radprof_massbin}.}
    \label{fig:cy6prof_massbin}
\end{figure}

The cylindrical profiles, or line-of-sight profiles, are perhaps of more value observationally than the radial profiles -- to create radial profiles from observations, it is necessary to deproject the line-of-sight observations. As such it is of use to consider these profiles alongside the radial profiles discussed in Section \ref{stn:Profiles}. To create these cylindrical profiles, for each cluster the central sphere of radius $R_\mathrm{200}$ was first extracted from the simulation\footnote{This causes some lack of precision very close to these edges as the number of particles in each bin becomes small.}, and then cylindrical shells were sliced from this sphere along six maximally spaced lines of sight through the core of the cluster. These 6 lines-of-sight cylindrical profiles were then averaged, to reduce the influence of inhomogeneity between the viewing angles in each cluster. We would expect the cylindrical profiles to have similar qualities to the radial profiles, albeit smoothed.

We can see this in Figure \ref{fig:cy6prof_massbin}. Here we once again observe that $T^y$ lies at systematically higher temperatures than the other temperature measures.  Curiously however, we see (especially at lower masses) $T^m$ and $T_\mathrm{sl}$ becoming somewhat indistinguishable. However, at larger radii, $T^m$ does always rise above $T_\mathrm{sl}$ which could lead to the observed volume averages. This is a result of the associated weightings in each temperature measure. $T_\mathrm{sl}$ has a $n_\mathrm{e}^2$ dependence, which in line-of-sight averages will substantially upweight the central hotter regions of the cluster making the overall line of sight appear far greater relative to $T^m$ than one would naively assume from the radial profiles. We can also see here that core-excision removes a dramatic turn down observed in the $T_\mathrm{sl}$ profile for higher mass clusters, which is not seen as clearly in the other two temperature measures.

Under redshift variation, these cylindrical profiles follow almost identical variation to that seen in the radial profiles, so while tabulated in the Appendix \ref{App:FitValues}, these are not discussed further here. By definition, in these cylindrical profiles we do not have the outer regions of the clusters so we cannot compare their behaviours as we could in the previous section for the radial profiles.

\section{Best Fit Values}
\label{App:FitValues}

In the following tables we display the fits for all of the relations mentioned above. For each scaling relationship, at each redshift, we bootstrap our fits with 5000 iterations to gain fits for the binned medians of our data and the 16$^\mathrm{th}$ and 84$^\mathrm{th}$ percentiles in each of these bins. Hence, the errors on each value are the bootstrapped errors in these median fits, 84$^\mathrm{th}$ and 16$^\mathrm{th}$ percentile bounding region edges. This allows the intercluster variance to be calculated -- that is, for example, at some mass, $M_{500}$, the median $y$-weighted temperature at redshift $z=0$ is given by Equation \eqref{eqn:MassScalingRelation}, using A, B and C given by the first row of Table \ref{apptab:5:TvM_fits}. However, the 68 per cent confidence region of that value, given by the intrinsic intercluster variation can be found through using Equation \eqref{eqn:MassScalingRelation} using parameters given by rows 4 and 7 of Table \ref{apptab:5:TvM_fits}.

\subsection{Volume Averages over $R_{500}$}
\label{App:VolAvFits}
Tables \ref{apptab:5:TvM_fits} to \ref{apptab:M5_sigmafits} show the temperature--mass and temperature--temperature volume averaged scalings for the sphere of radius $R_{500}$.

\begin{table}
\caption{The fit values for the medians, 84th and 16th percentiles of each temperature measure at each redshift. The errors are determined through bootstrap methods. The fit parameters correspond to those described in Equation \eqref{eqn:MassScalingRelation}.} \centering
\begin{tabular}{lrrr}
$M=M_{500}$& \multicolumn{1}{c}{A} & \multicolumn{1}{c}{B} & \multicolumn{1}{c}{C}\\
\hline\hline
\multicolumn{4}{l}{$z = 0.0$, median} \\
\hline
$T^y$ 		  & $4.763^{+0.015}_{-0.015}$ & $0.581^{+0.003}_{-0.002}$ & $0.013^{+0.001}_{-0.001}$\\
\noalign{\smallskip}
$T^m$ 		  & $4.248^{+0.013}_{-0.012}$ & $0.565^{+0.003}_{-0.002}$ & $0.002^{+0.001}_{-0.001}$\\
\noalign{\smallskip}
$T_\mathrm{{sl}}$ & $4.295^{+0.023}_{-0.025}$ & $0.514^{+0.012}_{-0.013}$ & $-0.039^{+0.005}_{-0.005}$\\
\hline
\multicolumn{4}{l}{84\%} \\
\hline
$T^y$ 		  & $5.147^{+0.020}_{-0.020}$ & $0.568^{+0.002}_{-0.004}$ & $0.017^{+0.001}_{-0.002}$\\
\noalign{\smallskip}
$T^m$ 		  & $4.544^{+0.015}_{-0.015}$ & $0.564^{+0.002}_{-0.008}$ & $0.005^{+0.001}_{-0.003}$\\
\hline
\end{tabular}
\label{apptab:5:TvM_fits}
\end{table}

\begin{table}
\caption{The fit values for the medians, 84th and 16th percentiles of each temperature measure against $T_{500}$ at each redshift. The errors are determined through bootstrap methods. The fit parameters correspond to those described in Equation \eqref{eqn:TvTscalingRelation}.} \centering
\begin{tabular}{lrrr}
$T_\mathrm{rel}=T_{500}$& \multicolumn{1}{c}{A} & \multicolumn{1}{c}{B} & \multicolumn{1}{c}{C}\\
\hline\hline
\multicolumn{4}{l}{$z = 0.0$, median} \\
\hline
$T^y$ 		  & $4.812^{+0.014}_{-0.013}$ & $0.889^{+0.003}_{-0.003}$ & $ 0.041^{+0.002}_{-0.002}$\\
\noalign{\smallskip}
$T^m$		  & $4.289^{+0.011}_{-0.011}$ & $0.873^{+0.004}_{-0.004}$ & $ 0.021^{+0.002}_{-0.002}$\\
\noalign{\smallskip}
$T_\mathrm{{sl}}$ & $4.293^{+0.023}_{-0.022}$ & $0.825^{+0.018}_{-0.018}$ & $-0.049^{+0.010}_{-0.011}$\\
\hline
\multicolumn{4}{l}{84\%} \\
\hline
$T^y$ 		  & $5.174^{+0.023}_{-0.021}$ & $0.865^{+0.005}_{-0.005}$ & $ 0.047^{+0.003}_{-0.003}$\\
\noalign{\smallskip}
$T^m$ 		  & $4.575^{+0.013}_{-0.013}$ & $0.859^{+0.004}_{-0.004}$ & $ 0.023^{+0.002}_{-0.002}$\\
\hline
\end{tabular}
\label{apptab:5:TvTd_fits}
\end{table}

\begin{table}
\caption{The fit values for the medians, 84th and 16th percentiles of each temperature measure against $T^m$ at each redshift. The errors are determined through bootstrap methods. The fit parameters correspond to those described in Equation \eqref{eqn:TvTscalingRelation}.} \centering
\begin{tabular}{lrrr}
$T_\mathrm{rel}=T^m_{500}$& \multicolumn{1}{c}{A} & \multicolumn{1}{c}{B} & \multicolumn{1}{c}{C}\\
\hline\hline
\multicolumn{4}{l}{$z = 0.0$, median} \\
\hline
$T^y$ 		  & $5.650^{+0.015}_{-0.014}$ & $1.028^{+0.008}_{-0.008}$ & $0.029^{+0.004}_{-0.004}$\\
\noalign{\smallskip}
$T_\mathrm{{sl}}$ & $4.946^{+0.033}_{-0.037}$ & $0.927^{+0.029}_{-0.037}$ & $-0.095^{+0.014}_{-0.017}$\\
\hline
\multicolumn{4}{l}{84\%} \\
\hline
$T^y$ 		  & $5.946^{+0.024}_{-0.022}$ & $1.013^{+0.008}_{-0.008}$ & $0.031^{+0.005}_{-0.006}$\\
\noalign{\smallskip}
$T_\mathrm{{sl}}$ & $5.323^{+0.036}_{-0.038}$ & $0.944^{+0.012}_{-0.057}$ & $-0.065^{+0.009}_{-0.026}$\\
\hline
\multicolumn{4}{l}{16\%} \\
\hline
\end{tabular}
\label{apptab:5:TvmT_fits}
\end{table}

\begin{table}
\caption{The fit values for the medians, 84th and 16th percentiles of each temperature measure for the Hot and Hot, Relaxed Samples against $M_{500}$ at each redshift. The errors are determined through bootstrap methods. The fit parameters correspond to those described in Equation \eqref{eqn:MassScalingRelation}, taking $C=0$.}  \centering
\begin{tabular}{lrr|rr}
&\multicolumn{2}{c}{Hot Sample}&\multicolumn{2}{c}{Hot, Relaxed Sample}\\
$M_{500}$& \multicolumn{1}{c}{A} & \multicolumn{1}{c}{B} & \multicolumn{1}{c}{A} & \multicolumn{1}{c}{B} \\
\hline\hline
\multicolumn{4}{l}{$z = 0.0$, median} \\
\hline
$T^y$ & $4.693^{+0.028}_{-0.028}$ & $0.633^{+0.009}_{-0.010}$ & $4.635^{+0.035}_{-0.036}$ & $0.626^{+0.010}_{-0.010}$\\
\noalign{\smallskip}
$T^m$ & $4.174^{+0.023}_{-0.025}$ & $0.598^{+0.019}_{-0.010}$ & $4.147^{+0.033}_{-0.034}$ & $0.593^{+0.019}_{-0.013}$\\
\noalign{\smallskip}
$T_\mathrm{{sl}}$ & $4.117^{+0.064}_{-0.053}$ & $0.531^{+0.043}_{-0.055}$ & $4.206^{+0.049}_{-0.051}$ & $0.531^{+0.051}_{-0.014}$\\
\hline
\multicolumn{4}{l}{84\%} \\
\hline
$T^y$ & $5.157^{+0.035}_{-0.036}$ & $0.622^{+0.010}_{-0.010}$ & $4.992^{+0.042}_{-0.040}$ & $0.647^{+0.010}_{-0.028}$\\
\hline
\end{tabular}
\label{apptab:5:HotRelaxed_fits}
\end{table}

\begin{table}
\caption{The fit values for the medians, 84th and 16th percentiles of $\sigma(T^y)$ at each redshift. The errors are determined through bootstrap methods. The fit parameters correspond to those described in Equation \eqref{eqn:MassScalingRelation}, with $C=0$.} \centering
\begin{tabular}{lrrr}
$(\sigma(T^y)/T^y)(M_{500})$& \multicolumn{1}{c}{A} & \multicolumn{1}{c}{B}\\
\hline\hline
\multicolumn{3}{l}{$z = 0.0$} \\
\hline
median & $0.386^{+0.002}_{-0.003}$ & $0.026^{+0.003}_{-0.004}$\\
\noalign{\smallskip}
84\%   & $0.450^{+0.003}_{-0.003}$ & $-0.013^{+0.003}_{-0.003}$\\
\noalign{\smallskip}
16\%   & $0.332^{+0.003}_{-0.003}$ & $0.031^{+0.004}_{-0.004}$\\
\hline\hline
\multicolumn{3}{l}{$z = 0.5$} \\
\hline
median & $0.446^{+0.003}_{-0.003}$ & $0.022^{+0.003}_{-0.003}$\\
\noalign{\smallskip}
84\%   & $0.504^{+0.004}_{-0.003}$ & $-0.020^{+0.003}_{-0.003}$\\
\noalign{\smallskip}
16\%   & $0.400^{+0.003}_{-0.003}$ & $0.042^{+0.004}_{-0.004}$\\
\hline\hline
\multicolumn{3}{l}{$z = 1.0$} \\
\hline
median & $0.489^{+0.003}_{-0.003}$ & $0.017^{+0.002}_{-0.003}$\\
\noalign{\smallskip}
84\%   & $0.519^{+0.005}_{-0.005}$ & $-0.040^{+0.005}_{-0.005}$\\
\noalign{\smallskip}
16\%   & $0.455^{+0.004}_{-0.004}$ & $0.049^{+0.004}_{-0.003}$\\
\hline
\end{tabular}
\label{apptab:M5_sigmafits}
\end{table}

\subsection{Volume Averages over $R_{200}$}
Tables \ref{apptab:2:TvM_fits} to \ref{apptab:M2_sigmafits} show the same as the previous section, but for the sphere of radius $R_{200}$.

\begin{table}
\caption{The fit values for the medians, 84th and 16th percentiles of each temperature measure at each redshift. The errors are determined through bootstrap methods. The fit parameters correspond to those described in Equation \eqref{eqn:MassScalingRelation}.} \centering
\begin{tabular}{lrrr}
$M=M_{200}$& \multicolumn{1}{c}{A} & \multicolumn{1}{c}{B} & \multicolumn{1}{c}{C}\\
\hline\hline
\multicolumn{4}{l}{$z = 0.0$, median} \\
\hline
$T^y$ 		  & $3.488^{+0.010}_{-0.009}$ & $0.595^{+0.002}_{-0.002}$ & $0.016^{+0.001}_{-0.001}$\\
\noalign{\smallskip}
$T^m$ 		  & $2.974^{+0.010}_{-0.009}$ & $0.588^{+0.003}_{-0.004}$ & $0.008^{+0.001}_{-0.002}$\\
\noalign{\smallskip}
$T_\mathrm{{sl}}$ & $3.128^{+0.051}_{-0.023}$ & $0.572^{+0.011}_{-0.023}$ & $-0.022^{+0.005}_{-0.014}$\\
\hline
\multicolumn{4}{l}{84\%} \\
\hline
$T^y$ 		  & $3.764^{+0.016}_{-0.014}$ & $0.585^{+0.003}_{-0.003}$ & $0.018^{+0.002}_{-0.001}$\\
\noalign{\smallskip}
$T^m$ 		  & $3.200^{+0.012}_{-0.012}$ & $0.587^{+0.003}_{-0.003}$ & $0.008^{+0.001}_{-0.002}$\\
\hline
\end{tabular}
\label{apptab:2:TvM_fits}
\end{table}

\begin{table}
\caption{The fit values for the medians, 84th and 16th percentiles of each temperature measure against $T_{200}$ at each redshift. The errors are determined through bootstrap methods. The fit parameters correspond to those described in Equation \eqref{eqn:TvTscalingRelation}.} \centering
\begin{tabular}{lrrr}
$T_\mathrm{rel}=T_{200}$& \multicolumn{1}{c}{A} & \multicolumn{1}{c}{B} & \multicolumn{1}{c}{C}\\
\hline\hline
\multicolumn{4}{l}{$z = 0.0$, median} \\
\hline
$T^y$ 		  & $4.650^{+0.013}_{-0.014}$ & $0.918^{+0.004}_{-0.006}$ & $0.038^{+0.002}_{-0.003}$\\
\noalign{\smallskip}
$T^m$ 		  & $3.946^{+0.010}_{-0.011}$ & $0.896^{+0.004}_{-0.003}$ & $0.020^{+0.002}_{-0.002}$\\
\noalign{\smallskip}
$T_\mathrm{{sl}}$ & $4.101^{+0.029}_{-0.030}$ & $0.846^{+0.010}_{-0.025}$ & $-0.041^{+0.005}_{-0.012}$\\
\hline
\multicolumn{4}{l}{84\%} \\
\hline
$T^y$ 		  & $5.014^{+0.025}_{-0.024}$ & $0.900^{+0.006}_{-0.004}$ & $0.039^{+0.003}_{-0.003}$\\
\noalign{\smallskip}
$T^m$ 		  & $4.243^{+0.016}_{-0.016}$ & $0.891^{+0.005}_{-0.006}$ & $0.018^{+0.002}_{-0.003}$\\
\hline
\end{tabular}
\label{apptab:2:TvTd_fits}
\end{table}

\begin{table}
\caption{The fit values for the medians, 84th and 16th percentiles of each temperature measure against $T^m$ at each redshift. The errors are determined through bootstrap methods. The fit parameters correspond to those described in Equation \eqref{eqn:TvTscalingRelation}.} \centering
\begin{tabular}{lrrr}
$T_\mathrm{rel}=T^m_{200}$& \multicolumn{1}{c}{A} & \multicolumn{1}{c}{B} & \multicolumn{1}{c}{C}\\
\hline\hline
\multicolumn{4}{l}{$z = 0.0$, median} \\
\hline
$T^y$ 		  & $5.936^{+0.017}_{-0.016}$ & $1.030^{+0.007}_{-0.005}$ & $0.023^{+0.003}_{-0.003}$\\
\noalign{\smallskip}
$T_\mathrm{{sl}}$ & $5.148^{+0.047}_{-0.066}$ & $0.935^{+0.017}_{-0.076}$ & $-0.072^{+0.008}_{-0.034}$\\
\hline
\multicolumn{4}{l}{84\%} \\
\hline
$T^y$ 		  & $6.275^{+0.025}_{-0.024}$ & $1.021^{+0.008}_{-0.008}$ & $0.024^{+0.004}_{-0.004}$\\
\noalign{\smallskip}
$T_\mathrm{{sl}}$ & $5.647^{+0.031}_{-0.037}$ & $0.927^{+0.013}_{-0.017}$ & $-0.063^{+0.006}_{-0.008}$\\
\hline
\multicolumn{4}{l}{16\%} \\
\hline
\end{tabular}
\label{apptab:2:TvmT_fits}
\end{table}

\begin{table}
\caption{The fit values for the medians, 84th and 16th percentiles of each temperature measure for the Hot and Hot, Relaxed Samples against $M_{200}$ at each redshift. The errors are determined through bootstrap methods. The fit parameters correspond to those described in Equation \eqref{eqn:MassScalingRelation}, taking $C=0$.}  \centering
\begin{tabular}{lrr|rr}
&\multicolumn{2}{c}{Hot Sample}&\multicolumn{2}{c}{Hot, Relaxed Sample}\\
$M_{200}$& \multicolumn{1}{c}{A} & \multicolumn{1}{c}{B} & \multicolumn{1}{c}{A} & \multicolumn{1}{c}{B} \\
\hline\hline
\multicolumn{4}{l}{$z = 0.0$, median} \\
\hline
$T^y$ & $3.406^{+0.028}_{-0.045}$ & $0.633^{+0.025}_{-0.010}$ & $3.422^{+0.051}_{-0.053}$ & $0.633^{+0.013}_{-0.012}$\\
\noalign{\smallskip}
$T^m$ & $2.915^{+0.042}_{-0.041}$ & $0.609^{+0.020}_{-0.025}$ & $2.976^{+0.037}_{-0.042}$ & $0.600^{+0.016}_{-0.013}$\\
\noalign{\smallskip}
$T_\mathrm{{sl}}$ & $3.103^{+0.084}_{-0.116}$ & $0.515^{+0.067}_{-0.041}$ & $3.337^{+0.122}_{-0.104}$ & $0.516^{+0.043}_{-0.050}$\\
\hline
\multicolumn{4}{l}{84\%} \\
\hline
$T^y$ & $3.741^{+0.040}_{-0.033}$ & $0.644^{+0.010}_{-0.012}$ & $3.713^{+0.092}_{-0.067}$ & $0.650^{+0.014}_{-0.024}$\\
\hline
\end{tabular}
\label{apptab:2:HotRelaxed_fits}
\end{table}

\begin{table}
\caption{The fit values for the medians, 84th and 16th percentiles of $\sigma(T^y)$ at each redshift. The errors are determined through bootstrap methods. The fit parameters correspond to those described in Equation \eqref{eqn:MassScalingRelation}, with $C=0$.} \centering
\begin{tabular}{lrrr}
$(\sigma(T^y)/T^y)(M_{200})$& \multicolumn{1}{c}{A} & \multicolumn{1}{c}{B}\\
\hline\hline
\multicolumn{3}{l}{$z = 0.0$} \\
\hline
median & $0.437^{+0.002}_{-0.002}$ & $0.008^{+0.004}_{-0.002}$\\
\noalign{\smallskip}
84\%   & $0.508^{+0.003}_{-0.003}$ & $-0.023^{+0.003}_{-0.003}$\\
\noalign{\smallskip}
16\%   & $0.388^{+0.002}_{-0.003}$ & $0.016^{+0.003}_{-0.004}$\\
\hline\hline
\multicolumn{3}{l}{$z = 0.5$} \\
\hline
median & $0.496^{+0.003}_{-0.003}$ & $0.011^{+0.004}_{-0.003}$\\
\noalign{\smallskip}
84\%   & $0.560^{+0.004}_{-0.004}$ & $-0.026^{+0.004}_{-0.004}$\\
\noalign{\smallskip}
16\%   & $0.451^{+0.002}_{-0.002}$ & $0.028^{+0.003}_{-0.003}$\\
\hline\hline
\multicolumn{3}{l}{$z = 1.0$} \\
\hline
median & $0.535^{+0.003}_{-0.003}$ & $0.007^{+0.003}_{-0.003}$\\
\noalign{\smallskip}
84\%   & $0.579^{+0.003}_{-0.003}$ & $-0.044^{+0.003}_{-0.003}$\\
\noalign{\smallskip}
16\%   & $0.498^{+0.003}_{-0.002}$ & $0.035^{+0.003}_{-0.002}$\\
\hline
\end{tabular}
\label{apptab:M2_sigmafits}
\end{table}

\subsection{Volume Averaged $Y$ Fits}
\label{App:YFits}
We display the $Y-M$ and $Y-T^y$ relations over spheres of both radii in tables \ref{apptab:Y5_Tfits} to \ref{apptab:M2_Yfits}.

\begin{table}
\caption{The fit values for the medians, 84th and 16th percentiles of $T^y$ to $Y_{500}$ at each redshift. The errors are determined through bootstrap methods. The fit parameters correspond to those described in Equation \eqref{eqn:YscalingRelation}. This is a replica of Table \ref{tab:Y5_Tfits} found in Section \ref{stn:T_T_scalings}.} \centering
\begin{tabular}{lrrr}
$T^Y-Y_{500}$& \multicolumn{1}{c}{A} & \multicolumn{1}{c}{B} & \multicolumn{1}{c}{C}\\
\hline\hline
\multicolumn{4}{l}{$z = 0.0$} \\
\hline
median & $5.017^{+0.012}_{-0.011}$ & $0.3749^{+0.0014}_{-0.0018}$ & $0.0044^{+0.0003}_{-0.0004}$\\
\noalign{\smallskip}
84 & $5.375^{+0.019}_{-0.019}$ & $0.3654^{+0.0021}_{-0.0021}$ & $0.0043^{+0.0005}_{-0.0005}$\\
\noalign{\smallskip}
16 & $4.732^{+0.012}_{-0.012}$ & $0.3796^{+0.0020}_{-0.0018}$ & $0.0046^{+0.0005}_{-0.0004}$\\
\hline\hline
\multicolumn{4}{l}{$z = 0.5$} \\
\hline
median & $5.745^{+0.020}_{-0.020}$ & $0.3707^{+0.0043}_{-0.0038}$ & $0.0034^{+0.0009}_{-0.0008}$\\
\noalign{\smallskip}
84 & $6.096^{+0.027}_{-0.027}$ & $0.3612^{+0.0035}_{-0.0039}$ & $0.0033^{+0.0008}_{-0.0009}$\\
\noalign{\smallskip}
16 & $5.423^{+0.021}_{-0.020}$ & $0.3772^{+0.0033}_{-0.0029}$ & $0.0038^{+0.0007}_{-0.0006}$\\
\hline\hline
\multicolumn{4}{l}{$z = 1.0$} \\
\hline
median & $6.639^{+0.037}_{-0.041}$ & $0.3693^{+0.0054}_{-0.0065}$ & $0.0016^{+0.0011}_{-0.0013}$\\
\noalign{\smallskip}
84 & $7.029^{+0.048}_{-0.047}$ & $0.3555^{+0.0054}_{-0.0052}$ & $0.0008^{+0.0011}_{-0.0010}$\\
\noalign{\smallskip}
16 & $6.254^{+0.036}_{-0.036}$ & $0.3837^{+0.0056}_{-0.0052}$ & $0.0038^{+0.0011}_{-0.0011}$\\
\hline
\end{tabular}
\label{apptab:Y5_Tfits}
\end{table}

\begin{table}
\caption{The fit values for the medians, 84th and 16th percentiles of $T^y$ to $Y_{200}$ at each redshift. The errors are determined through bootstrap methods. The fit parameters correspond to those described in Equation \eqref{eqn:YscalingRelation}.} \centering
\begin{tabular}{lrrr}
$T^Y-Y_{200}$& \multicolumn{1}{c}{A} & \multicolumn{1}{c}{B} & \multicolumn{1}{c}{C}\\
\hline\hline
\multicolumn{4}{l}{$z = 0.0$} \\
\hline
median & $4.197^{+0.011}_{-0.010}$ & $0.3801^{+0.0010}_{-0.0014}$ & $0.0045^{+0.0003}_{-0.0003}$\\
\noalign{\smallskip}
84 & $4.464^{+0.019}_{-0.017}$ & $0.3737^{+0.0021}_{-0.0016}$ & $0.0046^{+0.0005}_{-0.0004}$\\
\noalign{\smallskip}
16 & $3.954^{+0.011}_{-0.010}$ & $0.3831^{+0.0016}_{-0.0012}$ & $0.0048^{+0.0004}_{-0.0003}$\\
\hline\hline
\multicolumn{4}{l}{$z = 0.5$} \\
\hline
median & $4.822^{+0.017}_{-0.016}$ & $0.3839^{+0.0042}_{-0.0037}$ & $0.0043^{+0.0009}_{-0.0008}$\\
\noalign{\smallskip}
84 & $5.184^{+0.026}_{-0.025}$ & $0.3805^{+0.0035}_{-0.0040}$ & $0.0047^{+0.0008}_{-0.0009}$\\
\noalign{\smallskip}
16 & $4.538^{+0.014}_{-0.016}$ & $0.3876^{+0.0030}_{-0.0025}$ & $0.0045^{+0.0007}_{-0.0006}$\\
\hline\hline
\multicolumn{4}{l}{$z = 1.0$} \\
\hline
median & $5.637^{+0.033}_{-0.032}$ & $0.3918^{+0.0053}_{-0.0059}$ & $0.0040^{+0.0011}_{-0.0012}$\\
\noalign{\smallskip}
84 & $6.055^{+0.047}_{-0.044}$ & $0.3774^{+0.0049}_{-0.0047}$ & $0.0022^{+0.0010}_{-0.0010}$\\
\noalign{\smallskip}
16 & $5.278^{+0.028}_{-0.029}$ & $0.4028^{+0.0046}_{-0.0037}$ & $0.0058^{+0.0009}_{-0.0008}$\\
\hline
\end{tabular}
\label{apptab:Y2_Tfits}
\end{table}

\begin{table}
\caption{The fit values for the medians, 84th and 16th percentiles of $Y_{500}$ to $M_{500}$ at each redshift. The errors are determined through bootstrap methods. The fit parameters correspond to those described in Equation \eqref{eqn:MassScalingRelation}, with $Y$ in the place of $T$.} \centering
\begin{tabular}{lrrr}
$Y-M_{500}$& \multicolumn{1}{c}{A $[\times 10^{-4}]$} & \multicolumn{1}{c}{B} & \multicolumn{1}{c}{C}\\
\hline\hline
\multicolumn{4}{l}{$z = 0.0$} \\
\hline
median  & $2.528^{+0.014}_{-0.013}$ & $1.563^{+0.004}_{-0.003}$ & $0.0020^{+0.0017}_{-0.0013}$\\
\noalign{\smallskip}
84 	& $2.855^{+0.014}_{-0.014}$ & $1.559^{+0.003}_{-0.012}$ & $0.0028^{+0.0014}_{-0.0048}$\\
\noalign{\smallskip}
16 	& $2.236^{+0.011}_{-0.011}$ & $1.576^{+0.004}_{-0.004}$ & $0.0056^{+0.0015}_{-0.0015}$\\
\hline\hline
\multicolumn{4}{l}{$z = 0.5$} \\
\hline
median  & $2.201^{+0.011}_{-0.012}$ & $1.545^{+0.005}_{-0.005}$ & $-0.0056^{+0.0019}_{-0.0018}$\\
\noalign{\smallskip}
84 	& $2.448^{+0.015}_{-0.014}$ & $1.532^{+0.007}_{-0.008}$ & $-0.0066^{+0.0027}_{-0.0029}$\\
\noalign{\smallskip}
16 	& $2.002^{+0.009}_{-0.009}$ & $1.566^{+0.007}_{-0.005}$ & $-0.0021^{+0.0025}_{-0.0017}$\\
\hline\hline
\multicolumn{4}{l}{$z = 1.0$} \\
\hline
median & $1.925^{+0.012}_{-0.014}$ & $1.562^{+0.005}_{-0.006}$ & $-0.0039^{+0.0020}_{-0.0020}$\\
\noalign{\smallskip}
84 & $2.130^{+0.019}_{-0.021}$ & $1.520^{+0.011}_{-0.011}$ & $-0.0122^{+0.0034}_{-0.0035}$\\
\noalign{\smallskip}
16 & $1.749^{+0.015}_{-0.015}$ & $1.596^{+0.010}_{-0.010}$ & $0.0027^{+0.0033}_{-0.0032}$\\
\hline
\end{tabular}
\label{apptab:M5_Yfits}
\end{table}

\begin{table}
\caption{The fit values for the medians, 84th and 16th percentiles of $Y_{200}$ to $M_{200}$ at each redshift. The errors are determined through bootstrap methods. The fit parameters correspond to those described in Equation \eqref{eqn:MassScalingRelation}, with $Y$ in the place of $T$.} \centering
\begin{tabular}{lrrr}
$Y-M_{200}$& \multicolumn{1}{c}{A $[\times 10^{-4}]$} & \multicolumn{1}{c}{B} & \multicolumn{1}{c}{C}\\
\hline\hline
\multicolumn{4}{l}{$z = 0.0$} \\
\hline
median & $1.751^{+0.009}_{-0.009}$ & $1.587^{+0.003}_{-0.003}$ & $0.0096^{+0.0017}_{-0.0016}$\\
\noalign{\smallskip}
84 & $2.024^{+0.011}_{-0.012}$ & $1.580^{+0.004}_{-0.004}$ & $0.0034^{+0.0020}_{-0.0019}$\\
\noalign{\smallskip}
16 & $1.546^{+0.009}_{-0.008}$ & $1.592^{+0.004}_{-0.005}$ & $0.0121^{+0.0022}_{-0.0023}$\\
\hline\hline
\multicolumn{4}{l}{$z = 0.5$} \\
\hline
median & $1.552^{+0.007}_{-0.007}$ & $1.585^{+0.005}_{-0.012}$ & $0.0015^{+0.0021}_{-0.0053}$\\
\noalign{\smallskip}
84 & $1.757^{+0.010}_{-0.009}$ & $1.573^{+0.006}_{-0.007}$ & $-0.0044^{+0.0024}_{-0.0027}$\\
\noalign{\smallskip}
16 & $1.366^{+0.008}_{-0.008}$ & $1.589^{+0.009}_{-0.010}$ & $0.0044^{+0.0040}_{-0.0039}$\\
\hline\hline
\multicolumn{4}{l}{$z = 1.0$} \\
\hline
median & $1.354^{+0.012}_{-0.011}$ & $1.606^{+0.012}_{-0.018}$ & $0.0045^{+0.0047}_{-0.0065}$\\
\noalign{\smallskip}
84 & $1.568^{+0.013}_{-0.012}$ & $1.560^{+0.007}_{-0.012}$ & $-0.0154^{+0.0030}_{-0.0042}$\\
\noalign{\smallskip}
16 & $1.193^{+0.009}_{-0.009}$ & $1.629^{+0.012}_{-0.008}$ & $0.0122^{+0.0046}_{-0.0026}$\\
\hline
\end{tabular}
\label{apptab:M2_Yfits}
\end{table}

\subsection{Profile Fits}
\label{App:ProfFits}
In tables \ref{tab:RadProfz0p0} to \ref{tab:MomradProf}, we display the fit quantities for the radial profiles of the median temperature measures, ($T/T_{500}$) and variance, $\sigma(T^y)/T_{500}$. The same quantities for the cylindrical profiles are in tables \ref{tab:cy6Profz0p0} to \ref{tab:Momcy6Prof}.

These mass bins are organised so that the highest mass bin always corresponds to the MACSIS sample, hence the discrepency in mass bin sizes.

\begin{table*}
\caption{The fit values for the medians of the radial temperature profiles, $T/T_{500}$, at $z=0$. The errors are determined through bootstrap methods -- errors written as 0.00, correspond to very small values, $<10^{-9}$. The fit parameters correspond to those described in Equation \eqref{eqn:VikFits}. $m_{500} = M_{500}/M_\odot$.} \centering
\begin{tabular}{lrrrrrrrr}
$z=0$ & \multicolumn{1}{c}{$T_0$} & \multicolumn{1}{c}{$r_t$} & \multicolumn{1}{c}{$a$} & \multicolumn{1}{c}{$b$} & \multicolumn{1}{c}{$c$} & \multicolumn{1}{c}{$r_\mathrm{cool}$} & \multicolumn{1}{c}{$a_\mathrm{cool}$} & \multicolumn{1}{c}{$T_\mathrm{min}$} \\
\hline\hline
\multicolumn{9}{l}{$\mathrm{log}_{10}m_{500}<13.5$} \\
\hline
$T^y$ 		  & $2.44^{+0.89}_{-0.88}$ & $0.26^{+0.28}_{-0.11}$ & $0.27^{+0.31}_{-0.37}$ & $5.00^{+0.00}_{-1.37}$ & $0.33^{+0.39}_{-0.33}$ & $0.193^{+0.023}_{-0.055}$ & $2.52^{+0.48}_{-0.231}$ & $0.49^{+0.29}_{-0.12}$\\
\noalign{\smallskip}
$T^m$ 		  & $1.29^{+2.20}_{-0.09}$ & $0.69^{+0.08}_{-0.49}$ & $0.32^{+0.12}_{-0.28}$ & $5.00^{+0.00}_{-1.13}$ & $0.42^{+0.32}_{-0.11}$ & $0.161^{+0.112}_{-0.015}$ & $2.92^{+0.08}_{-0.875}$ & $0.39^{+0.29}_{-0.14}$\\
\noalign{\smallskip}
$T_\mathrm{{sl}}$ & $1.07^{+1.13}_{-0.19}$ & $1.00^{+0.17}_{-0.23}$ & $0.33^{+0.12}_{-0.31}$ & $2.33^{+1.01}_{-1.15}$ & $0.73^{+0.51}_{-0.16}$ & $0.159^{+0.014}_{-0.015}$ & $3.00^{+0.00}_{-0.239}$ & $0.33^{+0.46}_{-0.11}$\\
\hline\hline
\multicolumn{9}{l}{$13.5<\mathrm{log}_{10}m_{500}<14.0$} \\
\hline
$T^y$ 		  & $1.41^{+0.17}_{-0.05}$ & $0.50^{+0.04}_{-0.08}$ & $0.32^{+0.05}_{-0.15}$ & $5.00^{+0.00}_{-1.40}$ & $0.28^{+0.15}_{-0.05}$ & $0.103^{+0.008}_{-0.022}$ & $2.22^{+0.35}_{-0.20}$ & $0.32^{+0.11}_{-0.11}$\\
\noalign{\smallskip}
$T^m$ 		  & $1.11^{+0.72}_{-0.08}$ & $0.76^{+0.07}_{-0.34}$ & $0.56^{+0.09}_{-0.65}$ & $5.00^{+0.00}_{-3.01}$ & $0.24^{+0.67}_{-0.07}$ & $0.152^{+0.027}_{-0.071}$ & $1.69^{+1.31}_{-0.19}$ & $0.05^{+0.27}_{-0.05}$\\
\hline
\hline
\end{tabular}
\label{tab:RadProfz0p0}
\end{table*}

\begin{table*}
\caption{The fit values for the medians of the radial temperature profiles, $T/T_{500}$, at $z=0.5$. The errors are determined through bootstrap methods -- errors written as 0.00, correspond to very small values, $<10^{-9}$. The fit parameters correspond to those described in Equation \eqref{eqn:VikFits}.} \centering
\begin{tabular}{lrrrrrrrr}
$z=0.5$ & \multicolumn{1}{c}{$T_0$} & \multicolumn{1}{c}{$r_t$} & \multicolumn{1}{c}{$a$} & \multicolumn{1}{c}{$b$} & \multicolumn{1}{c}{$c$} & \multicolumn{1}{c}{$r_\mathrm{cool}$} & \multicolumn{1}{c}{$a_\mathrm{cool}$} & \multicolumn{1}{c}{$T_\mathrm{min}$} \\
\hline\hline
\multicolumn{9}{l}{$\mathrm{log}_{10}m_{500}<13.5$} \\
\hline
$T^y$ 		  & $1.53^{+0.20}_{-0.07}$ & $0.51^{+0.05}_{-0.10}$ & $0.23^{+0.09}_{-0.19}$ & $5.00^{+0.00}_{-1.48}$ & $0.38^{+0.18}_{-0.09}$ & $0.115^{+0.013}_{-0.022}$ & $2.80^{+0.20}_{-0.53}$ & $0.49^{+0.19}_{-0.18}$\\
\noalign{\smallskip}
$T^m$ 		  & $1.34^{+0.49}_{-0.23}$ & $0.65^{+0.19}_{-0.19}$ & $0.26^{+0.33}_{-0.35}$ & $3.19^{+1.81}_{-1.04}$ & $0.53^{+0.37}_{-0.32}$ & $0.126^{+0.050}_{-0.031}$ & $2.59^{+0.41}_{-0.80}$ & $0.29^{+0.50}_{-0.24}$\\
\noalign{\smallskip}
$T_\mathrm{{sl}}$ & $0.62^{+0.96}_{-0.12}$ & $1.83^{+0.40}_{-0.78}$ & $0.57^{+0.25}_{-0.47}$ & $1.77^{+3.23}_{-0.77}$ & $0.92^{+0.62}_{-0.52}$ & $0.184^{+0.072}_{-0.064}$ & $1.71^{+0.99}_{-0.16}$ & $0.014^{+0.391}_{-0.014}$\\
\hline\hline
\multicolumn{9}{l}{$13.5<\mathrm{log}_{10}m_{500}<14.0$} \\
\hline
$T^y$ 		  & $1.28^{+0.08}_{-0.05}$ & $0.50^{+0.04}_{-0.06}$ & $0.33^{+0.05}_{-0.07}$ & $5.00^{+0.00}_{-0.91}$ & $0.24^{+0.07}_{-0.05}$ & $0.103^{+0.006}_{-0.008}$ & $2.37^{+0.37}_{-0.23}$ & $0.40^{+0.13}_{-0.09}$\\
\noalign{\smallskip}
$T^m$ 		  & $0.86^{+0.09}_{-0.05}$ & $0.99^{+0.10}_{-0.12}$ & $0.61^{+0.07}_{-0.13}$ & $4.84^{+0.16}_{-1.72}$ & $0.22^{+0.13}_{-0.05}$ & $0.155^{+0.021}_{-0.023}$ & $1.70^{+0.29}_{-0.16}$ & $0.07^{+0.10}_{-0.04}$\\
\hline\hline
\end{tabular}
\label{tab:RadProfz0p5}
\end{table*}

\begin{table*}
\caption{The fit values for the medians of the radial temperature profiles, $T/T_{500}$, at $z=1$. The errors are determined through bootstrap methods --  errors written as 0.00, correspond to very small values, $<10^{-9}$. The fit parameters correspond to those described in Equation \eqref{eqn:VikFits}.}
\begin{tabular}{lrrrrrrrr}
$z=1$ & \multicolumn{1}{c}{$T_0$} & \multicolumn{1}{c}{$r_t$} & \multicolumn{1}{c}{$a$} & \multicolumn{1}{c}{$b$} & \multicolumn{1}{c}{$c$} & \multicolumn{1}{c}{$r_\mathrm{cool}$} & \multicolumn{1}{c}{$a_\mathrm{cool}$} & \multicolumn{1}{c}{$T_\mathrm{min}$} \\
\hline\hline
\multicolumn{9}{l}{$\mathrm{log}_{10}m_{500}<13.5$} \\
\hline
$T^y$ 		  & $1.36^{+0.18}_{-0.06}$ & $0.53^{+0.05}_{-0.10}$ & $0.36^{+0.06}_{-0.14}$ & $5.00^{+0.00}_{-1.34}$ & $0.24^{+0.14}_{-0.05}$ & $0.112^{+0.008}_{-0.016}$ & $2.30^{+0.47}_{-0.23}$ & $0.33^{+0.14}_{-0.11}$\\
\noalign{\smallskip}
$T^m$ 		  & $1.07^{+0.62}_{-0.14}$ & $0.80^{+0.16}_{-0.33}$ & $0.51^{+0.16}_{-0.49}$ & $3.73^{+1.27}_{-1.75}$ & $0.30^{+0.51}_{-0.15}$ & $0.144^{+0.040}_{-0.045}$ & $1.90^{+1.10}_{-0.33}$ & $0.11^{+0.69}_{-0.08}$\\
\noalign{\smallskip}
$T_\mathrm{{sl}}$ & $0.44^{+0.03}_{-0.04}$ & $2.18^{+0.34}_{-0.16}$ & $0.84^{+0.05}_{-0.11}$ & $4.88^{+0.12}_{-2.25}$ & $0.59^{+0.32}_{-0.07}$ & $0.252^{+0.028}_{-0.046}$ & $1.53^{+0.12}_{-0.05}$ & $0.004^{+0.014}_{-0.004}$\\
\hline\hline
\multicolumn{9}{l}{$13.5<\mathrm{log}_{10}m_{500}<14.0$} \\
\hline
$T^y$ 		  & $0.86^{+1.31}_{-0.06}$ & $0.91^{+0.13}_{-0.73}$ & $0.45^{+0.03}_{-0.93}$ & $5.00^{+0.00}_{-2.90}$ & $0.11^{+0.83}_{-0.03}$ & $0.115^{+0.140}_{-0.006}$ & $1.95^{+1.05}_{-0.24}$ & $0.18^{+2.80}_{-0.06}$\\
\noalign{\smallskip}
$T^m$ 		  & $0.59^{+0.04}_{-0.04}$ & $1.59^{+0.19}_{-0.14}$ & $0.57^{+0.04}_{-0.05}$ & $3.34^{+1.00}_{-0.84}$ & $0.41^{+0.11}_{-0.06}$ & $0.138^{+0.010}_{-0.010}$ & $1.73^{+0.14}_{-0.13}$ & $0.06^{+0.03}_{-0.03}$\\
\hline
\hline
\end{tabular}
\label{tab:RadProfz1p0}
\end{table*}

\begin{table*}
\caption{The fit values for the medians of the radial profiles of $\sigma(T^y)/T_{500}$ across all redshifts. The errors are determined through bootstrap methods --  errors written as 0.00, correspond to very small values, $<10^{-9}$. The fit parameters correspond to those described in Equation \eqref{eqn:VikFits}.} \centering
\begin{tabular}{lrrrrrrrr}
$\sigma(T^y)$ & \multicolumn{1}{c}{$T_0$} & \multicolumn{1}{c}{$r_t$} & \multicolumn{1}{c}{$a$} & \multicolumn{1}{c}{$b$} & \multicolumn{1}{c}{$c$} & \multicolumn{1}{c}{$r_\mathrm{cool}$} & \multicolumn{1}{c}{$a_\mathrm{cool}$} & \multicolumn{1}{c}{$T_\mathrm{min}$} \\
\hline\hline
\multicolumn{9}{l}{$z=0$} \\
\hline
$\mathrm{log}_{10}m_{500}<13.5$ & $0.401^{+0.016}_{-0.386}$ & $0.186^{+0.018}_{-0.010}$ & $-3.00^{+0.00}_{-0.00}$ & $5.00^{+0.00}_{-0.00}$ & $3.06^{+0.02}_{-0.94}$ & $0.118^{+0.300}_{-0.010}$    & $3.00^{+0.00}_{-1.33}$ & $0.00^{+0.47}_{-0.00}$\\
$13.5<\mathrm{log}_{10}m_{500}<14.0$ & $0.046^{+0.029}_{-0.012}$ & $0.093^{+0.004}_{-0.003}$ & $-3.00^{+0.00}_{-0.00}$ & $5.00^{+0.00}_{-0.00}$ & $2.58^{+0.12}_{-0.06}$ & $ 0.30^{+0.10}_{-0.042}$     & $1.27^{+0.10}_{-0.04}$ & $0.418^{+0.023}_{-0.020}$\\
$14.0<\mathrm{log}_{10}m_{500}<14.5$  & $0.286^{+0.025}_{-0.012}$ & $ 1.04^{+0.07}_{-0.14}$   & $-1.58^{+0.26}_{-0.21}$ & $3.83^{+0.73}_{-0.34}$ & $1.62^{+0.20}_{-0.21}$ & $0.075^{+0.022}_{-0.009}$    & $2.23^{+0.25}_{-0.10}$ & $36.7^{+41.8}_{-24.8}$\\
$14.5<\mathrm{log}_{10}m_{500}<14.78$  & $0.374^{+0.056}_{-0.022}$ & $ 1.09^{+0.12}_{-0.98}$   & $-0.61^{+0.13}_{-0.17}$ & $4.37^{+0.63}_{-1.62}$ & $0.87^{+0.18}_{-0.15}$ & $ 0.16^{+0.60}_{-0.03}$      & $2.67^{+0.33}_{-0.47}$ & $1.40^{+1.12}_{-1.06}$\\
$14.78<\mathrm{log}_{10}m_{500}$ & $ 1.40^{+0.25}_{-0.23}$   & $ 1.85^{+2.52}_{-0.86}$   & $-0.86^{+0.57}_{-0.26}$ & $1.00^{+0.17}_{-0.00}$ & $2.07^{+1.02}_{-0.16}$ & $0.0023^{+0.0100}_{-0.0007}$ & $1.27^{+0.67}_{-0.12}$ & $231^{+727}_{-226}$\\
\hline\hline
\multicolumn{9}{l}{$z=0.5$} \\
\hline\hline
\end{tabular}
\label{tab:MomradProf}
\end{table*}

\begin{table*}
\caption{The fit values for the medians of the cylindrical temperature profiles, $T/T_{500}$, at $z=0$. The errors are determined through bootstrap methods -- errors written as 0.00, correspond to very small values, $<10^{-9}$. The fit parameters correspond to those described in Equation \eqref{eqn:VikFits}.} \centering
\begin{tabular}{lrrrrrrrr}
$z=0$ & \multicolumn{1}{c}{$T_0$} & \multicolumn{1}{c}{$r_t$} & \multicolumn{1}{c}{$a$} & \multicolumn{1}{c}{$b$} & \multicolumn{1}{c}{$c$} & \multicolumn{1}{c}{$r_\mathrm{cool}$} & \multicolumn{1}{c}{$a_\mathrm{cool}$} & \multicolumn{1}{c}{$T_\mathrm{min}$} \\
\hline\hline
\multicolumn{9}{l}{$\mathrm{log}_{10}m_{500}<13.5$} \\
\hline
$T^y$ 		  & $0.94^{+0.02}_{-0.02}$ & $2.13^{+0.36}_{-0.26}$ & $ 0.24^{+0.02}_{-0.02}$ & $1.10^{+0.25}_{-0.09}$ & $0.00^{+0.00}_{-0.00}$ & $0.120^{+0.006}_{-0.007}$ & $2.39^{+0.26}_{-0.24}$ & $0.37^{+0.04}_{-0.05}$\\
\noalign{\smallskip}
$T^m$ 		  & $1.04^{+0.19}_{-0.11}$ & $0.38^{+0.14}_{-0.11}$ & $ 0.10^{+0.16}_{-0.21}$ & $4.23^{+5.80}_{-2.09}$ & $0.24^{+0.22}_{-0.14}$ & $0.092^{+0.057}_{-0.058}$ & $3.00^{+0.00}_{-1.41}$ & $0.64^{+0.43}_{-0.64}$\\
\noalign{\smallskip}
$T_\mathrm{{sl}}$ & $1.17^{+0.45}_{-0.22}$ & $0.40^{+0.26}_{-0.09}$ & $ 0.05^{+0.53}_{-0.27}$ & $2.85^{+2.13}_{-1.49}$ & $0.48^{+0.29}_{-0.40}$ & $0.100^{+0.184}_{-0.054}$ & $3.00^{+0.00}_{-1.63}$ & $0.68^{+0.62}_{-0.68}$\\
\hline\hline
\multicolumn{9}{l}{$13.5<\mathrm{log}_{10}m_{500}<14.0$} \\
\hline
$T^y$ 		  & $0.87^{+0.07}_{-0.05}$ & $1.03^{+0.14}_{-0.16}$ & $ 0.45^{+0.02}_{-0.02}$ & $1.02^{+0.10}_{-0.02}$ & $0.00^{+0.00}_{-0.00}$ & $0.119^{+0.006}_{-0.005}$ & $1.61^{+0.14}_{-0.13}$ & $0.18^{+0.04}_{-0.04}$\\
\noalign{\smallskip}
$T^m$ 		  & $1.31^{+0.77}_{-0.18}$ & $0.34^{+0.05}_{-0.07}$ & $-0.10^{+0.08}_{-0.15}$ & $1.68^{+0.46}_{-0.68}$ & $0.73^{+0.31}_{-0.14}$ & $0.053^{+0.477}_{-0.034}$ & $3.00^{+0.00}_{-0.36}$ & $0.15^{+1.81}_{-0.15}$\\
\hline\hline
\end{tabular}
\label{tab:cy6Profz0p0}
\end{table*}

\begin{table*}
\caption{The fit values for the medians of the cylindrical temperature profiles, $T/T_{500}$, at $z=0.5$. The errors are determined through bootstrap methods -- errors written as 0.00, correspond to very small values, $<10^{-9}$. The fit parameters correspond to those described in Equation \eqref{eqn:VikFits}.} \centering
\begin{tabular}{lrrrrrrrr}
$z=0.5$ & \multicolumn{1}{c}{$T_0$} & \multicolumn{1}{c}{$r_t$} & \multicolumn{1}{c}{$a$} & \multicolumn{1}{c}{$b$} & \multicolumn{1}{c}{$c$} & \multicolumn{1}{c}{$r_\mathrm{cool}$} & \multicolumn{1}{c}{$a_\mathrm{cool}$} & \multicolumn{1}{c}{$T_\mathrm{min}$} \\
\hline\hline
\multicolumn{9}{l}{$\mathrm{log}_{10}m_{500}<13.5$} \\
\hline
$T^y$ 		  & $0.85^{+0.04}_{-0.02}$ & $1.45^{+0.16}_{-0.20}$ & $0.38^{+0.02}_{-0.03}$ & $1.02^{+0.13}_{-0.02}$ & $0.00^{+0.00}_{-0.00}$ & $0.121^{+0.006}_{-0.006}$ & $1.86^{+0.19}_{-0.17}$ & $0.23^{+0.04}_{-0.04}$\\
\noalign{\smallskip}
$T^m$ 		  & $0.95^{+0.29}_{-0.53}$ & $0.47^{+1.58}_{-0.13}$ & $0.03^{+0.56}_{-0.16}$ & $3.97^{+3.77}_{-1.96}$ & $0.29^{+0.31}_{-0.29}$ & $0.206^{+0.334}_{-0.104}$ & $1.88^{+1.12}_{-0.66}$ & $0.76^{+0.63}_{-0.72}$\\
\noalign{\smallskip}
$T_\mathrm{{sl}}$ & $0.97^{+0.26}_{-0.36}$ & $0.59^{+0.25}_{-0.19}$ & $0.17^{+0.51}_{-0.34}$ & $3.40^{+6.38}_{-1.43}$ & $0.53^{+0.39}_{-0.40}$ & $0.178^{+0.218}_{-0.075}$ & $2.47^{+0.53}_{-1.10}$ & $0.68^{+0.86}_{-0.64}$\\
\hline\hline
\multicolumn{9}{l}{$13.5<\mathrm{log}_{10}m_{500}<14.0$} \\
\hline
$T^y$ 		  & $1.12^{+0.15}_{-0.39}$ & $0.47^{+0.80}_{-0.12}$ & $0.22^{+0.24}_{-0.10}$ & $2.74^{+110}_{-1.05}$  & $0.22^{+0.13}_{-0.22}$ & $0.090^{+0.019}_{-0.010}$ & $3.00^{+0.00}_{-1.60}$ & $0.74^{+0.19}_{-0.62}$\\ 
\noalign{\smallskip}
$T^m$ 		  & $0.74^{+0.23}_{-0.26}$ & $0.74^{+0.14}_{-0.44}$ & $0.20^{+0.48}_{-0.34}$ & $4.72^{+10.2}_{-2.86}$ & $0.27^{+0.52}_{-0.18}$ & $0.214^{+0.268}_{-0.103}$ & $1.44^{+1.56}_{-0.27}$ & $0.46^{+1.04}_{-0.42}$\\
\hline\hline
\end{tabular}
\label{tab:cy6Profz0p5}
\end{table*}

\begin{table*}
\caption{The fit values for the medians of the cylindrical temperature profiles, $T/T_{500}$, at $z=1$. The errors are determined through bootstrap methods --  errors written as 0.00, correspond to very small values, $<10^{-9}$. The fit parameters correspond to those described in Equation \eqref{eqn:VikFits}.} \centering
\begin{tabular}{lrrrrrrrr}
$z=1$ & \multicolumn{1}{c}{$T_0$} & \multicolumn{1}{c}{$r_t$} & \multicolumn{1}{c}{$a$} & \multicolumn{1}{c}{$b$} & \multicolumn{1}{c}{$c$} & \multicolumn{1}{c}{$r_\mathrm{cool}$} & \multicolumn{1}{c}{$a_\mathrm{cool}$} & \multicolumn{1}{c}{$T_\mathrm{min}$} \\
\hline\hline
\multicolumn{9}{l}{$\mathrm{log}_{10}m_{500}<13.5$} \\
\hline
$T^y$ 		  & $1.35^{+0.29}_{-0.34}$ & $0.34^{+0.37}_{-0.08}$ & $ 0.24^{+0.26}_{-0.28}$ & $2.54^{+75.1}_{-1.32}$ & $0.22^{+0.28}_{-0.22}$ & $0.088^{+0.039}_{-0.050}$ & $1.91^{+1.09}_{-0.68}$ & $0.20^{+0.82}_{-0.20}$\\
\noalign{\smallskip}
$T^m$ 		  & $1.03^{+0.83}_{-0.97}$ & $0.36^{+0.40}_{-0.18}$ & $-0.15^{+0.54}_{-0.12}$ & $1.73^{+15.0}_{-0.72}$ & $0.76^{+0.38}_{-0.66}$ & $0.289^{+1.170}_{-0.224}$ & $3.00^{+0.00}_{-1.85}$ & $1.27^{+0.77}_{-1.24}$\\
\noalign{\smallskip}
$T_\mathrm{{sl}}$ & $0.73^{+0.12}_{-0.31}$ & $0.88^{+0.70}_{-0.19}$ & $ 0.42^{+0.16}_{-0.39}$ & $3.38^{+1.52}_{-1.60}$ & $0.73^{+1.23}_{-0.30}$ & $0.119^{+0.120}_{-0.014}$ & $2.04^{+0.84}_{-0.44}$ & $0.27^{+0.77}_{-0.17}$\\
\hline\hline
\multicolumn{9}{l}{$13.5<\mathrm{log}_{10}m_{500}<14.0$} \\
\hline
$T^y$ 		  & $0.76^{+0.33}_{-0.34}$ & $0.91^{+0.35}_{-0.66}$ & $ 0.41^{+0.07}_{-0.54}$ & $10.5^{+454}_{-8.68}$  & $0.10^{+0.39}_{-0.06}$ & $0.113^{+0.601}_{-0.018}$ & $1.44^{+1.56}_{-0.31}$ & $0.23^{+1.22}_{-0.14}$\\
\noalign{\smallskip}
$T^m$ 		  & $0.57^{+0.41}_{-0.40}$ & $0.92^{+0.32}_{-0.21}$ & $-0.12^{+0.08}_{-0.07}$ & $3.61^{+4.89}_{-2.14}$ & $0.72^{+0.80}_{-0.49}$ & $0.335^{+0.072}_{-0.101}$ & $1.62^{+0.95}_{-0.39}$ & $1.46^{+0.45}_{-0.38}$\\
\hline\hline
\end{tabular}
\label{tab:cy6Profz1p0}
\end{table*}

\begin{table*}
\caption{The fit values for the medians of the cylindrical profiles of $\sigma(T^y)/T_{500}$ across all redshifts. The errors are determined through bootstrap methods --  errors written as 0.00, correspond to very small values, $<10^{-9}$. The fit parameters correspond to those described in Equation \eqref{eqn:VikFits}.} \centering
\begin{tabular}{lrrrrrrrr}
$\sigma(T^y)$ & \multicolumn{1}{c}{$T_0$} & \multicolumn{1}{c}{$r_t$} & \multicolumn{1}{c}{$a$} & \multicolumn{1}{c}{$b$} & \multicolumn{1}{c}{$c$} & \multicolumn{1}{c}{$r_\mathrm{cool}$} & \multicolumn{1}{c}{$a_\mathrm{cool}$} & \multicolumn{1}{c}{$T_\mathrm{min}$} \\
\hline\hline
\multicolumn{9}{l}{$z=0$} \\
\hline
$\mathrm{log}_{10}m_{500}<13.5$ & $0.44^{+0.99}_{-0.08}$ & $0.27^{+0.09}_{-0.19}$ & $-3.00^{+0.00}_{-0.00}$ & $3.76^{+0.98}_{-1.94}$    & $2.76^{+0.35}_{-0.18}$ & $0.066^{+1.360}_{-0.005}$ & $2.71^{+0.29}_{-0.18}$ & $26.5^{+36.0}_{-25.2}$\\
\noalign{\smallskip}
$13.5<\mathrm{log}_{10}m_{500}<14.0$ & $0.30^{+0.49}_{-0.16}$ & $0.64^{+0.74}_{-0.50}$ & $-1.15^{+0.12}_{-0.11}$ & $3.9^{+20.0}_{-1.9}$      & $0.85^{+0.81}_{-0.65}$ & $0.133^{+0.044}_{-0.021}$ & $1.76^{+0.44}_{-0.10}$ & $5.5^{+16.5}_{-4.0}$\\
\noalign{\smallskip}
$14.0<\mathrm{log}_{10}m_{500}<14.5$ & $0.20^{+0.53}_{-0.08}$ & $0.96^{+0.14}_{-0.89}$ & $-0.94^{+0.36}_{-0.33}$ & $6.7^{+18.9}_{-5.1}$      & $0.41^{+0.67}_{-0.14}$ & $0.053^{+0.931}_{-0.020}$ & $1.45^{+0.82}_{-0.13}$ & $10.2^{+31.8}_{-9.6}$\\
\noalign{\smallskip}
$14.5<\mathrm{log}_{10}m_{500}<14.78$ & $0.32^{+0.04}_{-0.17}$ & $1.01^{+0.12}_{-0.11}$ & $-0.29^{+0.16}_{-0.87}$ & $16^{+621}_{-10}$         & $0.34^{+0.27}_{-0.12}$ & $0.078^{+0.070}_{-0.064}$ & $1.37^{+0.97}_{-0.37}$ & $1.1^{+98.9}_{-0.6}$\\
\noalign{\smallskip}
$14.78<\mathrm{log}_{10}m_{500}$ & $0.40^{+0.04}_{-0.01}$ & $0.96^{+0.25}_{-0.10}$ & $-0.20^{+0.02}_{-0.02}$ & $8.97^{+48.2}_{-4.86}$    & $0.34^{+0.40}_{-0.08}$ & $0.007^{+0.001}_{-0.001}$ & $1.00^{+0.00}_{-0.00}$ & $1.91^{+0.37}_{-0.32}$\\
\hline\hline
\multicolumn{9}{l}{$z=0.5$} \\
\hline\hline
\end{tabular}
\label{tab:Momcy6Prof}
\end{table*}

\bsp
\label{lastpage}
\end{document}